\documentclass{emulateapj}

\usepackage{graphicx}

\newcommand{\hcn}{HCN(J=1--0)}
\newcommand{\hhcn}{HCN(J=2--1)}
\newcommand{\hhhcn}{HCN(J=3--2)}
\newcommand{\hcop}{HCO$^+$(J=1--0)}
\newcommand{\hhhcop}{HCO$^+$(J=3--2)}
\newcommand{\kms}{km\,s$^{-1}$}

\begin{document}



\title{A multi-transition HCN and HCO$^+$ study of 12 nearby active
galaxies:\\ AGN versus SB environments}

\shorttitle{A multi-transition line study of active galaxies}  
\shortauthors{Krips et al.}

\author{M. Krips\altaffilmark{1}, R. Neri\altaffilmark{2},
  S. Garc\'{\i}a-Burillo\altaffilmark{3},
  S. Mart\'{\i}n\altaffilmark{1}, F. Combes\altaffilmark{4},
  J. Graci\'a-Carpio\altaffilmark{3}, and A. Eckart\altaffilmark{5}}

\altaffiltext{1}{Harvard-Smithsonian Center for Astrophysics, SMA
project, 60 Garden Street, MS 78 Cambridge, MA 02138, USA,
mkrips@cfa.harvard.edu, smartin@cfa.harvard.edu}

\altaffiltext{2}{Institut de Radio Astronomie Millim\'etrique, 
Saint Martin d'H\`eres, F-38406, France; neri@iram.fr}

\altaffiltext{3}{Observatorio Astron\'omico Nacional (OAN) -
Observatorio de Madrid, C/ Alfonso XII 3, 28014 Madrid, Spain;
s.gburillo@oan.es,j.gracia@oan.es}

\altaffiltext{4}{Observatoire de Paris, LERMA, 61 Av. de
l'Observatoire, 75014 Paris, France;
Francoise.Combes@obspm.fr}

\altaffiltext{5}{Universit\"at zu K\"oln, I.Physikalisches Institut,
Z\"ulpicher Str. 77, 50937 K\"oln, Germany; eckart@ph1.uni-koeln.de}

\begin{abstract}
Recent studies have indicated that the HCN-to-CO(J=1--0) and
HCO$^+$-to-HCN(J=1--0) ratios are significantly different between
galaxies with AGN (active galactic nucleus) and SB (starburst)
signatures. In order to study the molecular gas properties in active
galaxies and search for differences between AGN and SB environments,
we observed the \hcn, (J=2--1), (J=3--2), \hcop\, and \hhhcop\,
emission with the IRAM 30m in the centre of 12 nearby active galaxies
which either exhibit nuclear SB and/or AGN signatures. Consistent with
previous results, we find a significant difference of the
\hhcn-to-\hcn, \hhhcn-to-\hcn, \hhhcop-to-\hcop\, and HCO$^+$-to-HCN
intensity ratios between the sources dominated by an AGN and those
with an additional or pure central SB: the HCN, HCO$^+$ and
HCO$^+$-to-HCN intensity ratios tend to be higher in the galaxies of
our sample with a central SB as opposed to the pure AGN cases which
show rather low intensity ratios. Based on an LVG analysis of these
data, i.e., assuming purely collisional excitation, the (average)
molecular gas densities in the SB dominated sources of our sample seem
to be systematically higher than in the AGN sources.  The LVG analysis
seems to further support systematically higher HCN and/or lower
HCO$^+$ abundances as well as similar or higher gas temperatures in
AGN compared to the SB sources of our sample.  Also, we find that the
HCN-to-CO ratios decrease with increasing rotational number J for the
AGN while they stay mostly constant for the SB sources.

\end{abstract}

\keywords{galaxies: active --- galaxies: ISM --- radio lines: galaxies
--- galaxies: individual (NGC~1068, NGC~5194, NGC~4826, NGC~3627,
NGC~4569, NGC~6951, NGC~6946, NGC~2140, M82, NGC6240, Mrk231, Arp220)}

\section{Introduction}

Activity in galaxies can be attributed to two main phenomena, highly
active star formation, also called starburst (SB), and mass accretion
onto a supermassive black hole, often simply referred to as active
galactic nucleus (AGN).  Obviously, molecular gas plays not only a key
role as fuel in the activity process but should also, in turn, be
strongly affected by the activity. Depending on the type, degree and
evolutionary phase of the activity, different physical processes can
be involved in changing the excitation conditions and chemical layout
of the molecular gas, whether it is through strong ultra-violet (UV)
or X-ray radiation fields or kinematical processes such as galaxy
interaction, large scale shocks, gas out- or inflow.  Knowing the
composition and characteristics of the molecular gas in active
environments is thus essential for the understanding of the activity
itself, its evolution and possible differences between AGN and SB
activity. Because of the differences in the radiation fields
accompanying AGN and SB activity, AGN are suspected to create
excitation and chemical conditions for the surrounding molecular gas
significantly different from those in SB environments.  Indeed,
several recent studies, mainly based on molecular gas tracers such as
CO, HCN, and HCO$^+$(J=1--0), appear to support this hypothesis: the
HCN-to-CO(J=1--0) intensity ratios appear to be significantly higher
and the HCO$^+$-to-HCN(J=1--0) intensity ratios significantly lower in
AGN (e.g., NGC~1068, NGC~6951, M51) than in SB (e.g., M82, NGC~6946)
environments \citep[e.g.,][]{ste94,ste96,koh99,koh01,koh03,koh05}. The
difference in the intensity ratios can have various origins such as a)
systematically different gas densities, b) systematically different
gas temperatures, c) different radiation fields (UV vs.\ X-rays)
eventually yielding different HCN, HCO$^+$ and/or CO abundances
\citep[e.g.,][]{tie85,bla87,ste95,lep96,ma96}, d) shocks, e) the
evolutionary stage of the activity, particularly important for
starburst, f) additional non-collisional excitation of the gas through
IR pumping by UV/X-ray heated dust \citep[e.g.][]{gar06,weiss07}, and
g) supernova explosions (SNe), especially important for the HCO$^+$
excitation.

In each case, the thermal and chemical structures of the gas should
significantly differ between SB and AGN dominated regions. We thus
carried out IRAM 30m observations of three HCN and two HCO$^+$
transitions in 12 nearby active galaxies (see Table~\ref{tab1}) to
study the excitation conditions in SB and AGN dominated regions and
their differences mainly as function of their gas densities,
temperatures and molecular abundances.  The sources in this sample
have been selected according to the following criteria: 1.)  presence
of either SB and/or AGN activity; 2.) previously detected \hcn\,
emission; 3.)  available information on the CO emission; 4.)  a
declination above $-$20$^\circ$ so that they are observable from the
IRAM 30m telescope.

\section{Observations}
\label{obs}
\begin{figure*}[!t]
   \centering
\resizebox{5.4cm}{!}{\rotatebox{-90}{\includegraphics{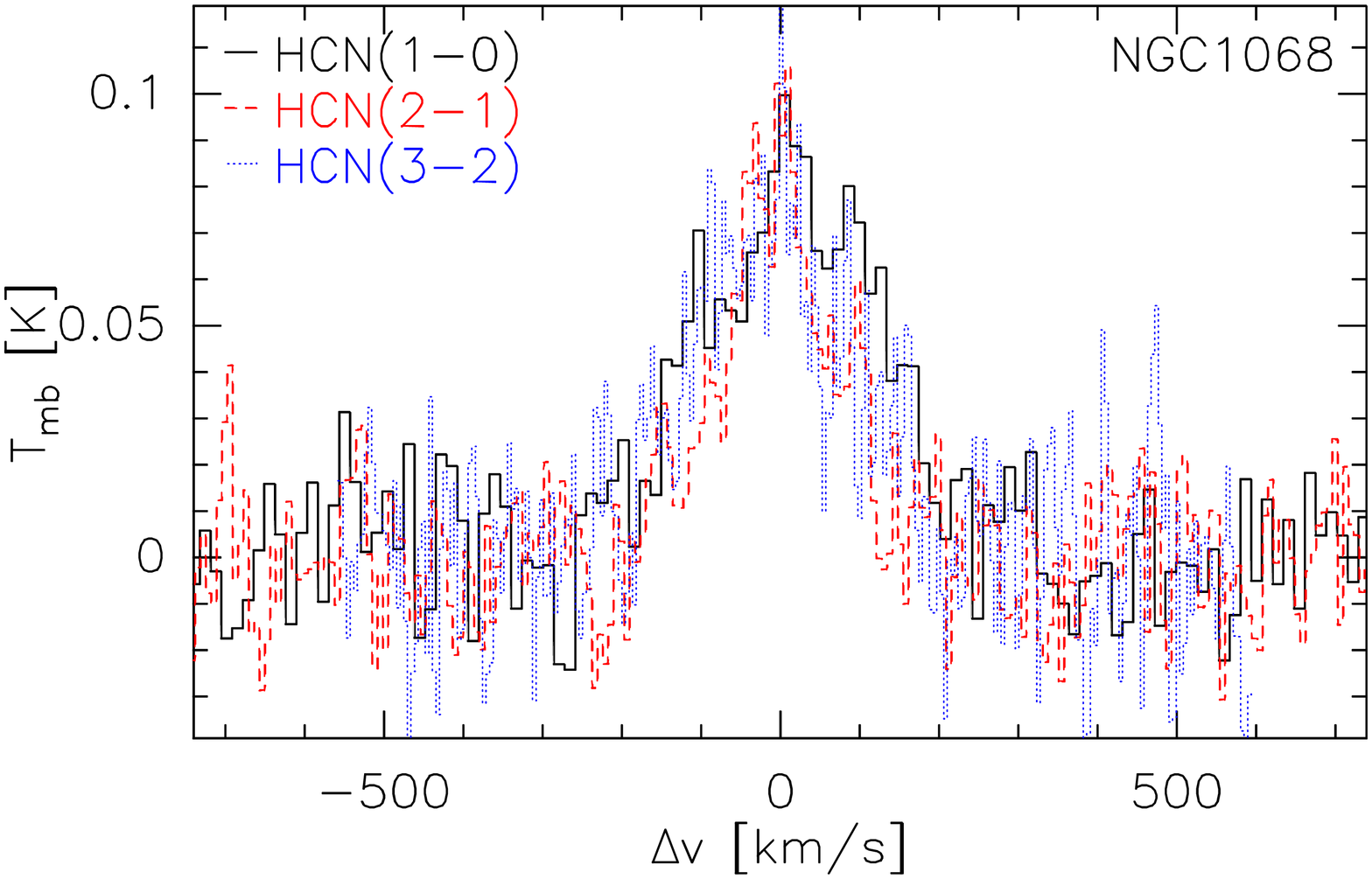}}}
\resizebox{5.4cm}{!}{\rotatebox{-90}{\includegraphics{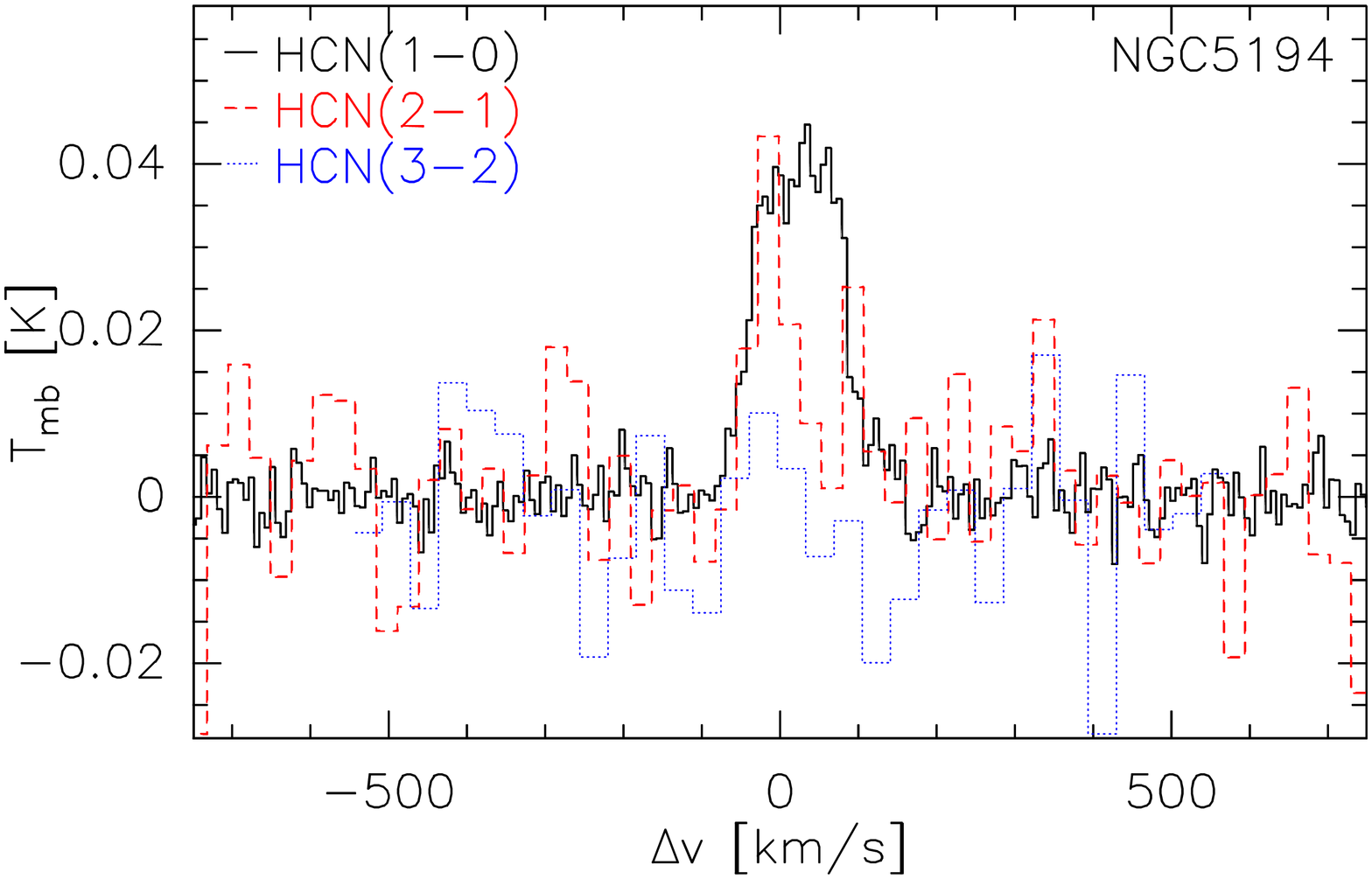}}}
\resizebox{5.4cm}{!}{\rotatebox{-90}{\includegraphics{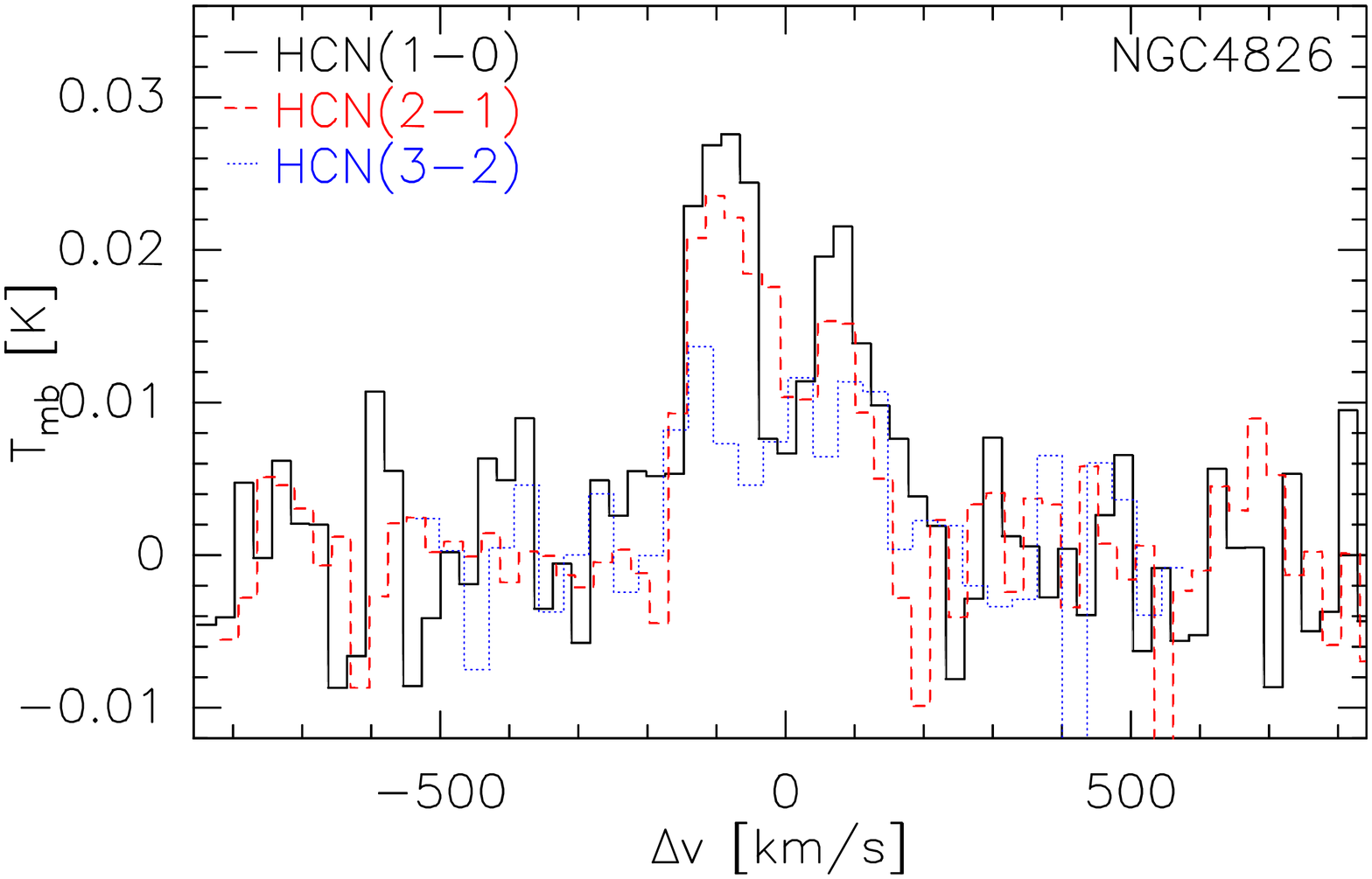}}}
\resizebox{5.3cm}{!}{\rotatebox{-90}{\includegraphics{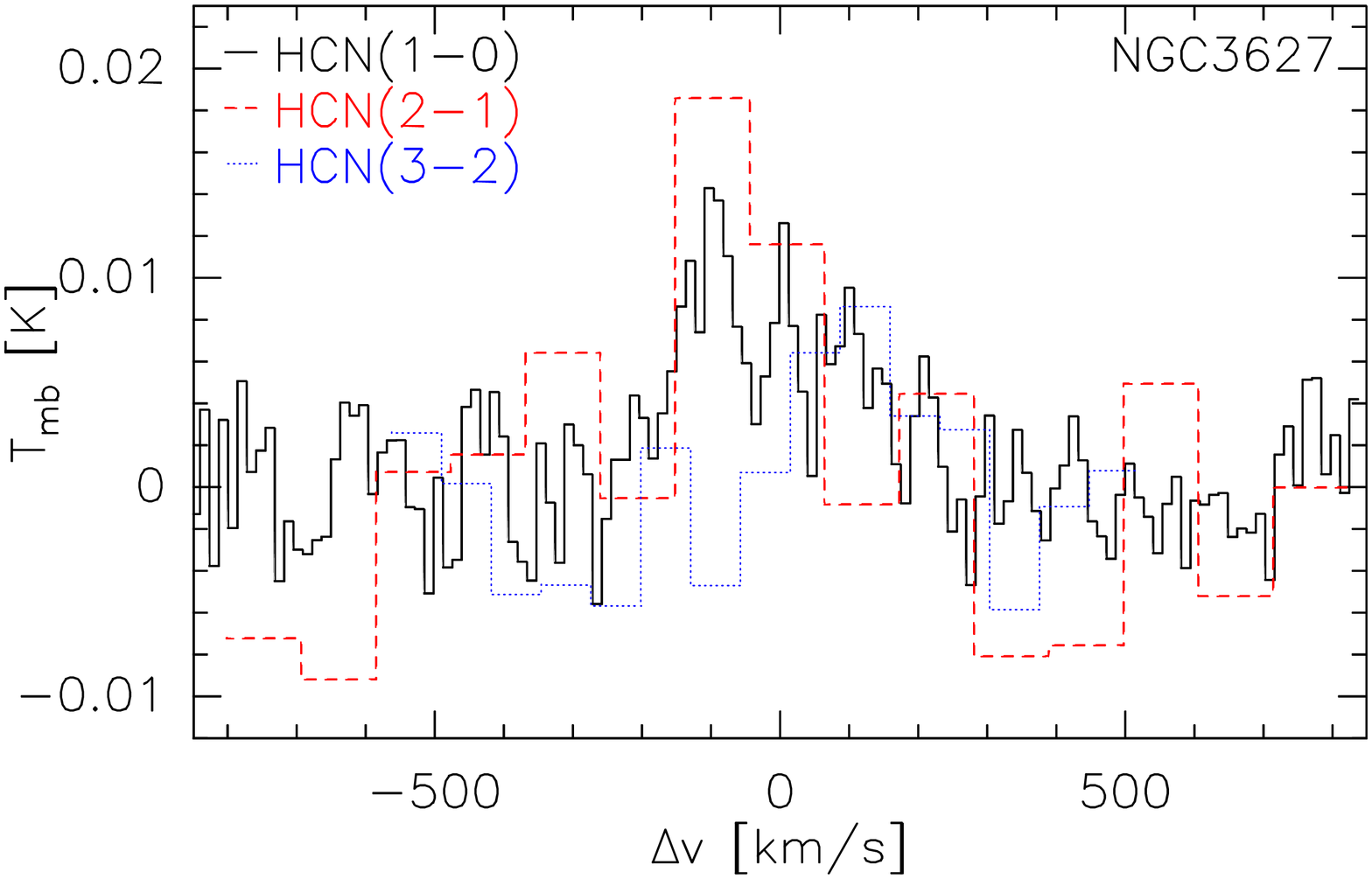}}}
\resizebox{5.4cm}{!}{\rotatebox{-90}{\includegraphics{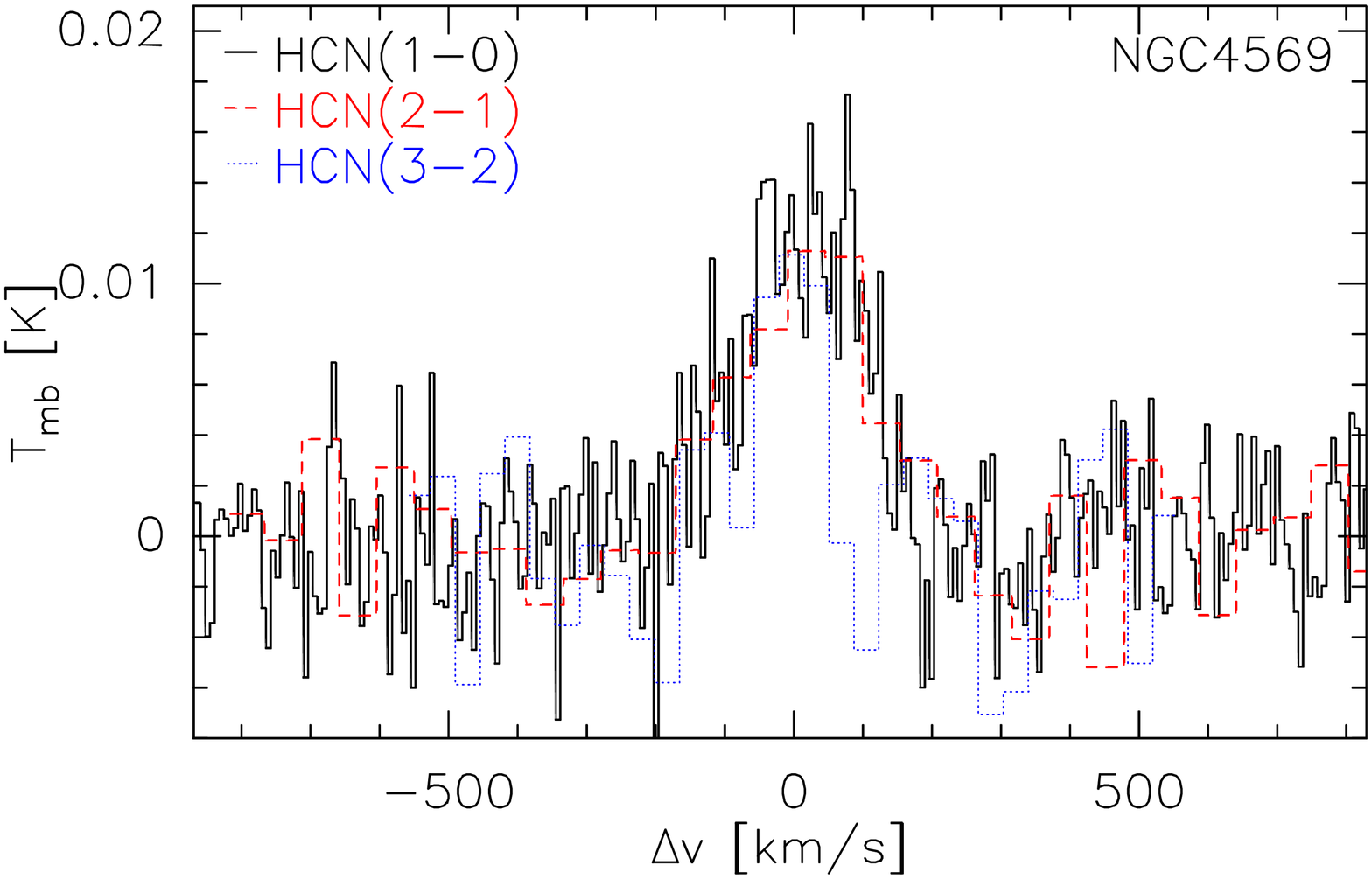}}}
\resizebox{5.4cm}{!}{\rotatebox{-90}{\includegraphics{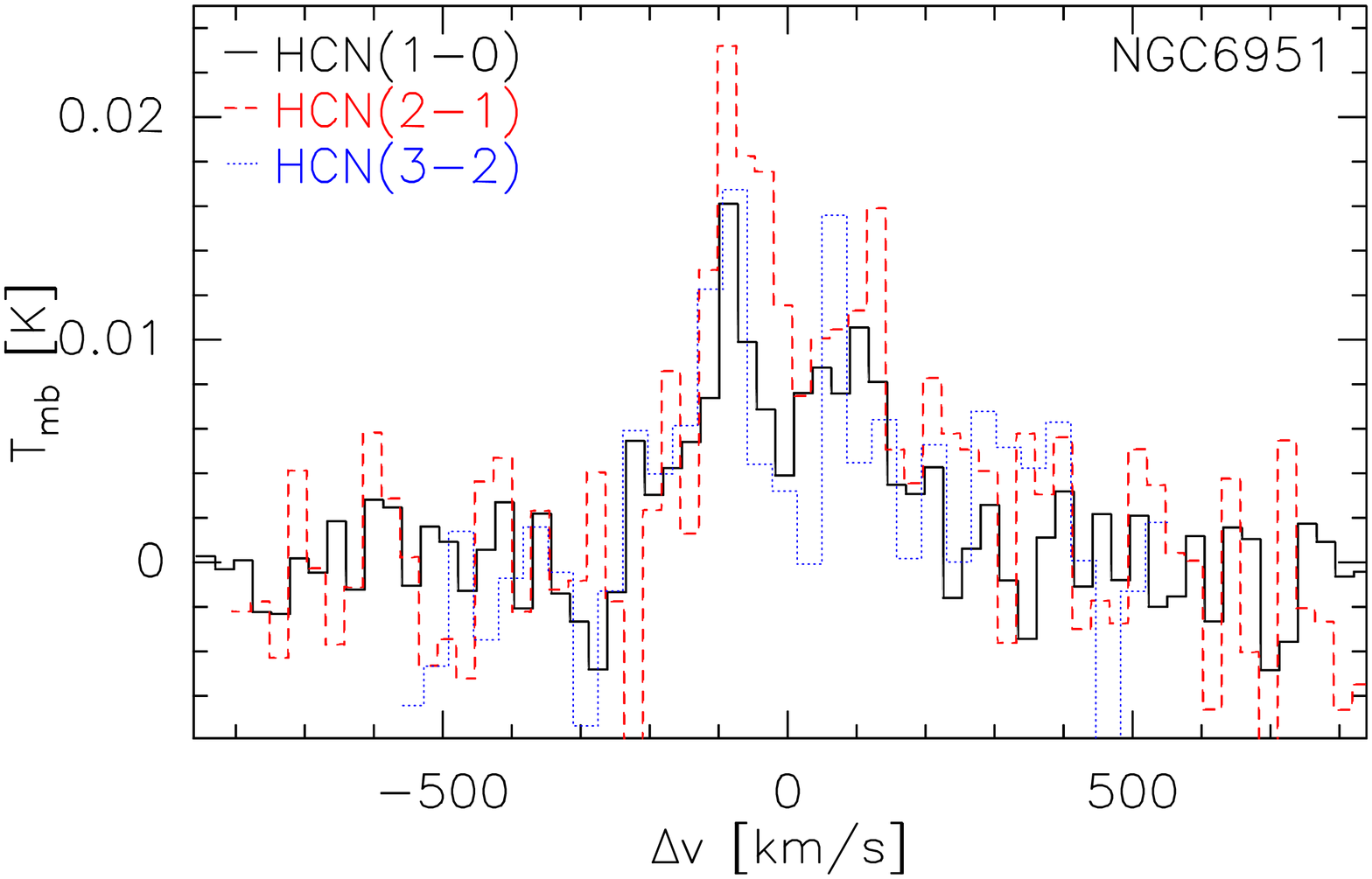}}}
\resizebox{5.4cm}{!}{\rotatebox{-90}{\includegraphics{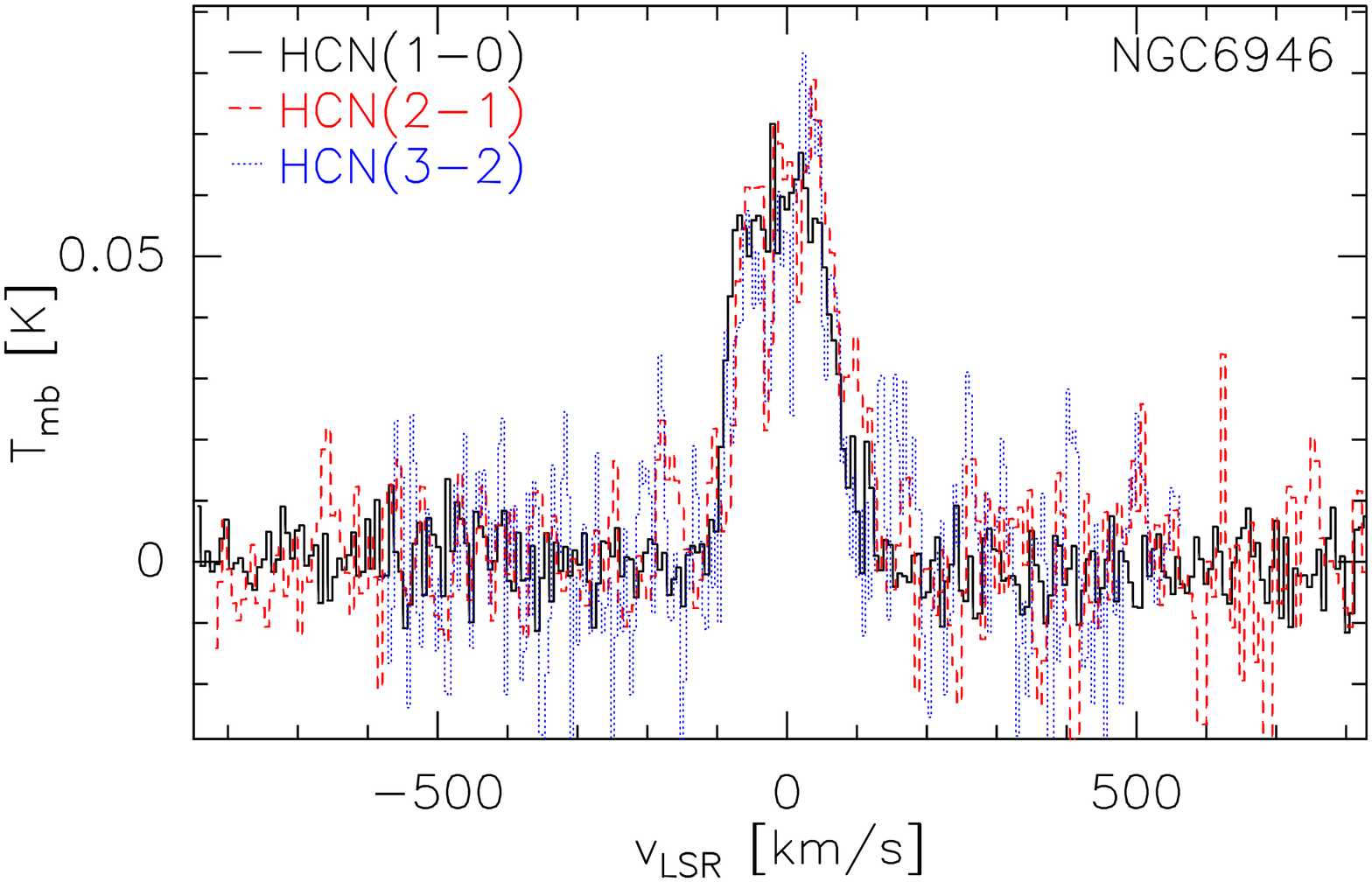}}}
\resizebox{5.4cm}{!}{\rotatebox{-90}{\includegraphics{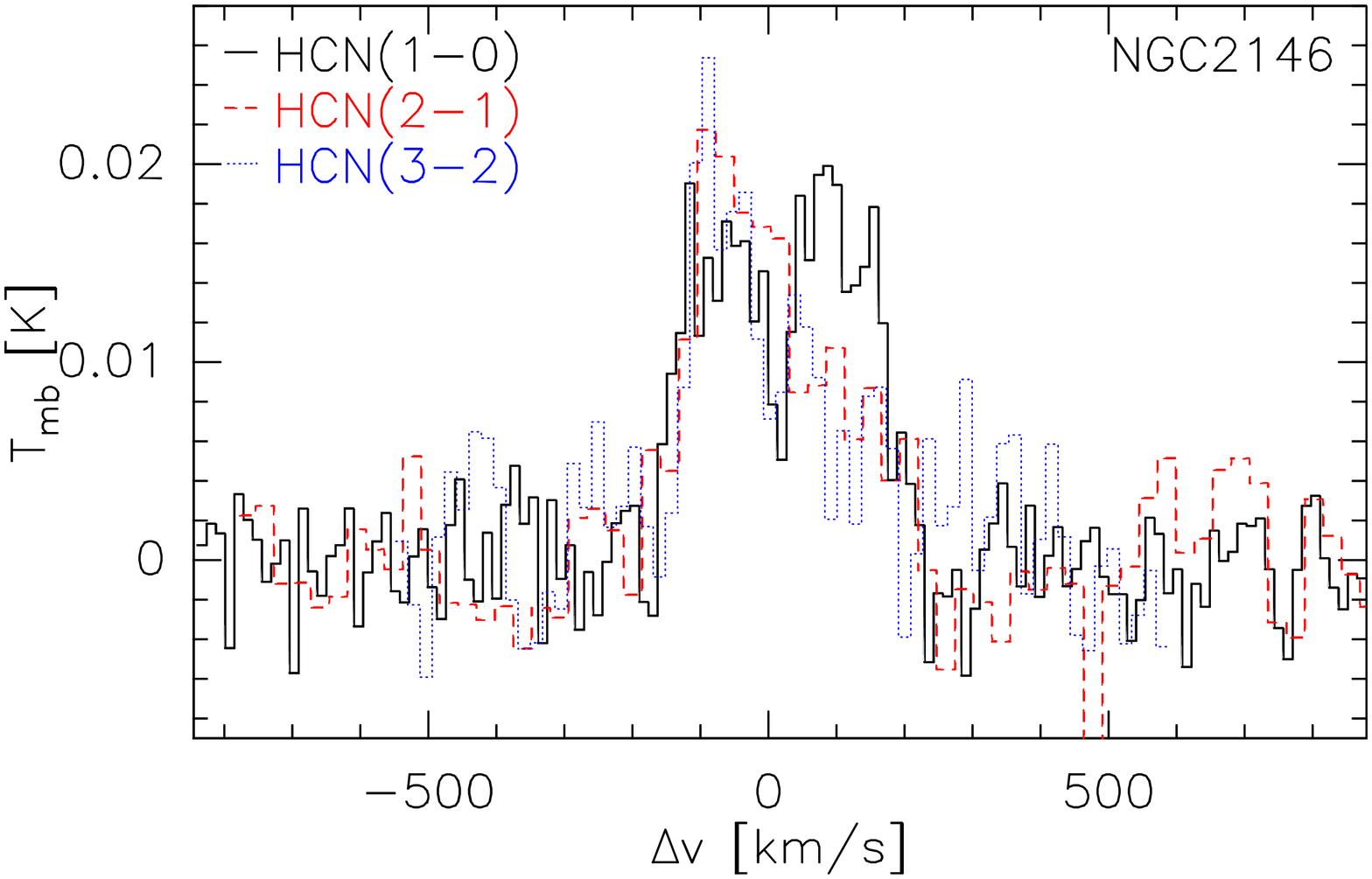}}}
\resizebox{5.4cm}{!}{\rotatebox{-90}{\includegraphics{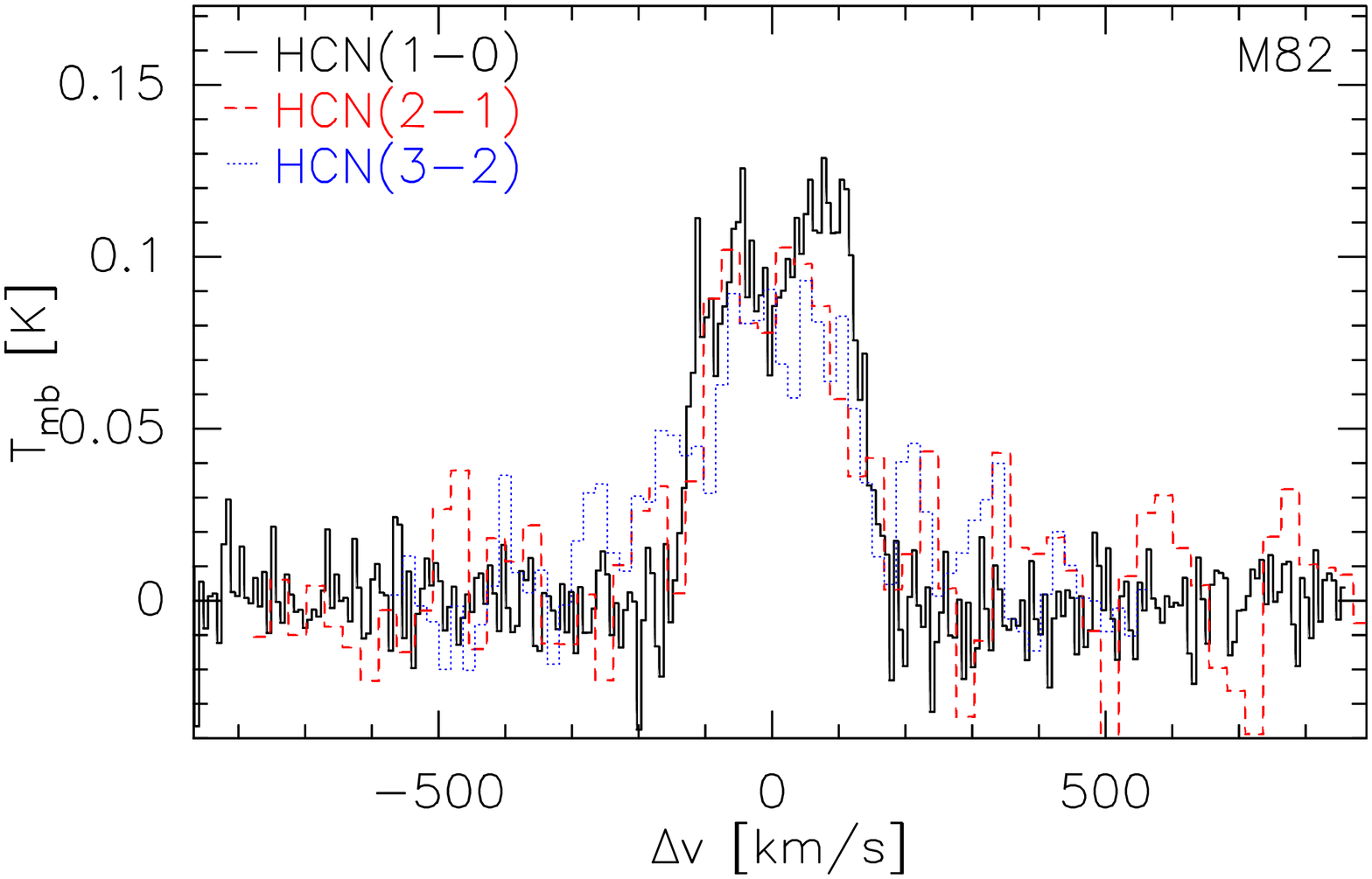}}}
\resizebox{5.4cm}{!}{\rotatebox{-90}{\includegraphics{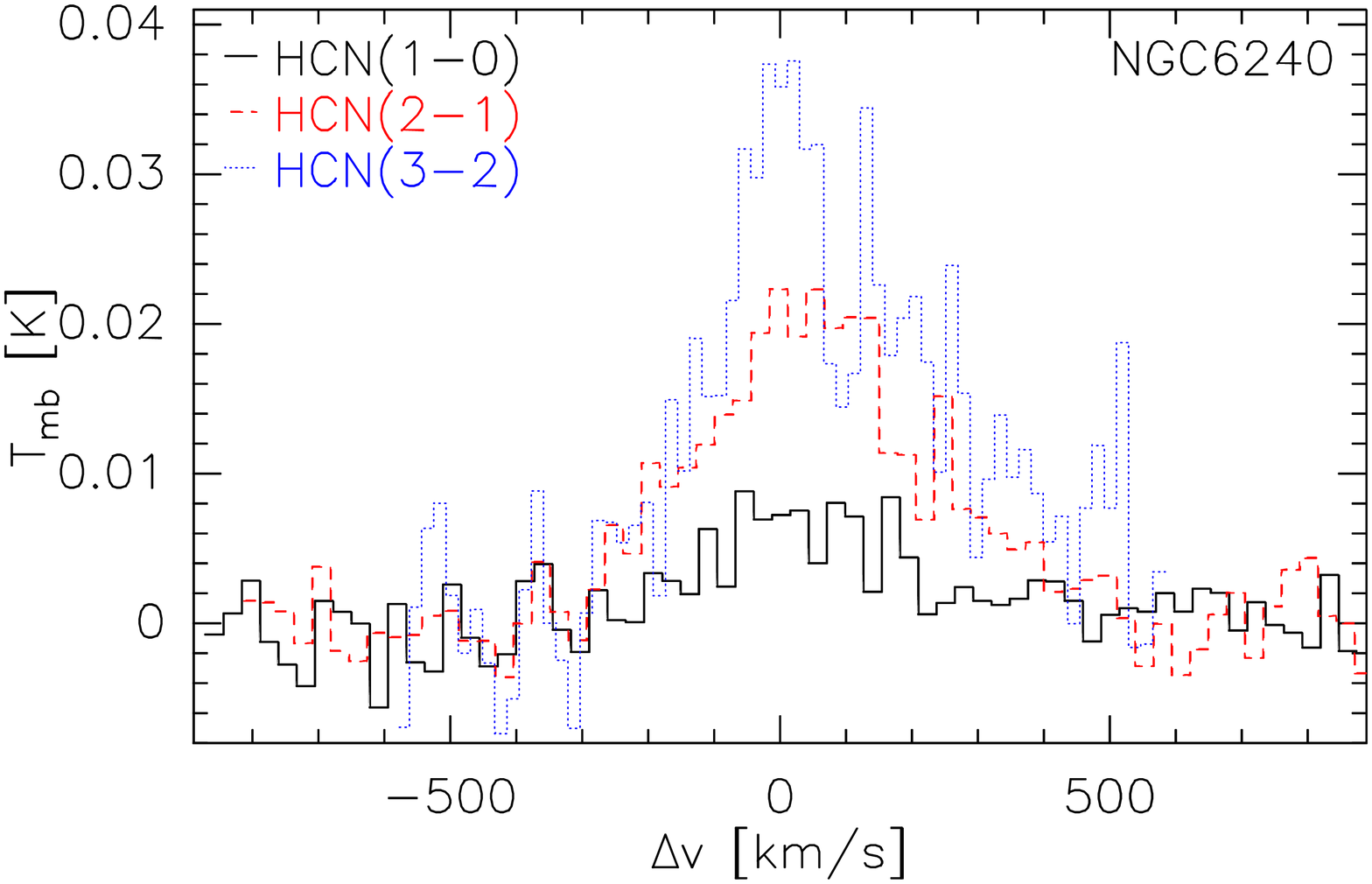}}}
\resizebox{5.4cm}{!}{\rotatebox{-90}{\includegraphics{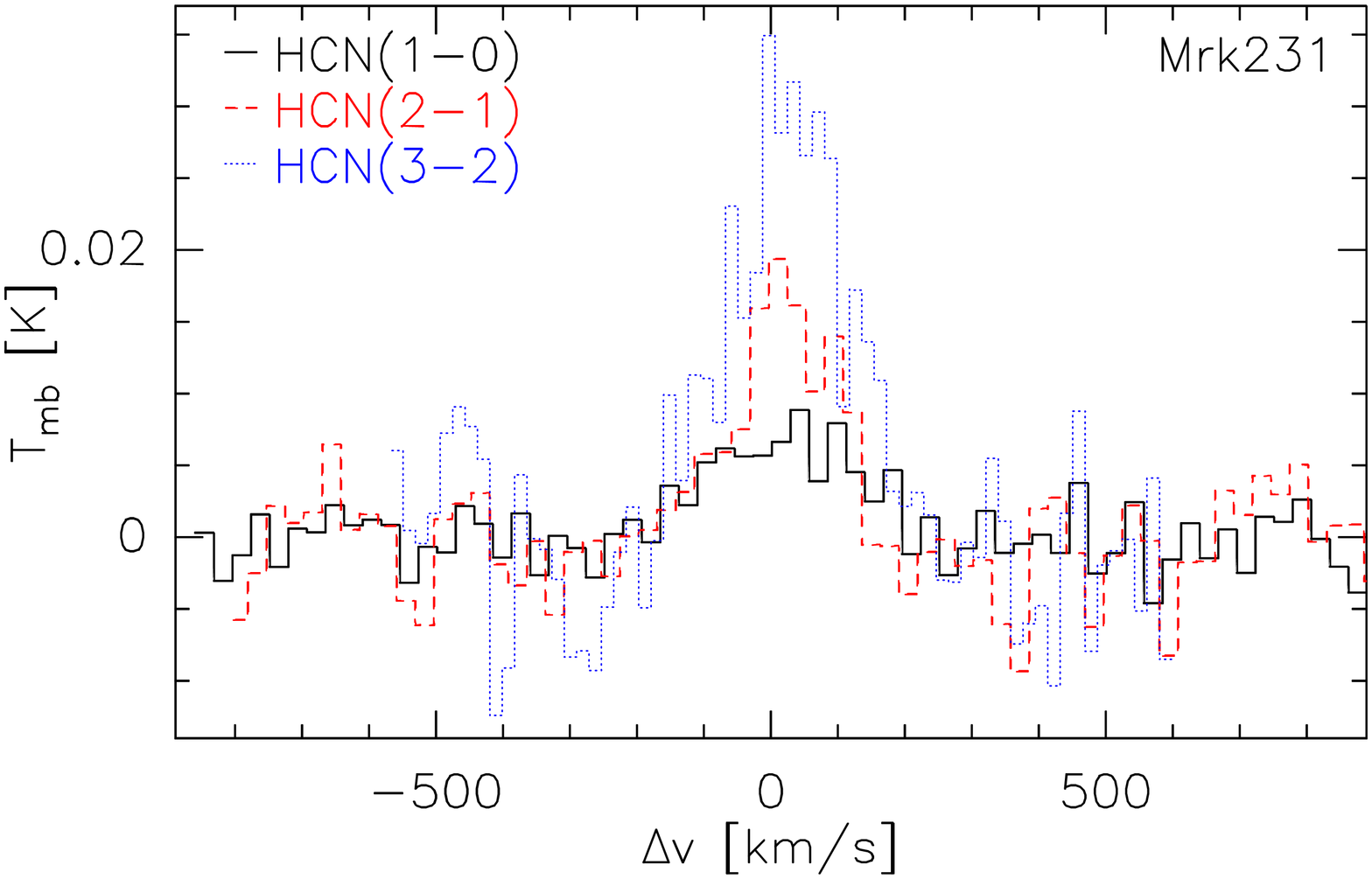}}}
\resizebox{5.4cm}{!}{\rotatebox{-90}{\includegraphics{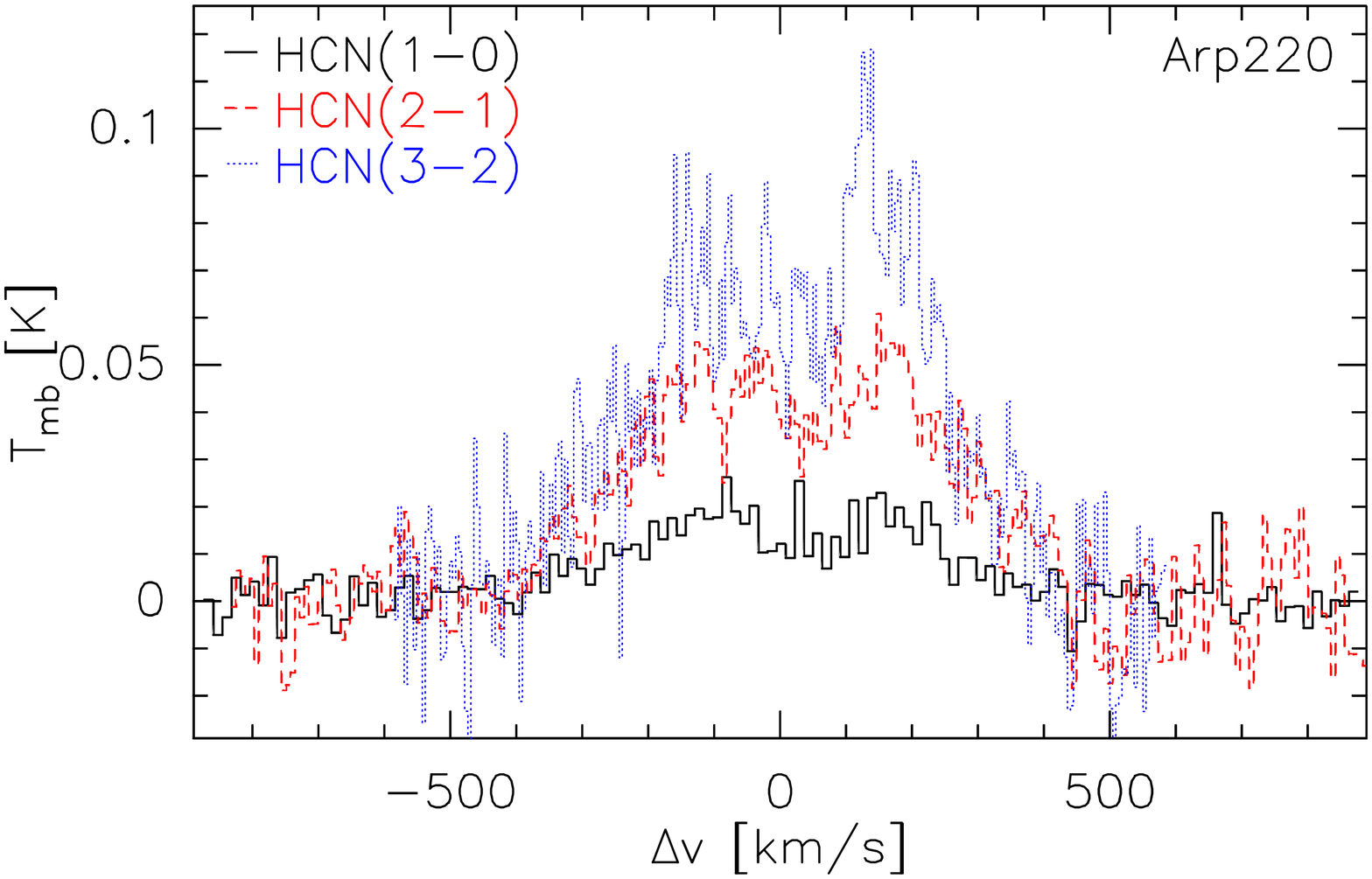}}}
\caption{Line spectra of \hcn\, ({\it solid black}), \hhcn\, ({\it
    dashed red}; please note that color figures are only available in the
    online version of the paper) and \hhhcn\, ({\it dotted blue})
  obtained at the centre of each of the 12 galaxies with the IRAM 30m
  telescope. The velocity scale is relative to the LSR velocity of the
  respective galaxy. Temperature scale is in main beam temperature
  (Kelvin) and has not been corrected for beam filling factors.}
  \label{spec}
\end{figure*}

\begin{figure*}[!t]
   \centering
\resizebox{5.4cm}{!}{\rotatebox{-90}{\includegraphics{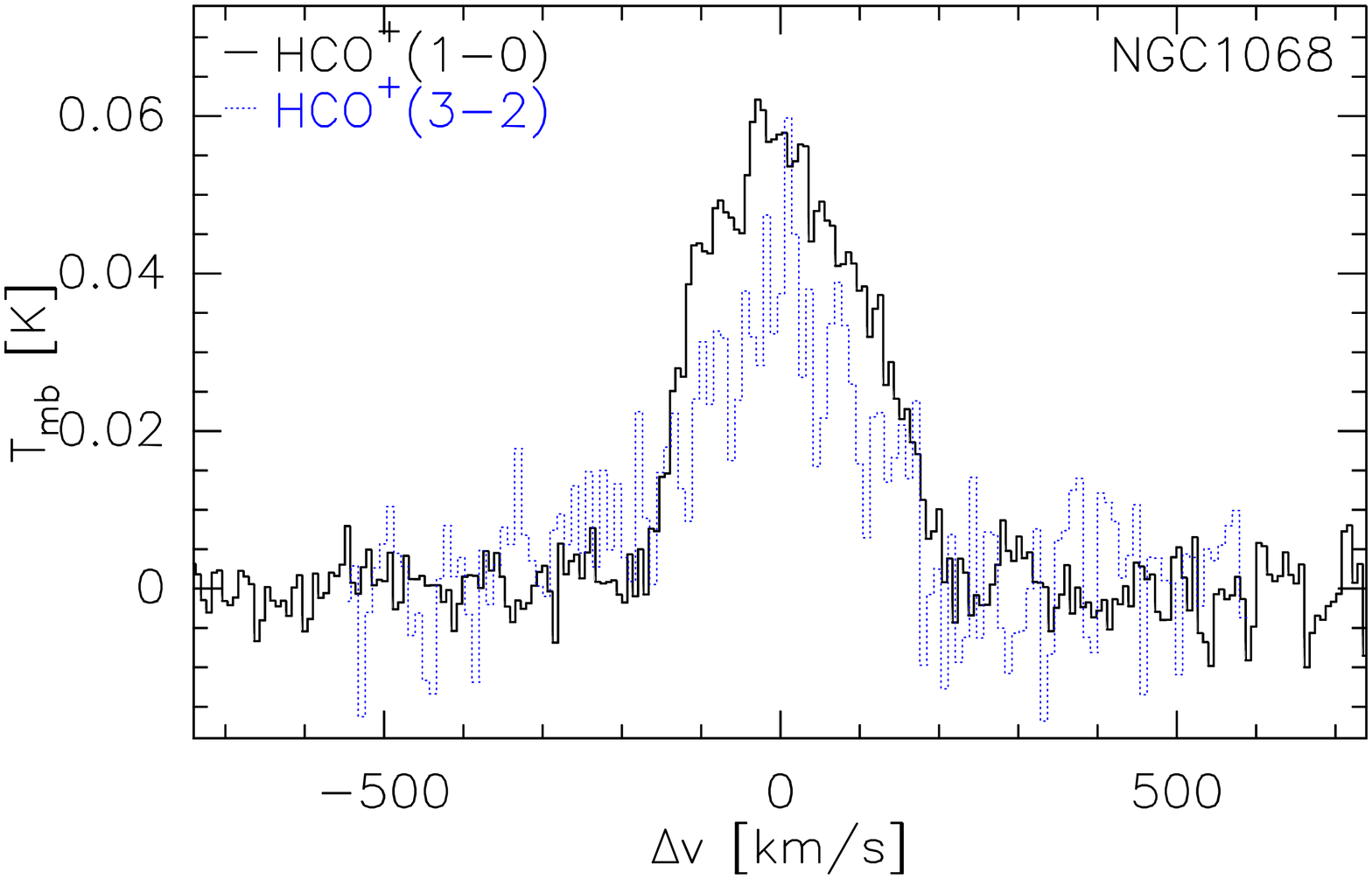}}}
\resizebox{5.4cm}{!}{\rotatebox{-90}{\includegraphics{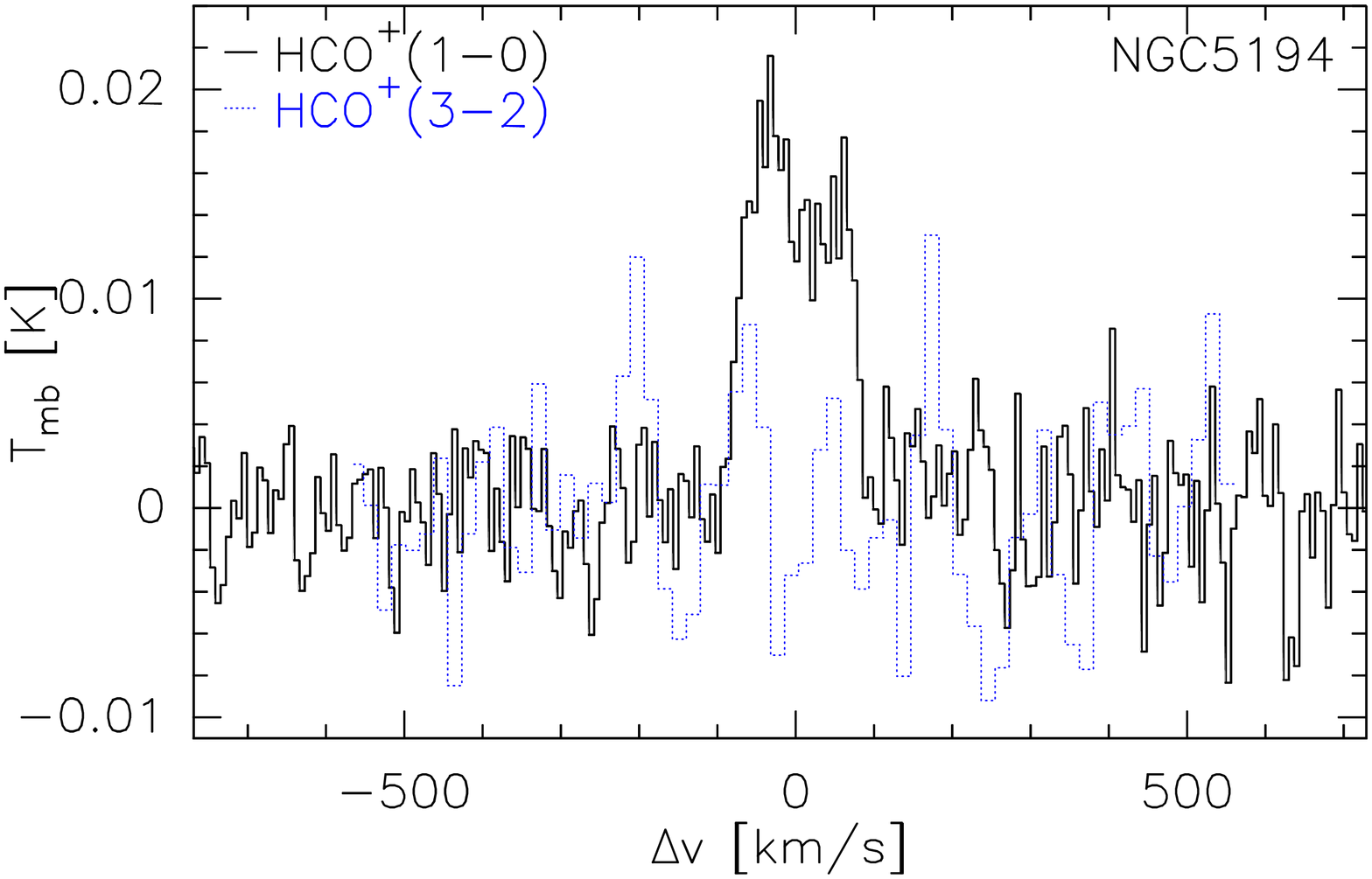}}}
\resizebox{5.4cm}{!}{\rotatebox{-90}{\includegraphics{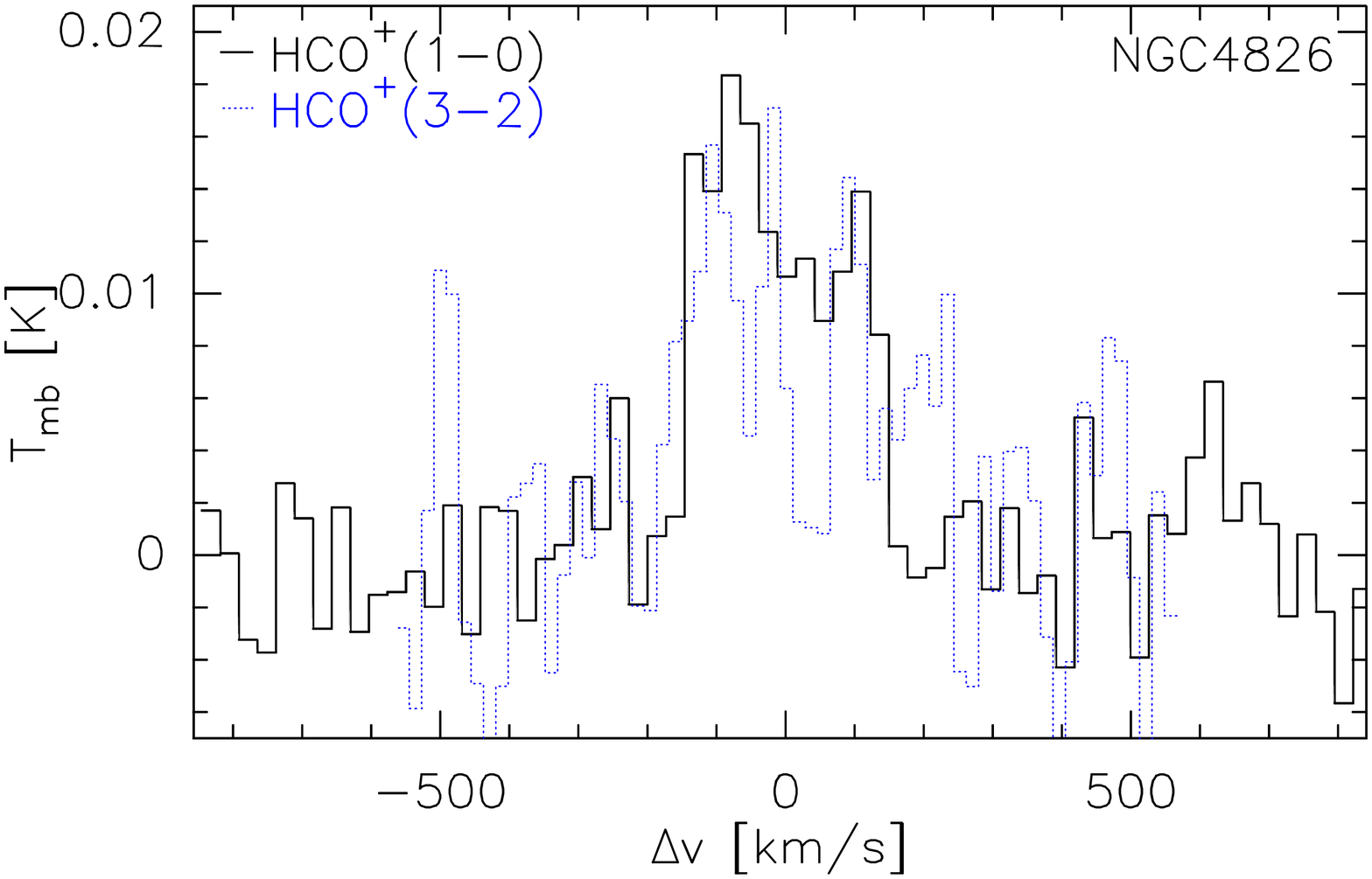}}}
\resizebox{5.3cm}{!}{\rotatebox{-90}{\includegraphics{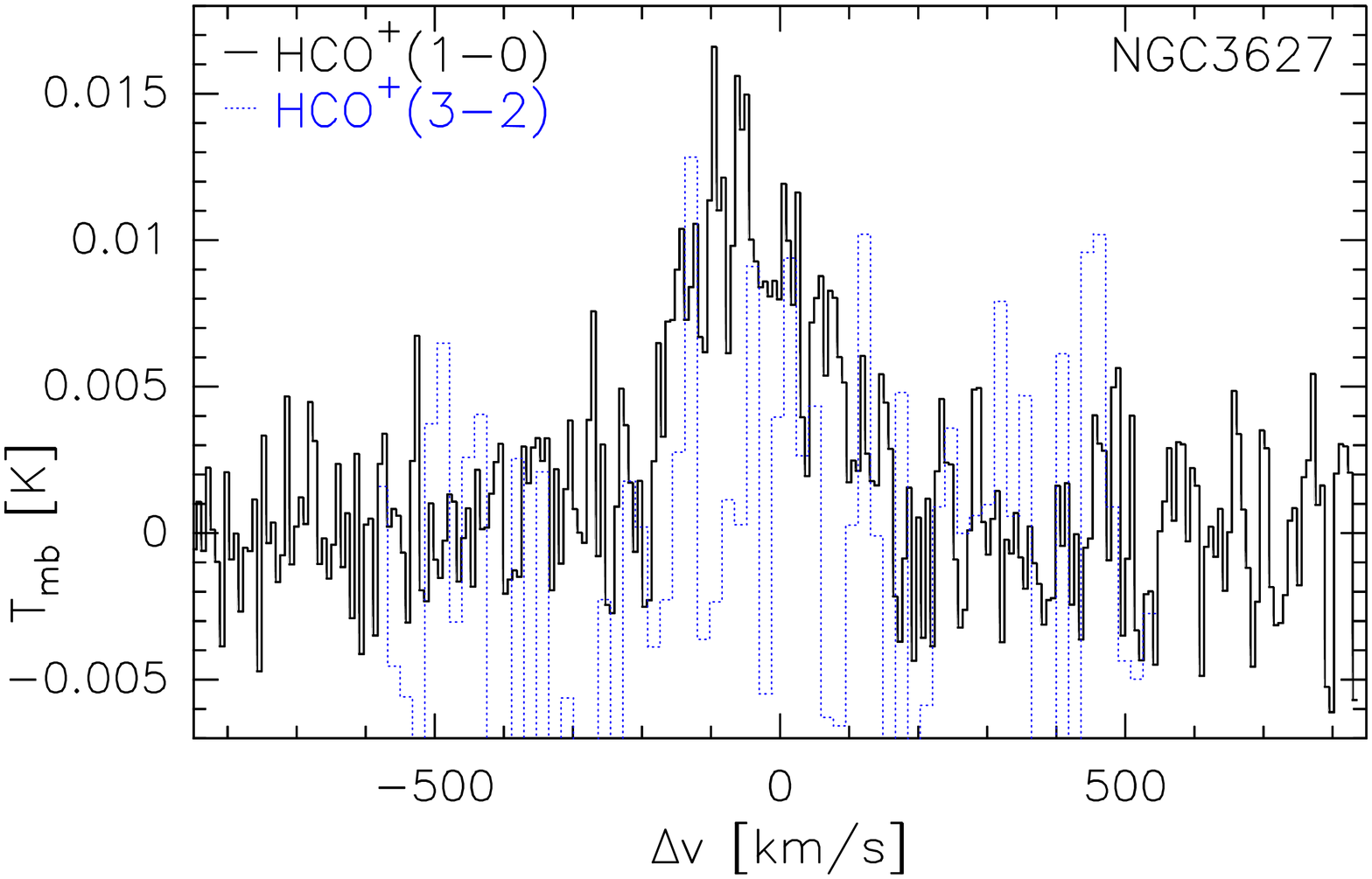}}}
\resizebox{5.4cm}{!}{\rotatebox{-90}{\includegraphics{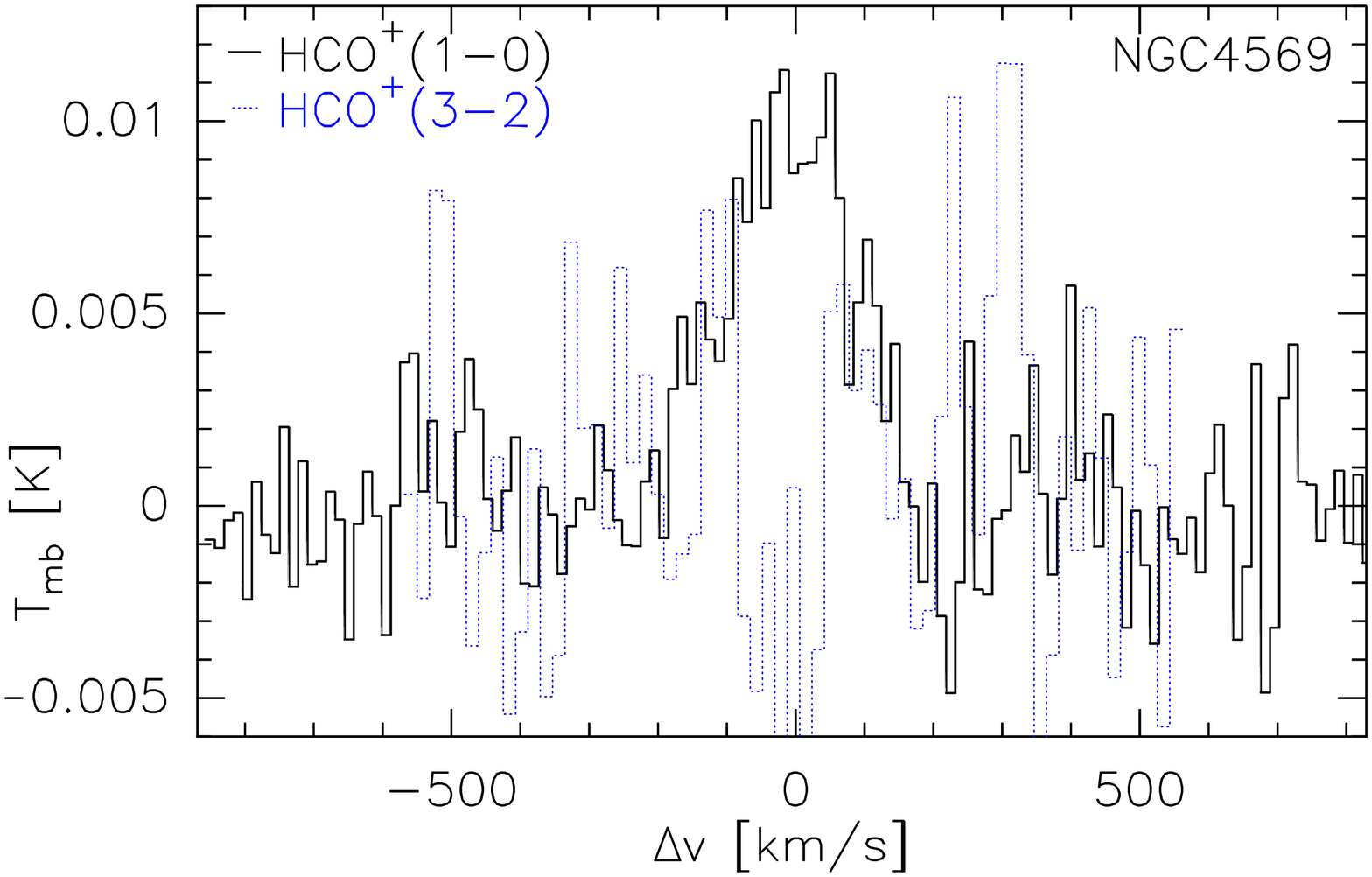}}}
\resizebox{5.4cm}{!}{\rotatebox{-90}{\includegraphics{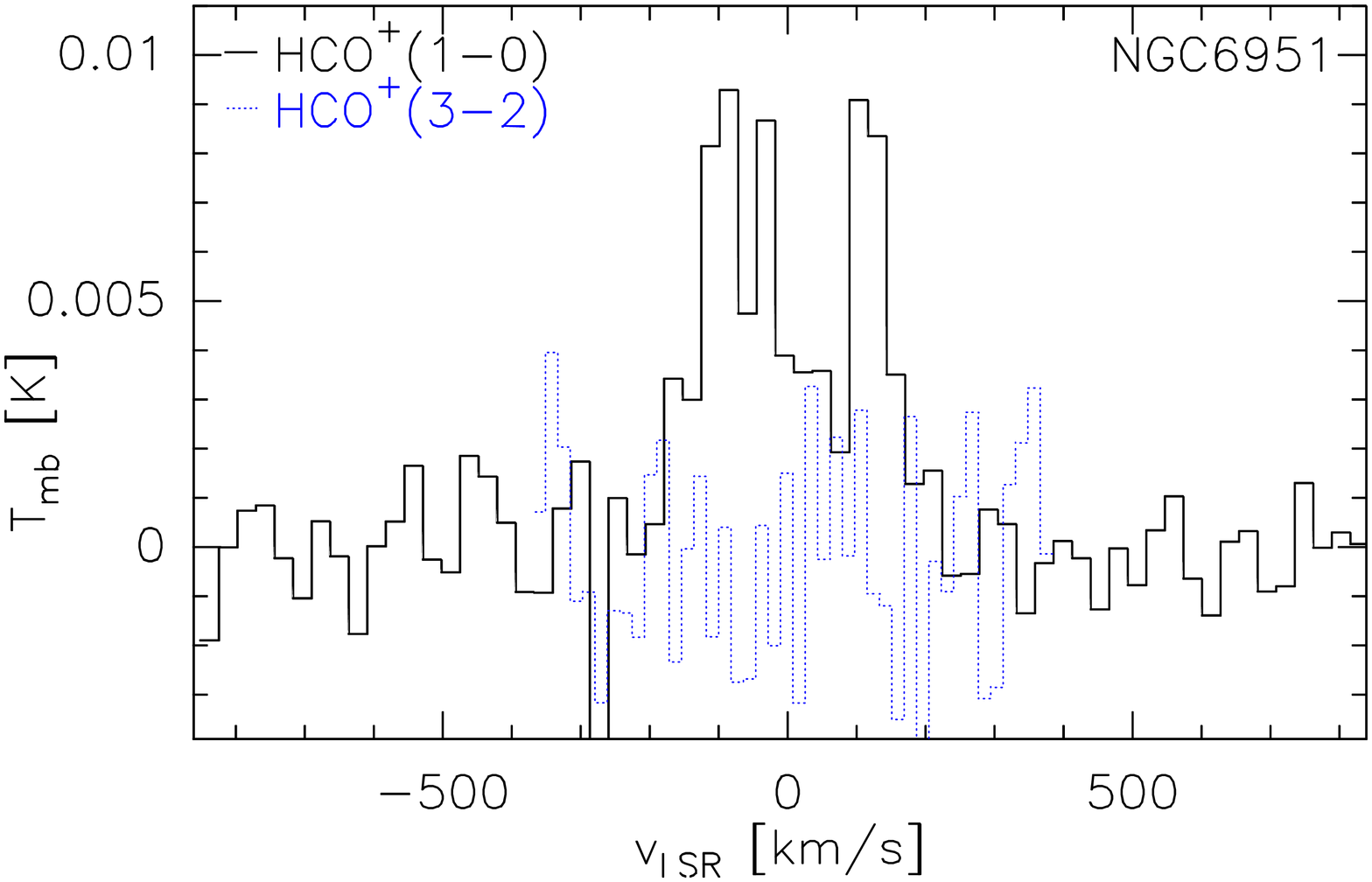}}}
\resizebox{5.4cm}{!}{\rotatebox{-90}{\includegraphics{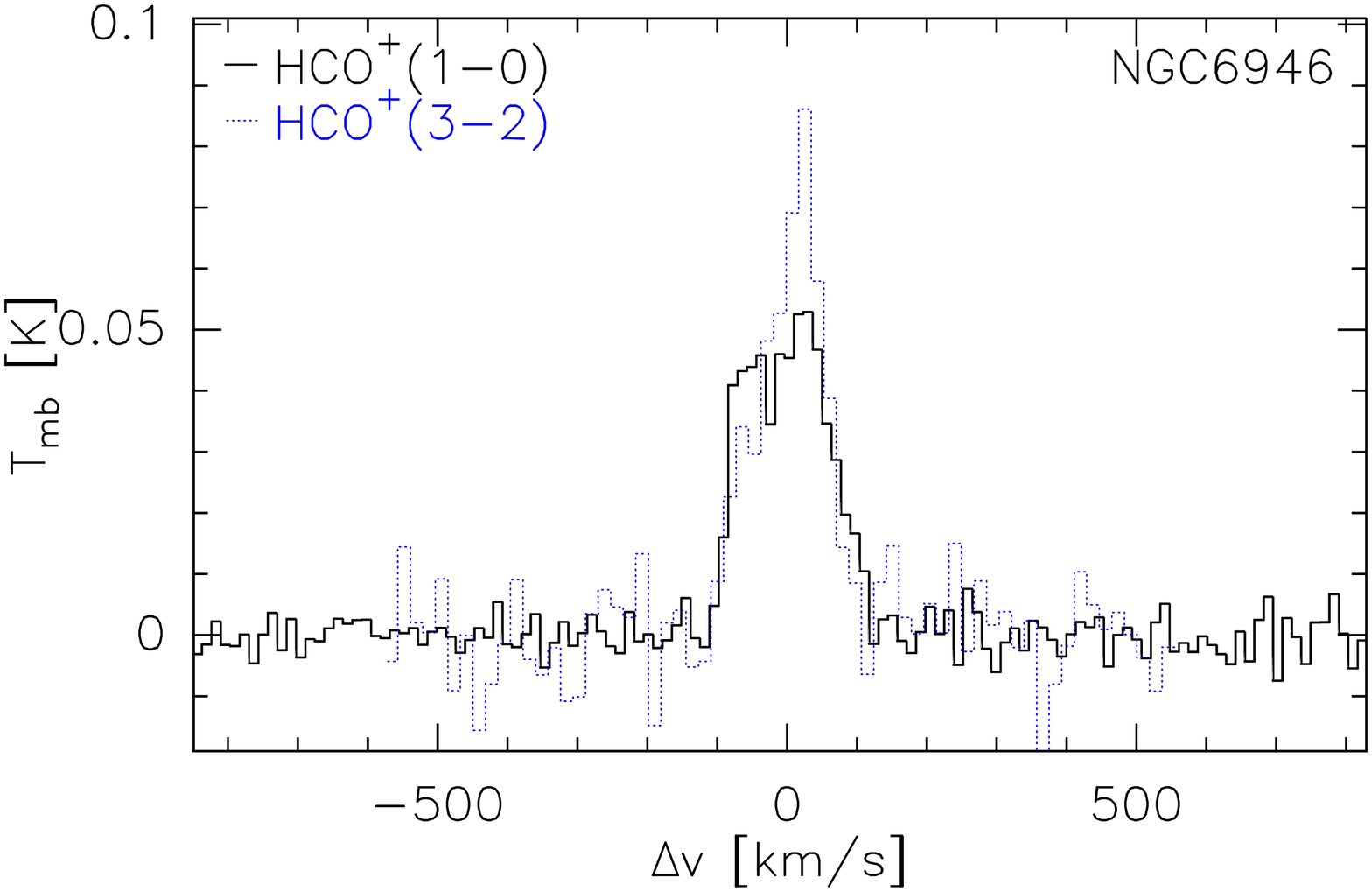}}}
\resizebox{5.4cm}{!}{\rotatebox{-90}{\includegraphics{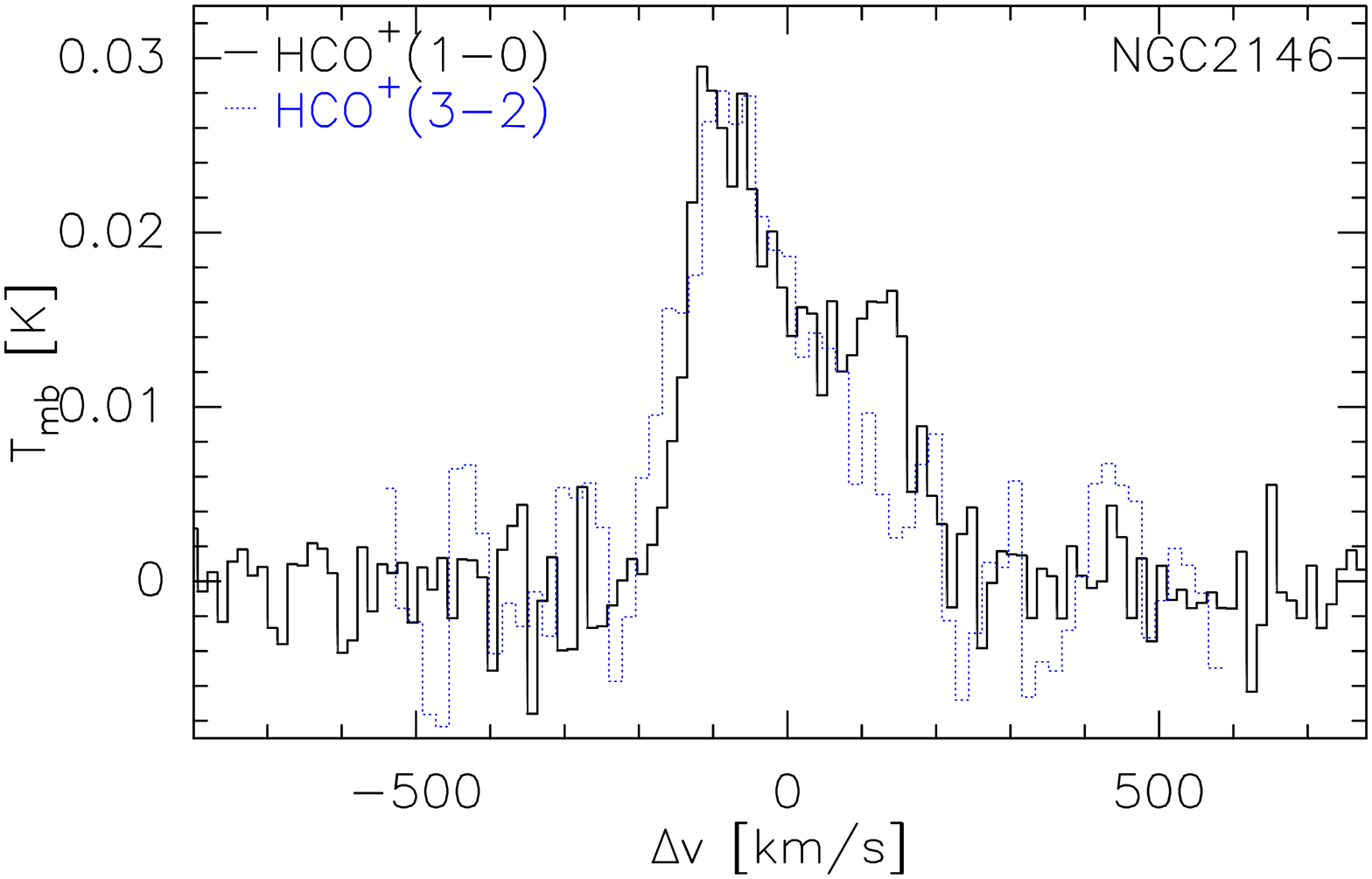}}}
\resizebox{5.4cm}{!}{\rotatebox{-90}{\includegraphics{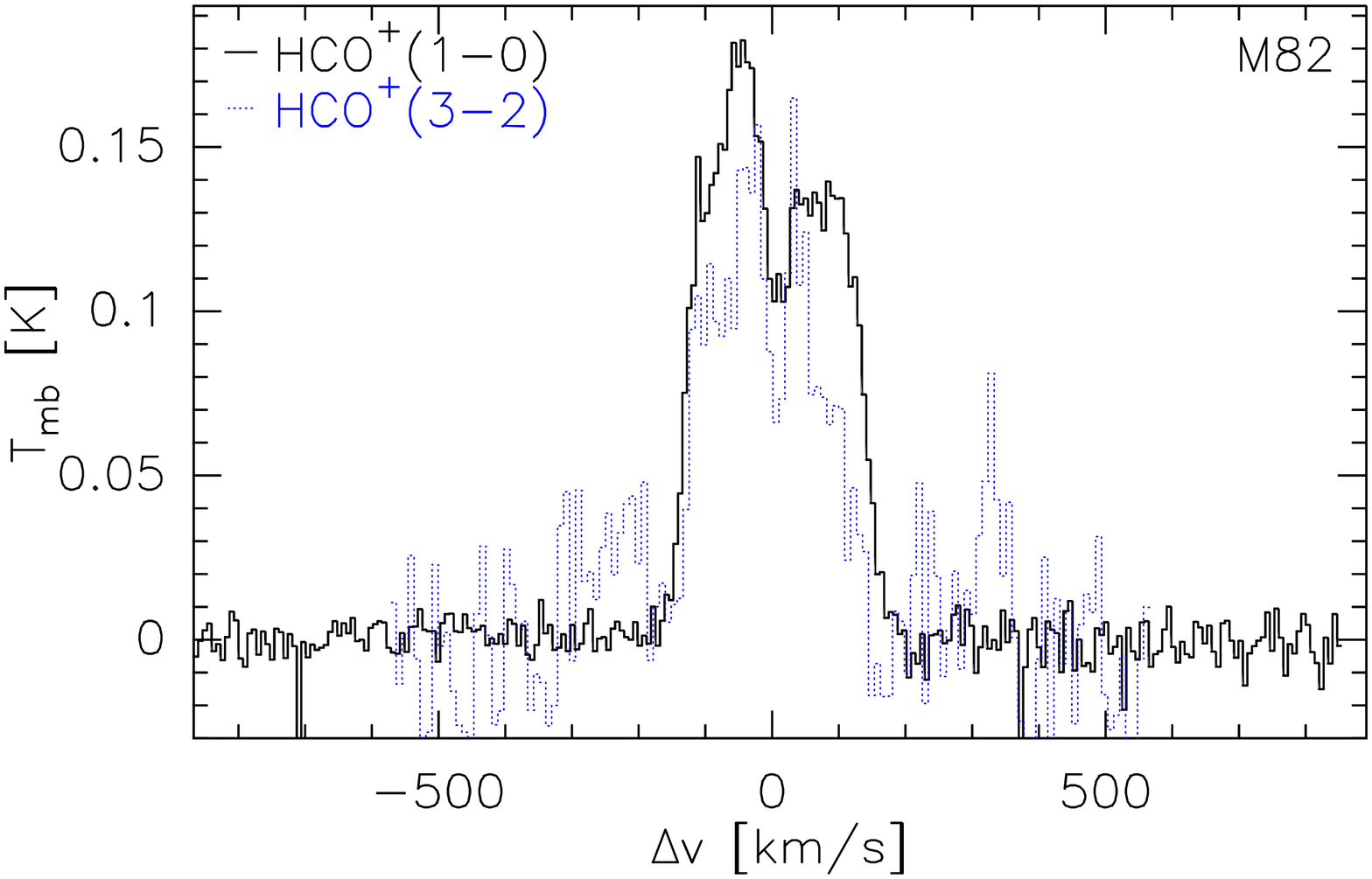}}}
\caption{Line spectra of \hcop\, ({\it solid black}) and \hhhcop\,
({\it dotted blue}) obtained at the centre in 9 galaxies of our sample
with the IRAM 30m telescope. The velocity scale corresponds to the LSR
velocity of the respective galaxy. Flux scale is in main beam
temperature (Kelvin) and has not been corrected for beam filling
factors. The HCN and HCO$^+$ intensities of NGC~6240, Mrk231 and
Arp220 were taken from Graci\'a-Carpio et al.\ (2006) and their
spectra are thus not shown here. } \label{spec2}
\end{figure*}

We observed \hcn, \hhcn, \hhhcn, \hcop\, and \hhhcop\, in the centre
($<$30$''$) of 12 nearby active galaxies\footnote{The HCO$^+$ emission
from the three ULIRGs in our sample, namely Arp~220, Mrk~231 and
NGC~6240, were taken from Graci\'a-Carpio et al.\ (2006).}
(Table~\ref{tab1}, Fig.~\ref{spec} and \ref{spec2}) with the IRAM 30m
telescope at Pico Veleta (Spain) during January (HCN) and August 2006
(HCN+HCO$^+$). In the January run, the AD set of SIS receivers was
tuned in single side band mode to the redshifted frequencies of \hcn\
at 3\,mm, \hhcn\ at 2\,mm and \hhhcn\ at 1\,mm. In the August run, the
AD set of SIS receivers were tuned in single side band mode to the
redshifted frequencies of \hcop\ at 3\,mm, \hhcn\ at 2\,mm and \hhhcn\
\& \hhhcop\ at 1\,mm.  We used the 1~MHz backends with an effective
total bandwidth of 512~MHz at 3mm and the 4~MHz backends with an
effective total bandwidth of 1024~MHz at 2\,mm and 1\,mm. We spent
$\sim$1-4~hours on each target resulting in line detections with good
to excellent signal-to-noise ratios (SNRs$\geq$5) for most of the
sources. The atmospheric opacity at 225~GHz ranged between
$\sim$0.1-0.2 in $\sim$80\% of the time in January and between
$\sim$0.2-0.3 in August 2006. The (redshifted) \hhcn\ line is still
far enough in frequency from the atmospheric water absorption line at
183.3~GHz to be detectable without any contamination. Unfortunately,
the same water absorption line prevents a reliable observation of the
HCO$^+$(J=2--1) transition,which is $\sim$1~GHz closer to the water
line. Also, the 2\,mm receiver performance in the 177.5-183 GHz window
is severely reduced, yielding untolerably high system temperatures
($>$1500~K). We regularly checked the pointing on a nearby planet
and/or bright quasar resulting in a pointing accuracy within a few
arcseconds (i.e., $\sim$2-4$''$).

\section{Data analysis}
\label{ana}
Throughout the paper, the temperature scale is equivalent to T$_{\rm
mb}$, i.e., main beam brightness temperature, which is defined as
T$_{\rm mb}$=T$_{\rm a}^\star$$\cdot$F$_{\rm eff}$/B$_{\rm eff}$. The
beam ($\equiv$B$_{\rm eff}$) and forward efficiencies ($\equiv$F$_{\rm
eff}$) together with the beam sizes are given in Table~\ref{tab0}. As
beam filling effects are a crucial point for the analysis of our data,
especially since we are mainly interested in the intensity ratios
between different line transitions, all derived intensity ratios have
been very carefully corrected using beam filling factors (see also
Table~\ref{tab2}). We use the following definitions and relations:
\begin{eqnarray}
\begin{array}{cc}
\rm R^{mol}_{J_u,J_l/10} = \frac{f_{J_u,J_l}^{-1}\cdot I^{\rm
mol}_{J_u,J_l}} {f_{10}^{-1}\cdot I^{mol}_{10}}, & \hspace*{0.3cm}
J_u=3,2;\,\,\, J_l=J_u-1;\,\,\, \\ & \rm mol=HCN~or~HCO^+\\
\end{array}
\end{eqnarray}
with R$^{\rm mol}_{\rm J_u,J_l/10}$ being the line intensity ratio
between the same molecule (i.e., HCN or HCO$^+$) at transition
(J=J$_u$$\rightarrow$J$_l$) and transition (J=1$\rightarrow$0),
I$^{\rm mol}_{J_u,J_l}$ the intensity (see Equation~\ref{int}) of the
molecule at transition (J=J$_u$$\rightarrow$J$_l$) and f$_{\rm
J_u,J_l}$ the filling factor (see Equation~\ref{fillfa}). The
HCO$^+$-to-HCN intensity ratio at transition
(J=J$_u$$\rightarrow$J$_l$) is defined as:
\begin{eqnarray}
\begin{array}{cc}
\rm R^{HCO^+/HCN}_{J_u,J_l} = \frac{I^{HCO^+}_{J_u,J_l}}{I^{HCN}_{J_u,J_l}},
& \hspace*{0.3cm} J_u=3,1;~J_l=J_u-1; \\ 
& \rm (f^{HCN}_{J_u,J_l}=f^{HCO^+}_{J_u,J_l})
\end{array}
\end{eqnarray}
%
%
%
I$^{\rm mol}_{J_u,J_l}$ is the intensity of the molecular line in
K~\kms\, at transition (J=J$_u$$\rightarrow$J$_l$) and defined as
\begin{eqnarray}
\begin{array}{cc}
I^{\rm mol}_{\rm J_u,J_l} = \int T_{\rm mb}^{\rm mol(J=J_u - J_l)}dv,
& \hspace*{0.3cm} J_u=3,2,1;\,\,\, \\ & \rm J_l=J_u-1;\,\,\,\\ & {\rm
mol}={\rm HCN~or~HCO^+}
\end{array}
\label{int}
\end{eqnarray}
f$_{\rm J_u,J_l}$ is the beam filling factor and defined as
\begin{eqnarray}
\begin{array}{cc}
{\rm f}_{\rm J_u,J_l}=\left(
\frac{(\theta_s^{J_u,J_l})^2}
{(\theta_s^{J_u,J_l})^2+(\theta_b^{J_u,J_l})^2}\right)^a, &
\hspace*{0.3cm} J_u=3,2,1;\,\,\, \\  
& \rm J_l=J_u-1
\end{array}
\label{fillfa}
\end{eqnarray}
with $\theta_s^{\rm J_u,J_l}$$\equiv$size of the HCN(J=$\rm J_u -
J_l$) emission region ($\simeq$size of the HCO$^+$(J=$\rm J_u - J_l$)
emission region), $\theta_b^{\rm J_u,J_u-1}$$\equiv$ beamsize at the
HCN(J=$\rm J_u - J_l$) ($\simeq$HCO$^+$(J=$\rm J_u - J_l$)) frequency
(see Table~\ref{tab0}); $a$ is either 1 in case of a circular source
or 0.5 in case of an elliptical source, i.e., one that fills the beam
in one direction (in the elliptical case we thus assume the minor axis
of the emission as estimate for $\theta_s$ to derive the beam filling
factor while we use the Full Width at Half Maximum (FWHM) in the
circular case). 

Some of the more distant sources in our sample are significantly
smaller ($\leq$5$''$) than the beamsizes resulting in smaller filling
factors than for the more nearby galaxies (see Table~\ref{tab2}). The
size of the HCN and HCO$^+$ emission region for each galaxy has been
estimated from interferometric and/or single-dish HCN and/or HCO$^+$
maps where present or as a conservative upper limit from CO
maps. Uncertainties of the order of 1-5$''$ in the assumed source
sizes translate into a $\leq$20\% uncertainty in the intensity
ratios. For most of the sources, we assume that the size of the HCN
(HCO$^+$) emission region is similar in all three (two) transitions,
i.e., $\theta_s^{\rm 10}\approx\theta_s^{\rm 21}\approx\theta_s^{\rm
  32}$. However, if this assumption were invalid, i.e., the size of
the emission in the higher transitions is actually smaller by up to a
factor of 50\% than in the ground state, the estimated filling factor
ratios between the different transitions could be too high by up to a
factor of 4. Thus, in the case of NGC~1068, we explicitly take into
account that the \hhhcn, \hhcn\, and \hhhcop\, emission are more
compact (by a factor of $\sim$1.5) than the \hcn\, and \hcop\,
emission as indicated by recent SMA observation of the \hhhcn\ and
\hhhcop\, emission in NGC~1068 (Krips et al.\ in prep.). As a
conservative approach, we also use a lower size of the \hhhcn, \hhcn\,
and \hhhcop\, line emission for NGC~5194 whose interferometric maps
also indicate a decreasing size of the emitting region with increasing
rotational number \citep[Table~\ref{tab2}][]{mat05}. If we did not
account for these differences in source sizes at different
transitions, we might underestimate the line ratios for the AGN
sources. Consequently, the so determined intensity ratios can be
regarded as conservative upper limits for these two cases. 

As we observed a large on the fly map ($\sim$2$'$) for M82 in HCN and
HCO$^+$, we were able to average the emission in all line transitions
over the same region resulting in identical filling factors. We thus
assume a filling factor of 1 for M82.

\section{Results}
\label{res}

\subsection{The sample}
\begin{figure}[!t]
\epsscale{0.9} \rotatebox{-90}{\plotone{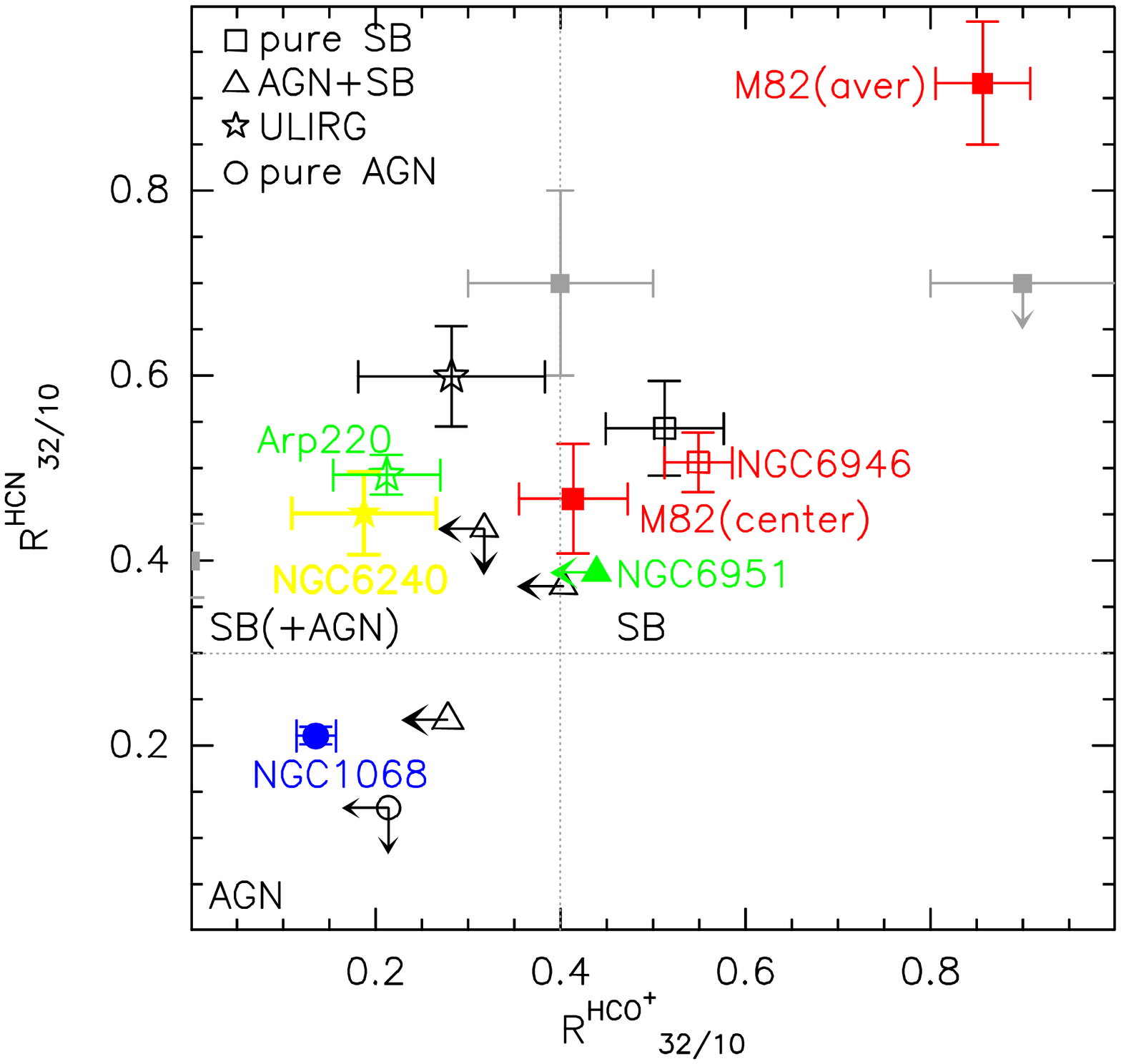}}
\caption{HCO$^+$(J=3--2)-to-HCO$^+$(J=1--0) and
HCN(J=3--2)-to-HCN(J=1--0) intensity ratios of all 12 sources (this
paper, Table~\ref{tab3}) and NGC~253 and NGC~4945 ({\it filled grey
box}; taken from the literature, Table~\ref{tab3}). NGC~1068 ({\it
filled blue circle}), NGC~6951 ({\it filled green triangle}), Arp220
({\it open green triangle}), NGC~6240 ({\it filled yellow star}) and
M82 ({\it filled red box}) are highlighted. The dotted grey lines
should guide the readers eyes and indicate the putative different
locations of the SB and AGN dominated sources in this diagram.}
\label{ratios2}
\end{figure}

\begin{figure}[!t]
\epsscale{0.9} \rotatebox{-90}{\plotone{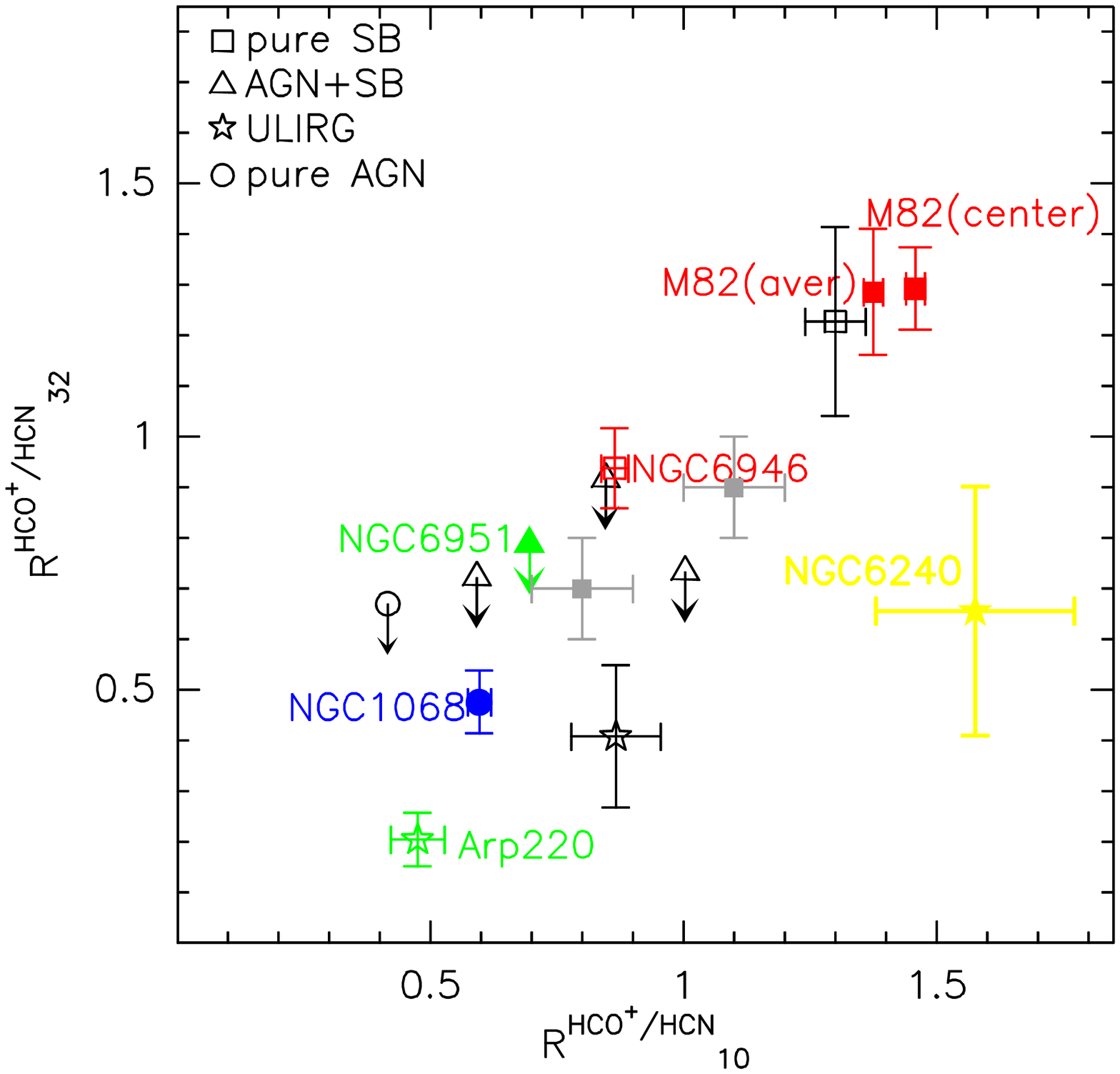}}
\caption{HCO$^+$-to-HCN(J=1--0) and HCO$^+$-to-HCN(J=3--2) intensity
ratios of all 12 sources (this paper, Table~\ref{tab3}) and NGC~253
and NGC~4569 ({\it filled grey box}; taken from the literature,
Table~\ref{tab3}). NGC~1068 ({\it filled blue circle}), NGC~6951 ({\it
filled green triangle}), Arp220 ({\it open green triangle}), NGC~6240
({\it filled yellow star}) and M82 ({\it filled red box}) are
highlighted.}
\label{ratios3}
\end{figure}

We detect all twelve galaxies in \hcn\, and \hcop, eleven in \hhcn,
ten in \hhhcn, and seven in \hhhcop\, (Table~\ref{tab2},
Fig.~\ref{spec} \& \ref{spec2}). The HCN intensity ratios are listed
in Table~\ref{tab3} and plotted in
Fig.~\ref{ratios}-\ref{ratios3}. 

The diagrams (Fig.~\ref{ratios}-\ref{ratios3}) clearly indicate
significant differences in the intensity ratios between the different
activity types in our sample (see also discussion in the next
sub-Sections):
\begin{itemize}
\item[1.)] R$^{\rm HCN}_{\rm J_u,J_l}$ is {\it low} (i.e.,
  $\lesssim$0.4) in the 'pure' AGN sources of our sample and {\it
  high} in those with a dominant SB (i.e., $\geq$0.4), suggesting an
  increasing R$^{\rm HCN}_{\rm J_u,J_l}$ with an increasing SB
  contribution. The most extreme examples of R$^{\rm HCN}_{\rm
    J_u,J_l}$ are: NGC~1068 ({\it low} R$^{\rm HCN}_{\rm J_u,J_l}$;
  AGN), NGC~6951 \& Arp220 \& NGC~6946 ({\it moderate} R$^{\rm
    HCN}_{\rm J_u,J_l}$; AGN+SB), and M82 ({\it high} R$^{\rm
    HCN}_{\rm J_u,J_l}$; SB).
\item[2.)] R$^{\rm HCO^+}_{32/10}$ is {\it low} in the 'pure' AGN
  sources (i.e., $\leq$0.3) but also in the composite (AGN+SB)
  sources, while it is {\it high} for the 'pure' SB sources (i.e.,
  $>$0.4).  In combination with R$^{\rm HCN}_{32/10}$, this creates so
  three different regions, separating the composite sources from the
  'pure' SBs and AGN. The most extreme examples for each group are:
  NGC~1068 ({\it low} R$^{\rm HCO^+}_{32/10}$ \& R$^{\rm HCN}_{32/10}$
  ; AGN), Arp~220 \& NGC~6951 ({\it moderate} R$^{\rm HCO^+}_{32/10}$
  \& {\it low} R$^{\rm HCN}_{32/10}$; AGN+SB), and M82 ({\it high}
  R$^{\rm HCO^+}_{32/10}$ \& R$^{\rm HCN}_{32/10}$; SB).
\item[3.)]  R$^{\rm HCO^+/HCN}_{J_u,J_l}$ is {\it low to moderate} in
the 'pure' AGN and almost all composite sources (i.e., $<$1), while it
is {\it moderate to high} for the 'pure' starbursts (i.e.,
$\geq$1). The most extreme examples are: NGC~1068 ({\it low} R$^{\rm
HCO^+/HCN}_{J_u,J_l}$; AGN), Arp220 ({\it low} R$^{\rm
HCO^+/HCN}_{J_u,J_l}$; AGN+SB), NGC~6240 ({\it high} R$^{\rm
HCO^+/HCN}_{J_u,J_l}$; AGN+SB), NGC~6946 \& M82 ({\it high} R$^{\rm
HCO^+/HCN}_{J_u,J_l}$; SB).

\end{itemize}
The apparent grouping of the two dominant activity types in our sample
into different intensity ratios supports fundamental differences
between the excitation conditions of the two main activity types. It
seems highly unlikely that biases in filling factors could lead to
such a systematic trend.

A comparison to CO data taken either from the literature or from
previous IRAM 30m observations reveals no similar separation effect in
the CO line transition ratios for our sample (see
Table~\ref{tab3}). However, as a very interesting result, the
HCN-to-CO luminosity ratios appear to decrease with increasing
rotational number J in the AGN sources (Table~\ref{tab4}), while those
of the SB sources remain more or less constant or even slightly
increase.

\subsection{Individual sources}
\label{dis}
In this Section, we will discuss some individual sources of our
sample, each representing a good example of one of the ratio-extremes
described in the previous Section.

\subsubsection{M82 - SB dominated galaxy}
M82 is the best testcase for a ``pure'' (evolved) SB in our sample. We
mapped the entire central disk in M82 with the IRAM 30m at 3~mm and
obtained a number of discrete pointings along the disk at 2~mm and
1~mm. The HCN and HCO$^+$ data are in good agreement with previous
measurements \citep[e.g.,][]{ngu92}. The gas disk in M82 is known to
house two giant PDRs \citep[e.g.,][]{gar02}, probably also a central
massive black hole in formation \citep[e.g.,][]{mat00,pat06} and a
superbubble emerging from a past SNe \citep[e.g.][]{weiss99,kron81}. A
different gas chemistry could hence be at play in the centre than in
the two PDRs. Thus, the HCN and HCO$^+$ intensity ratios, discussed
here, have been averaged over the entire map, and taken at the eastern
PDR position and at the centre to allow for a better comparison. The
30m observations indicate some variations especially of R$^{\rm
  HCN}_{\rm 21/10}$, R$^{\rm HCN}_{\rm 32/10}$ and R$^{\rm
  HCO^+}_{32/10}$ between the position of the PDR and the nucleus (see
Table~\ref{tab3}), while R$^{\rm HCO^+/HCN}_{\rm J_u,J_l}$ seems to be
quite similar between the PDR and the center.  The averaged ratios are
very similar to the PDR ones indicating that PDR chemistry may
dominate the overall emission in the disk.  The PDR position in M82
shows the highest R$^{\rm HCN}_{\rm J_u,J_l/10}$, R$^{\rm
  HCO^+}_{32/10}$ and R$^{\rm HCO^+/HCN}_{\rm J_u,J_l}$ of all the
sources in our sample (see filled red box in
Fig.~\ref{ratios}-\ref{ratios3}). The center in M82, however, seems to
be, overall, similar to the other SB sources of our sample for most
line ratios (compare next sub-Sections).

\subsubsection{NGC~6946 - SB dominated galaxy}
NGC~6946 is a local galaxy whose starburst activity is assumed to be
much younger than that in M82. NGC~6946 is not part of a galaxy-merger
or -interaction, as it is the case for M82 \citep[e.g.,][]{pis00}. No
signs of any significant PDR have yet been found in this galaxy and
also large-scale, high-velocity shocks do not seem to play a major
role yet \citep[e.g.,][]{sch07}. This difference to M82 in its SB
properties may explain its location in the diagrams
(Fig.~\ref{ratios}-\ref{ratios3}) with respect to M82, i.e., in the
middle of the diagram, representing thus eventually a different type
of starburst activity. As such, it might set tight constraints to our
comparison between AGN and SB and underline the importance of the
evolutionary stage of the starburst.

\subsubsection{NGC~1068 - AGN dominated galaxy}
NGC~1068 is the best example in our sample for housing a pure AGN in a
central radius of 1~kpc ($\equiv$14$''$)\footnote{Please note, that
  $\sim$10-20\% of the overall HCN emission in NGC~1068 is located in
  the spiral arms \citep[][]{tac94} which are possibly dominated by
  star formation. However, even for the largest beam-size of 29$''$
  the spiral arms should only marginally ($\sim$10\%) contribute to
  the detected HCN emission in the 30m beam. The same is true for
  HCO$^+$.}. No strong evidence for any significant {\it nuclear}
starburst has been reported so far (e.g., MIR: Laurent et al.\ 2000;
NIR(PAH): Imanishi 2002; Optical/Near-UV: Cid-Fernandes et al.\ 2001).
\cite{mar03} estimate that a compact nuclear starburst would
contribute less than 1\% to the total IR luminosity. Thus, NGC~1068
represents the best counter-part to M82 in terms of activity
type. NGC~1068 appears to be always located in the opposite part of
the diagrams in Fig.~\ref{ratios}-\ref{ratios3} with respect to M82,
supporting the differences in the excitation conditions of the
molecular gas suspected between AGN and SB environments. Moreover,
\cite{use04} have discussed the possibility of NGC~1068 harbouring a
giant XDR in its nucleus that is used to explain the suprisingly high
HCN-to-CO and low HCO$^+$-to-HCN luminosity ratios (see also
Table~\ref{tab4}). The potential prototypical XDR nature of the AGN in
NGC~1068 classifies this source as an ideal counter-part to the
PDR-dominated galaxy M82, in terms of effects of the radiation field
onto the surrounding gas chemistry.

\subsubsection{NGC~5194 - AGN dominated galaxy}
Besides NGC~1068, the centre in NGC~5194 is most likely dominated by
an AGN as well. This source is assumed to be in a post SB stage in
which the massive star formation has already disappeared
\citep[e.g.,][]{thr91,sau96,gre98}; its nuclear activity is assumed to
be caused by a low-luminosity active galactic nucleus of LINER type
\citep[e.g.,][]{ho97}. NGC~5194 is located close to NGC~1068 in all
diagrams (Fig.~\ref{ratios}-\ref{ratios3}), substantiating the
differences between AGN and SBs in our sample. It is also one of the
few sources for which a high HCN-to-CO but low
HCO$^+$-to-HCN(J=1--0)$^+$ intensity ratio has been found (see
Table~\ref{tab4}).

\subsubsection{NGC~6951 - AGN+SB galaxy}
\begin{figure}[!t]
\epsscale{0.9} \rotatebox{-90}{\plotone{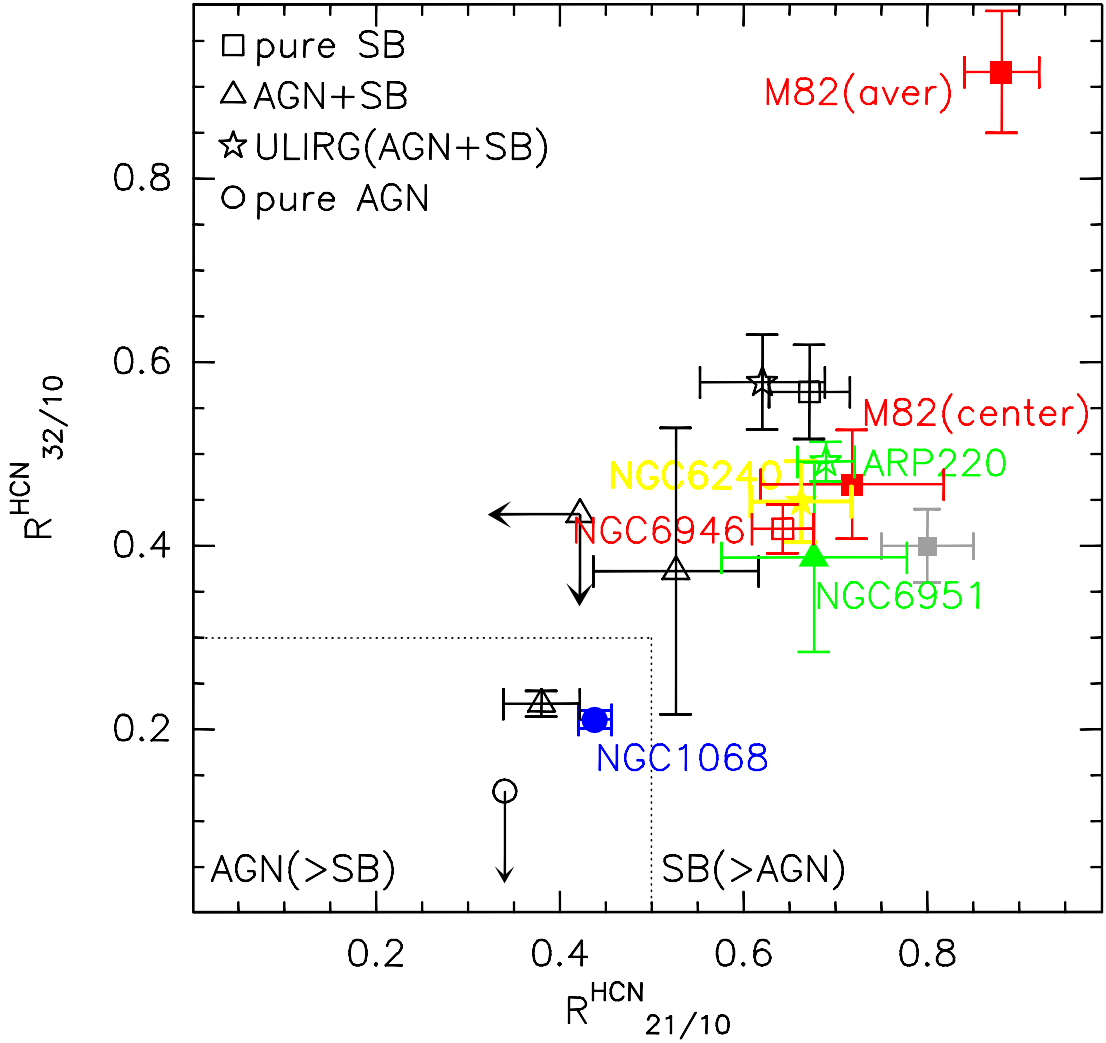}}
\caption{HCN(J=2--1)-to-HCN(J=1--0) and HCN(J=3--2)-to-HCN(J=1--0)
  intensity ratios of all 12 sources (this paper, Table~\ref{tab3})
  and IC~342 ({\it filled grey box}; data taken from the literature,
  Table~\ref{tab4}). NGC~1068 ({\it filled blue circle}), NGC~6951
  ({\it filled green triangle}), Arp220 ({\it open green star}),
  NGC~6240 ({\it filled yellow star}) and M82 ({\it filled red box})
  are highlighted. The dotted grey lines should guide the readers eyes
  and indicate the putative different locations of the SB and AGN
  dominated sources in this diagram.}
\label{ratios}
\end{figure}

The best example of a composite source in our sample likely is
NGC~6951 (highlighted with a filled green triangle in
Fig.~\ref{ratios}-\ref{ratios3}). It is known to house a prominent SB
ring (e.g., optical: M\'arquez \& Moles 1993; Wozniak et al. 1995;
Rozas, Beckman, \& Knapen 1996; Gonz\'alez-Delgado et al. 1997; radio:
Vila et al. 1990; Saikia et al. 1994) as well as a Seyfert
type~2/LINER nucleus in its central 20$''$
\citep[e.g.,][]{boe93,ho97}.  \cite{koh99} have mapped this source in
\hcn\, with the Nobeyama array revealing that most of the HCN emission
is concentrated in the SB ring, similar to the CO emission. Probably
due to missing sensitivity and angular resolution ($\sim3-4''$), they
fail to detect \hcn\, emission in the very centre (inner
2$''$). However, recent high angular resolution observations of \hcn\,
in NGC~6951 carried out in the extended configuration of the IRAM PdBI
show compact HCN emission in the centre as well \citep[][]{kra07}
indicating a high HCN-to-CO(J=1--0) luminosity ratio of $\sim$1. The
interferometric maps suggest that the HCN emission in the SB ring
dominates the lines measured with the IRAM 30m telescope while the
central HCN (and HCO$^+$) emission probably contributes only
marginally (less than 10\%) to the observed brightness temperatures
with the IRAM 30m. This explains the location of NGC~6951 in
Fig.~\ref{ratios}, close to the SB dominated sources in our sample. It
has to be mentioned, though, that we might miss part (i.e.,
$\sim$30-40\%) of the \hhhcn\, and \hhhcop\, emission from the SB ring
because of the lower beam size at these frequencies as recent SMA
observation of the \hhhcn\, emission in NGC~6951 indicate (Krips et
al., in prep.). However, this only slightly changes the position of
NGC~6951 in Fig.~\ref{ratios}-\ref{ratios3}.  The HCO$^+$ emission
separates it significantly from the evolved starburst M82, similar to
NGC~6240 and Arp~220. This may be either a consequence of the
evolutionary stage of the starburst or the additional existence of an
AGN.

\subsubsection{Arp~220 - ULIRG (1 AGN + 1 SB nucleus)}
\label{arp220} 
Arp~220 is the prototypical ULIRG and as such represents well the
higher-activity ULIRG population in our sample.  A SB is outweighing
the centre of Arp220 \citep[e.g.,][and references therein]{ris06} but
an AGN may be present in at least one of the two nuclei though this is
still controversial \citep[e.g.,][]{san88,haa01,ima06,downes07}.  The
central (i.e., 2-4$''$) molecular gas emission in Arp~220 is
concentrated in two peaks, very similar thus in its gas morphology to
NGC~1068.  Despite the differences in origin for the gas morphology of
these two galaxies, i.e., two nuclei in Arp~220 \citep[e.g.,][]{sak99}
versus a (probably) warped disk in NGC~1068 \citep[e.g.,][]{sch00},
the beam filling factor ratios in the 30m beam should be very similar
for the two sources (when assuming similar source sizes in all
transitions), making the differences between their spectra even more
pronounced, especially in the HCN transitions (Fig.~\ref{spec}). Even
if we account for smaller source sizes for the higher transitions in
NGC~1068 (which actually leads to different filling factors between
NGC~1068 and Arp~220; see Table~\ref{tab2}), R$^{\rm HCN}_{\rm
  J_u,J_l/10}$ is still significantly lower in NGC~1068 than in
Arp~220. Interestingly, the situation for R$^{\rm HCO^+}_{32/10}$ and
R$^{\rm HCO^+/HCN}_{\rm J_u,J_l}$ is much different: here the ratios
of both sources are very similar to each other. This potentially
differentiates Arp~220 significantly from M82 and the composite
sources and, together with NGC~6240 (see next sub-Section), may
indicate that the ULIRGs have to be eventually handled as a different
'class' in our sample, i.e., separate from the local starbursts and
composite sources. Also, Arp~220 is located close to NGC~6946 in the
R$^{\rm HCN}_{\rm J_u,J_l/10}$ diagram (Fig.~\ref{ratios}) but farther
away from it in those including HCO$^+$. Its intensity ratios seem to
be more like those of NGC~6951, which has a dominating starburst but
also a central weak AGN (see previous Section). \cite{aal07} recently
discussed the possibility of an XDR changing the molecular gas
chemistry as an alternative to IR pumping to explain the high
HNC-to-HCN(J=3--2) intensity ratios detected in Arp~220 by them.

\subsubsection{NGC~6240 - ULIRG (2 AGN)}
The two nuclei in NGC~6240 have been recently found to {\it both}
harbor an AGN in addition to the very pronounced SB in this evolved
merger \citep[e.g.,][]{kom03}.  Most of the molecular gas as well as
dust appears to be located between the two nuclei
\citep[e.g.,][]{bry99,tac99,nak05,ion07} as opposed to Arp220 in which
two gas disks may still be present
\citep[e.g.,][]{sak99}. \cite{ion07} also find evidence for a gas
outflow/inflow that either is connected to starburst superwinds or
outflows from the AGN. The existence of these and the two AGN in
NGC~6240 may create exceptional and very extreme conditions for the
molecular gas in the central region of NGC~6240.  Similar to Arp~220,
it is located close to NGC~6946 in the HCN diagram but closer to
NGC~6951 in the diagram displaying R$^{\rm HCO^+}_{32/10}$
(Fig.~\ref{ratios2}). However, in the R$^{\rm HCO^+/HCN}_{\rm
  J_u,J_l}$ diagram (Fig.~\ref{ratios3}), it seems to populate a very
own region, quite separate from the rest, which may be linked to the
extreme conditions in its centre.

\section{LVG simulations}
\label{lvg}
We have run simulations of the excitation conditions for HCN and
HCO$^+$ using the Large Velocity Gradient (LVG) approximation in
MIRIAD to connect the observed intensity ratios with physical
parameters such as kinetic gas temperature, gas density and molecular
abundances.  Although we find differences in the values obtained for
HCO$^+$ between the LVG code used in MIRIAD and RADEX, we have
concentrated our analysis on the MIRIAD code. The differences between
the two codes seem to be only present for gas regions with very high
HCO$^+$ column densities ($\gg$10$^{15}$cm$^{-2}$~km$^{-1}$~s) that
lie, however outside the range studied in this paper. For lower column
densities, RADEX and the LVG code in MIRIAD produce almost identical
results. We think that the difference could be either a consequence of
a different 'parameter sampling' between the codes or different
assumptions that start to fail for higher column densities in one of
the two codes.

We assume a one component model for the LVG analysis, i.e., all HCN as
well as HCO$^+$ transitions originate from the same region underlying
the same gas temperature and density. We carried out a reduced
$\chi^2$-test to constrain the above mentioned parameters; the
$\chi^2$-test includes constraints based on R$^{\rm
  HCN}_{J_u,J_l/10}$, R$^{\rm HCO^+/HCN}_{\rm 10/10}$ and R$^{\rm
  HCO^+}_{32/10}$ (Table~\ref{tab3}).

We chose 4 exemplary sources in our sample, each representing one of
the activity types and one of the ratio-groups: NGC~1068 as example of
a pure AGN, NGC~6951 as example of a composite source, M82 as example
of a pure SB, and NGC~6240 as the example with the most extreme
intensity ratios among the three ULIRGs in our sample. For the LVG
analysis, we varied the gas temperature in a range of T$_{\rm
k}$=20-240~K using steps of 20~K, the H$_2$ volume densities in a
range of n(H$_2$)= 10$^{1..7}$cm$^{-3}$, the HCN(J=1--0) column
densities per velocity interval in a range of
N(HCN(J=1--0))/dv$\simeq$10$^{11..19}$cm$^{-2}$\ km$^{-1}$~s and the
HCO$^+$-to-HCN abundance ratios in a range of [HCN]/[HCO$^+$]=0.01-50.
We also base the discussion and simulations on the following
definitions and relations:
\begin{equation}
{\rm Z(HCN)}\equiv[{\rm HCN}]/[{\rm H_2}]\equiv {\rm n(HCN)/n(H_2)} \\
\end{equation}
with Z(HCN) being the fractional abundance of HCN, [X] the molecular
abundance of the molecule X and n(X) the volume density of X in units
of cm$^{-3}$. We then have
\begin{eqnarray}
\begin{array}{c}
{\rm N(HCN)/dv\,=\,Z(HCN)/(dv/dr)\times n(H_2);} \\ 
{\rm \hspace*{0.2cm} with\,\,\,N(HCN)=n(HCN)\times dr}
\end{array}
\end{eqnarray}
where N(HCN)/dv is the HCN column density per velocity interval in
units of cm$^{-2}$/(\kms), and dv/dr the ``velocity gradient'' in
units of \kms\,/pc.

\begin{figure*}[!t]
   \centering
\resizebox{5.9cm}{!}{\rotatebox{-90}{\includegraphics{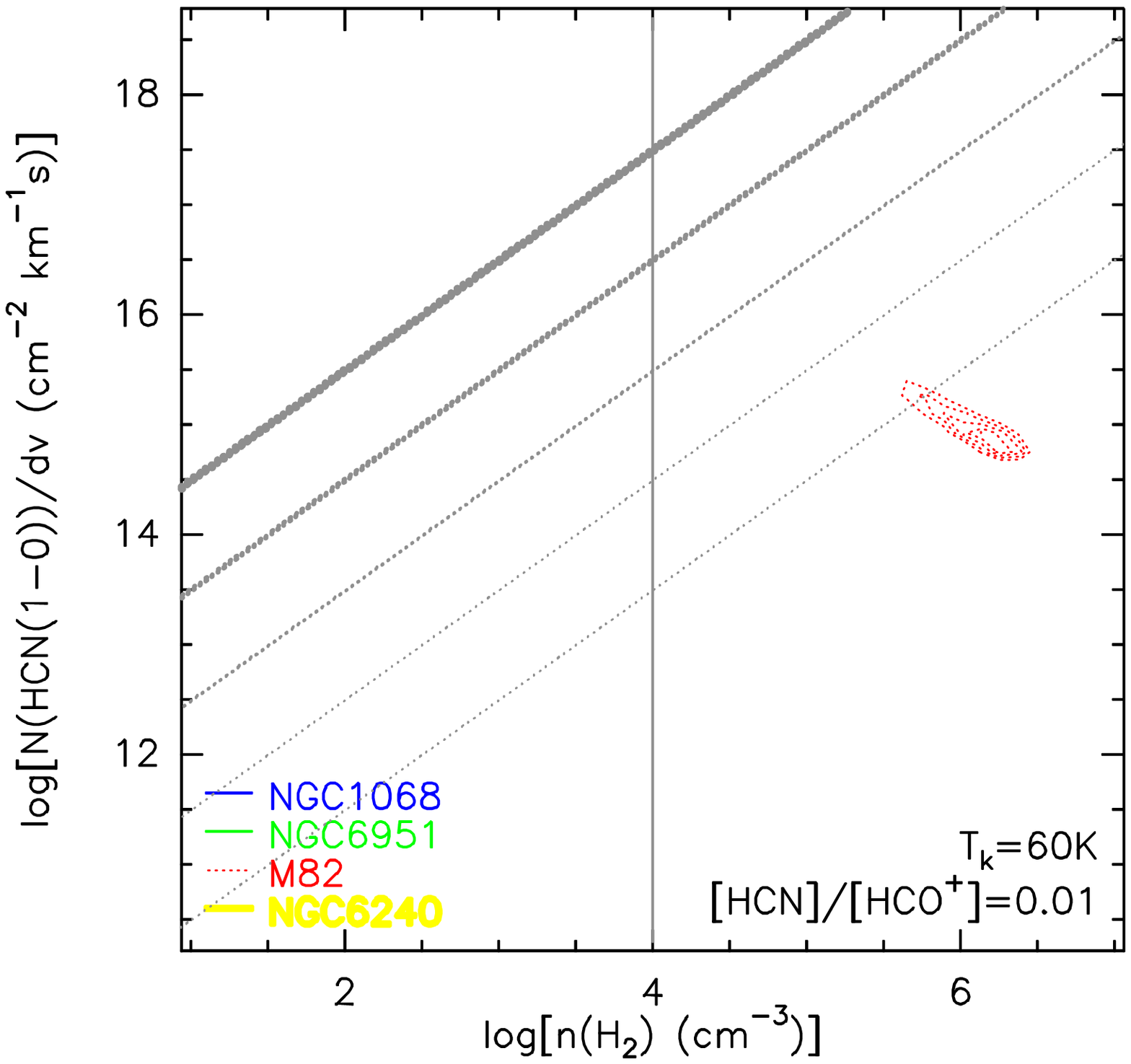}}}
\resizebox{5.9cm}{!}{\rotatebox{-90}{\includegraphics{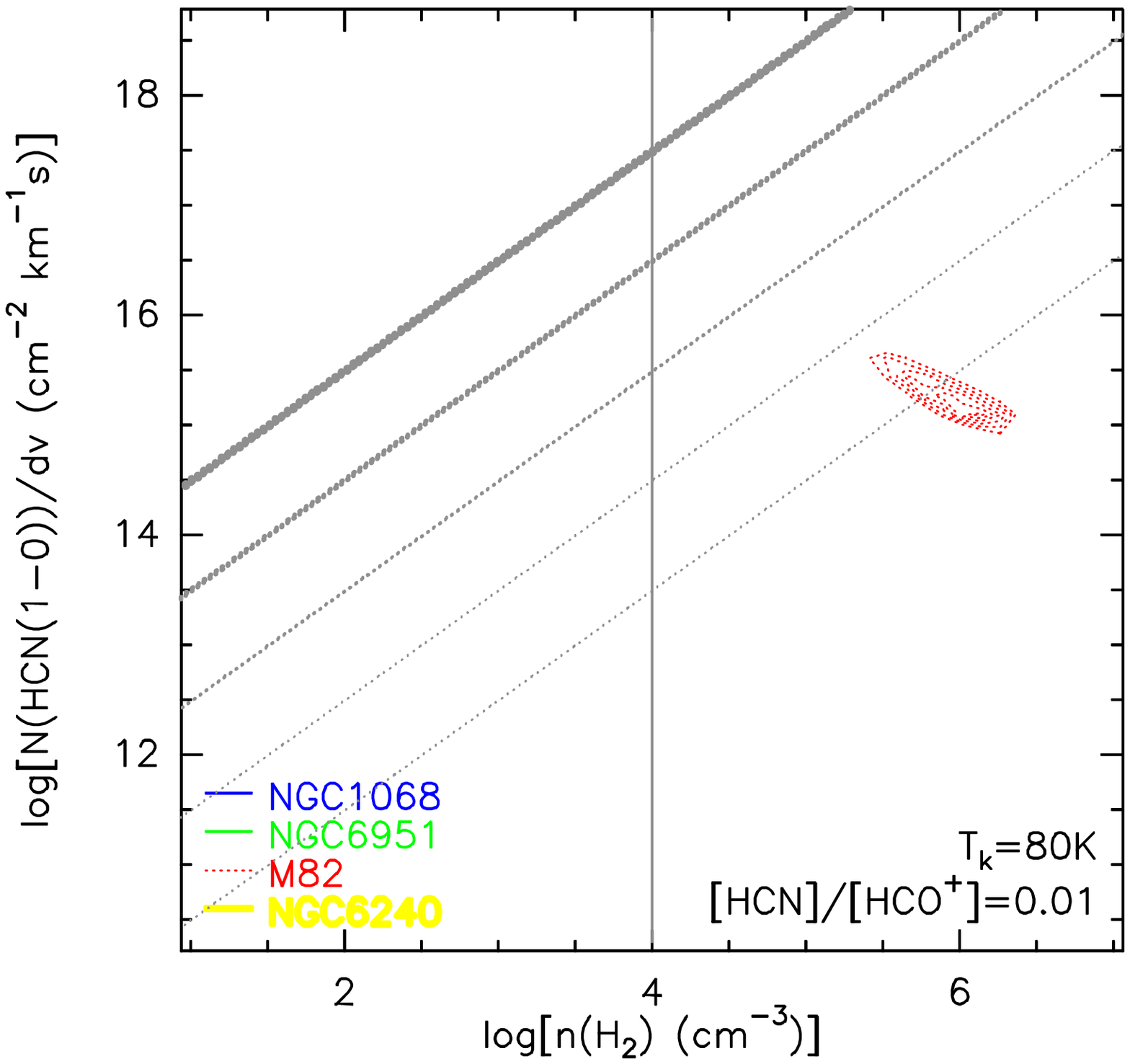}}}
\resizebox{5.9cm}{!}{\rotatebox{-90}{\includegraphics{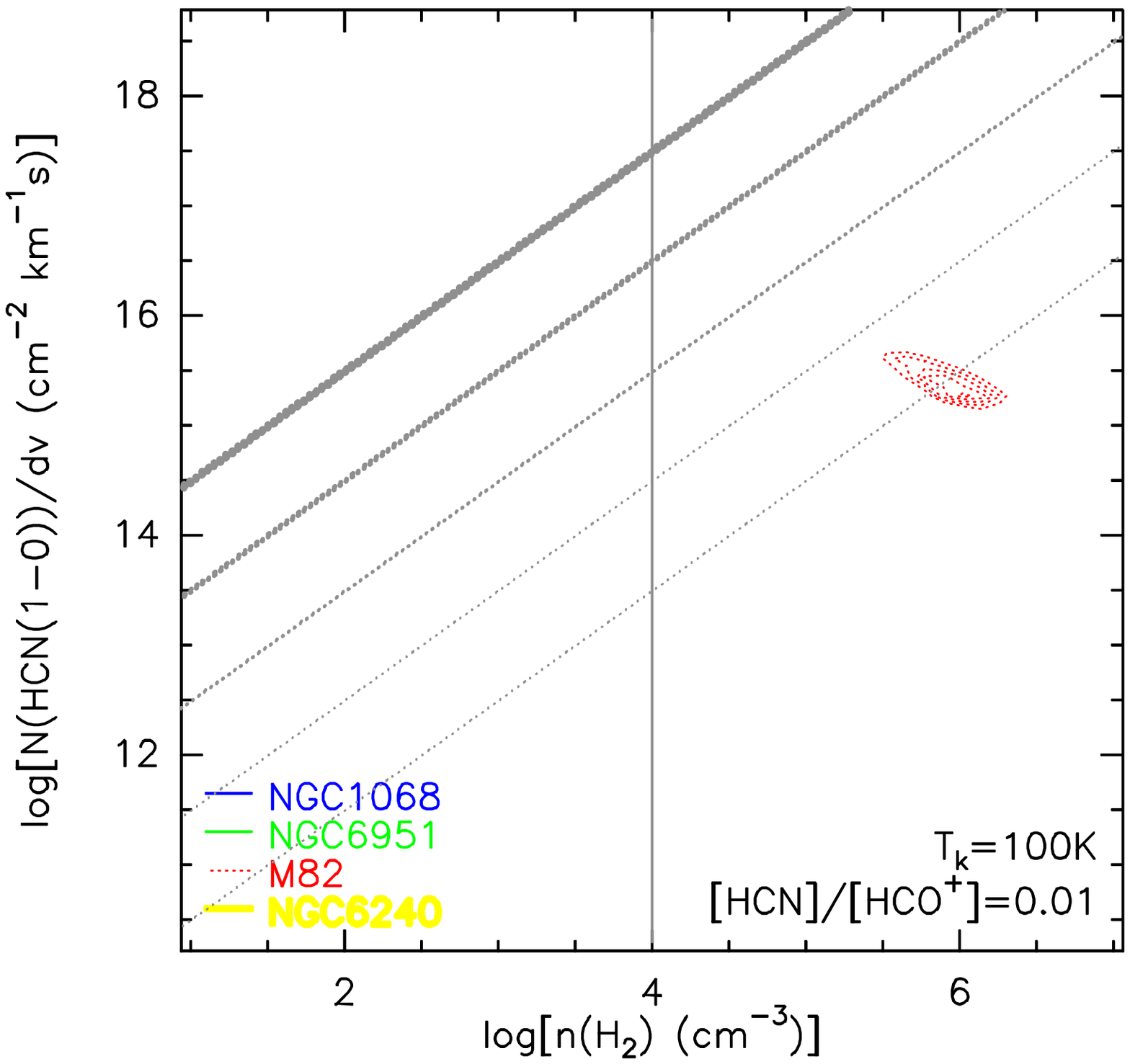}}}
\resizebox{5.9cm}{!}{\rotatebox{-90}{\includegraphics{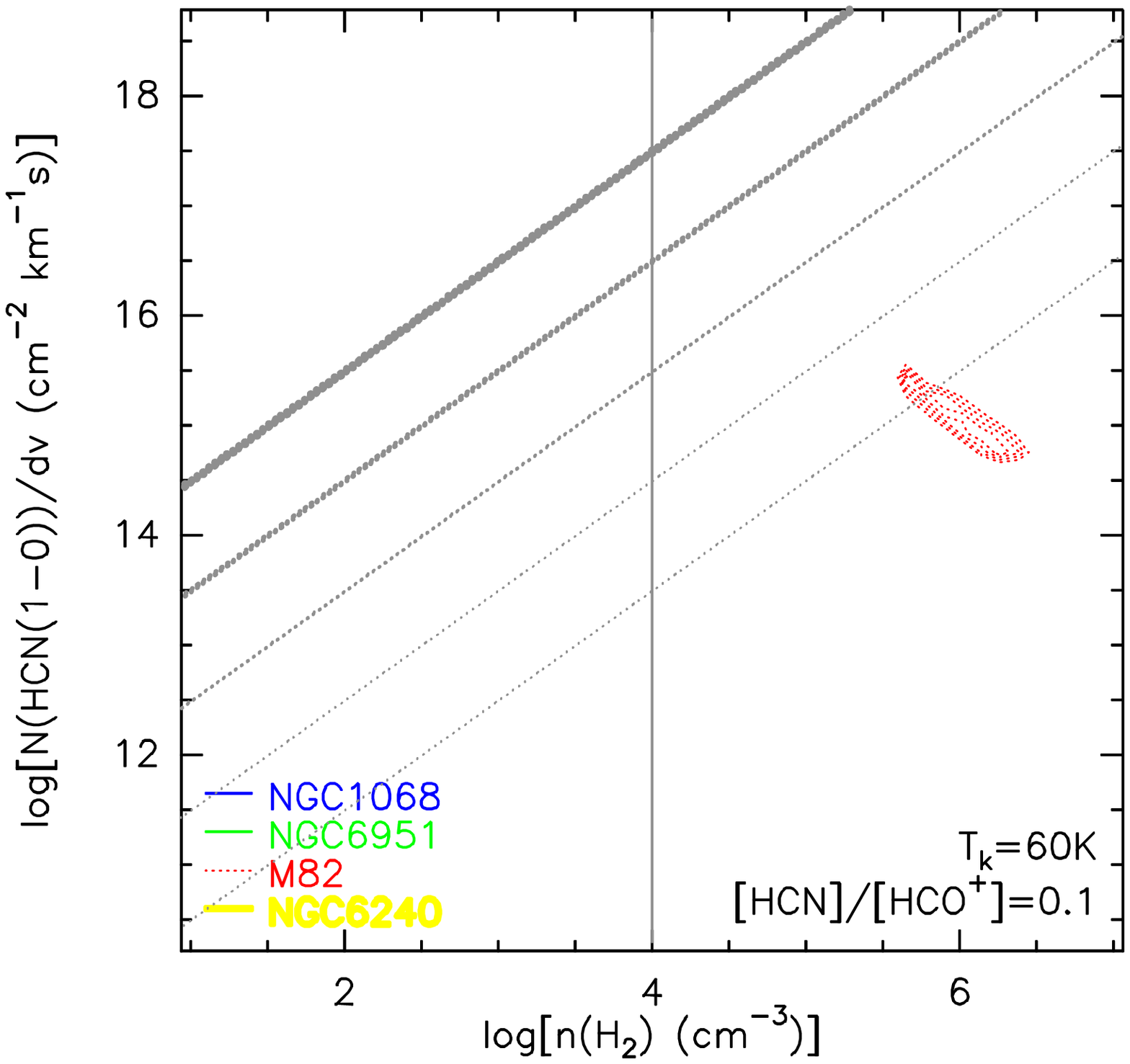}}}
\resizebox{5.9cm}{!}{\rotatebox{-90}{\includegraphics{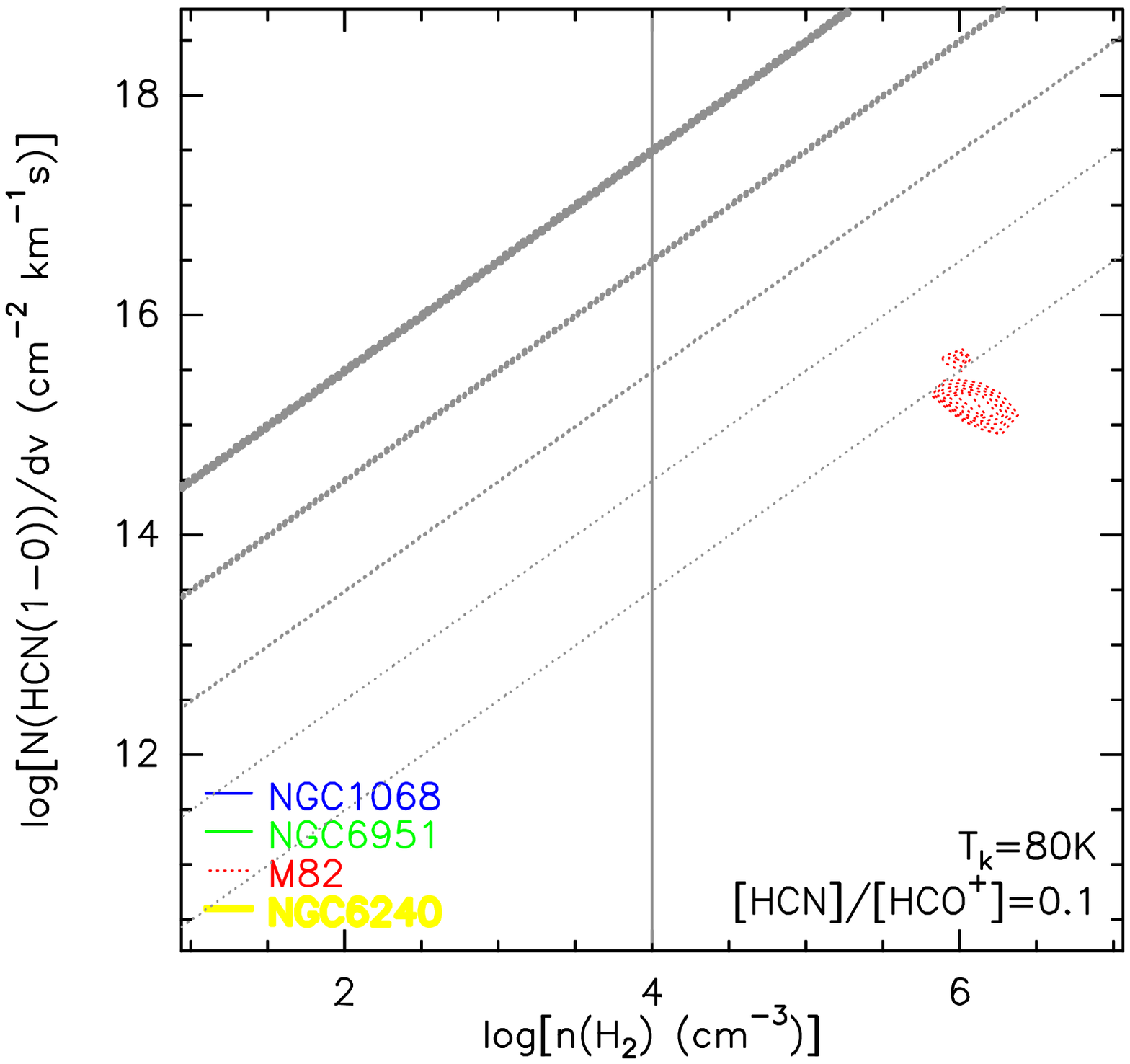}}}
\resizebox{5.9cm}{!}{\rotatebox{-90}{\includegraphics{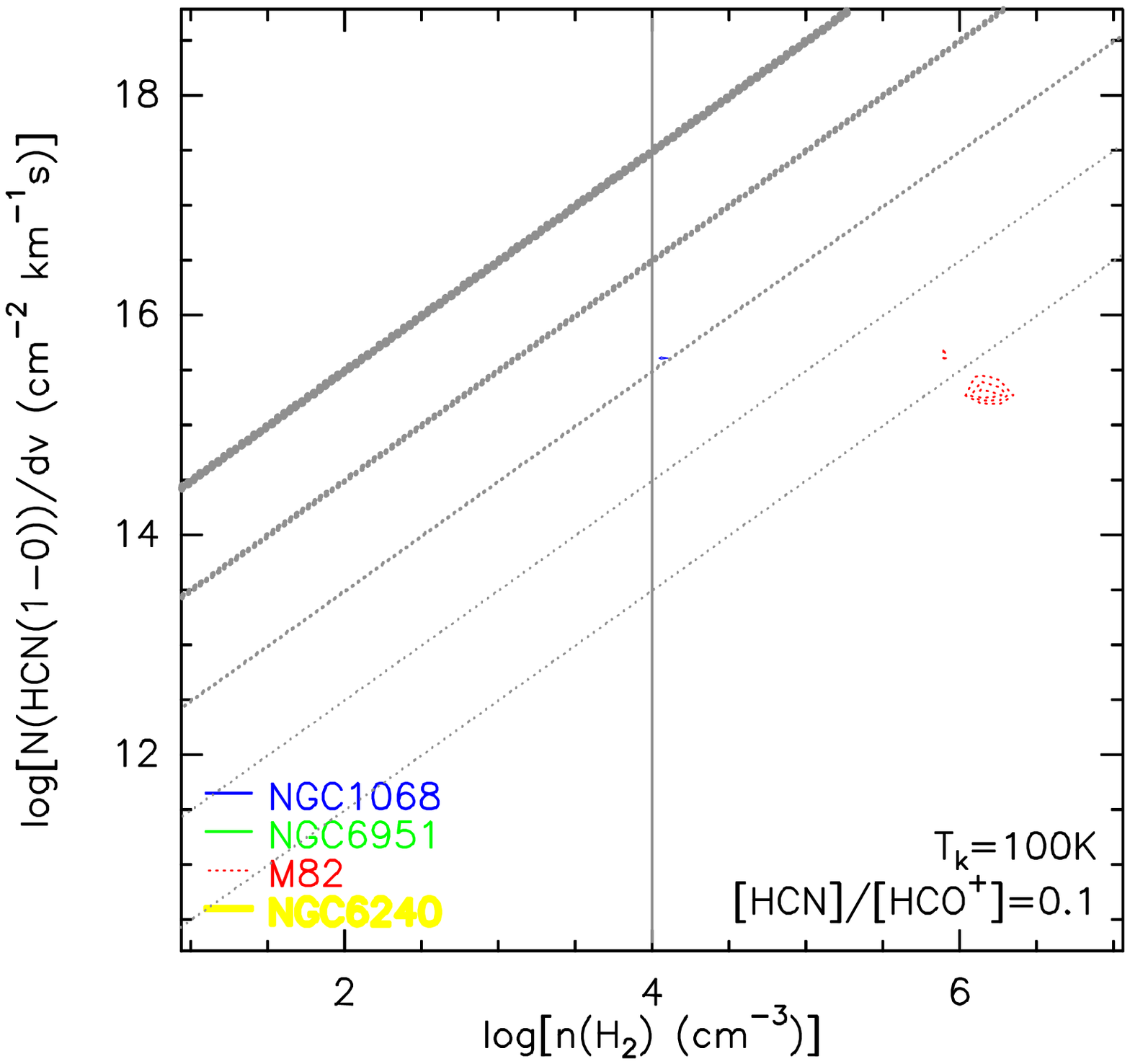}}}
\resizebox{5.9cm}{!}{\rotatebox{-90}{\includegraphics{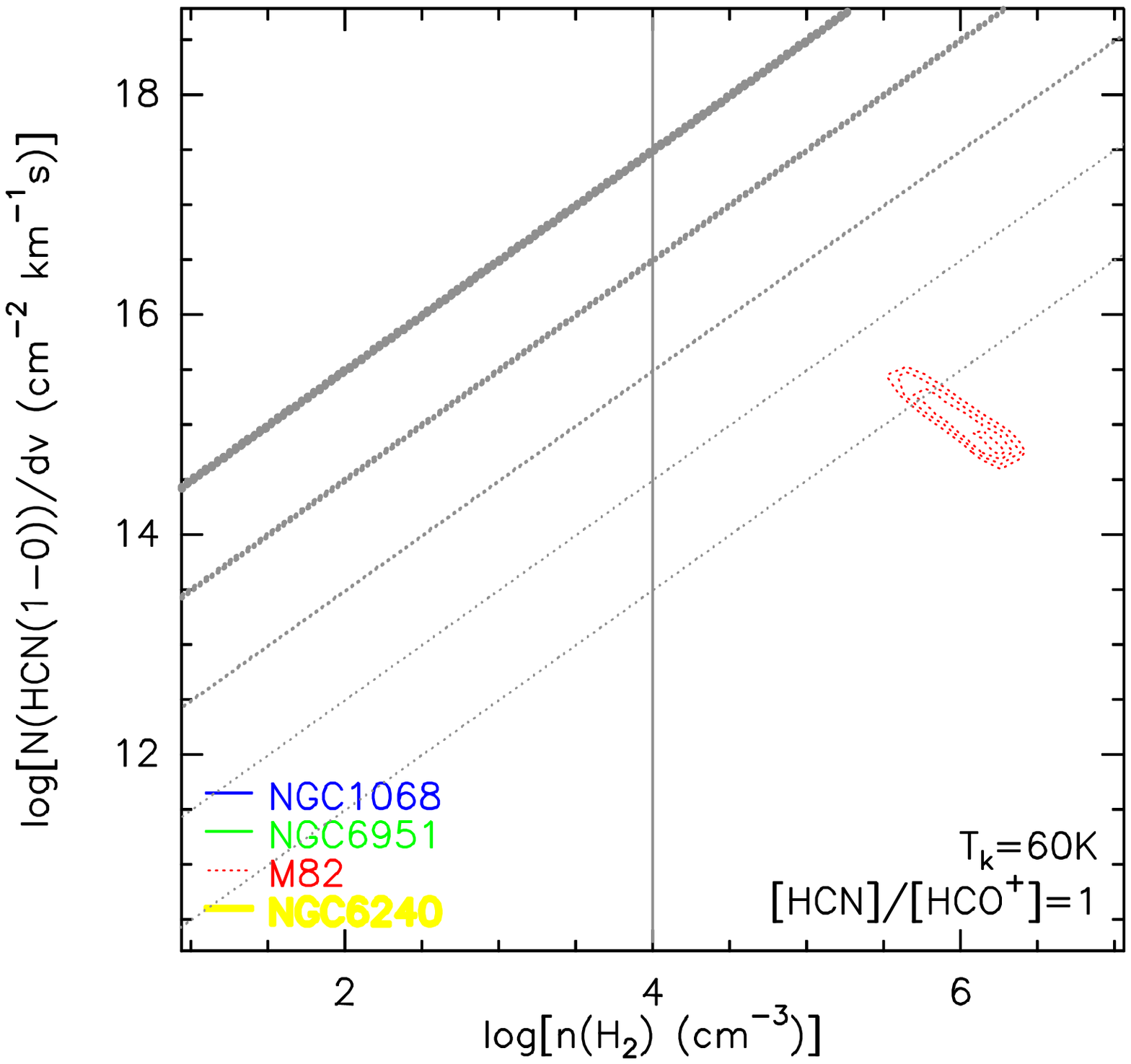}}}
\resizebox{5.9cm}{!}{\rotatebox{-90}{\includegraphics{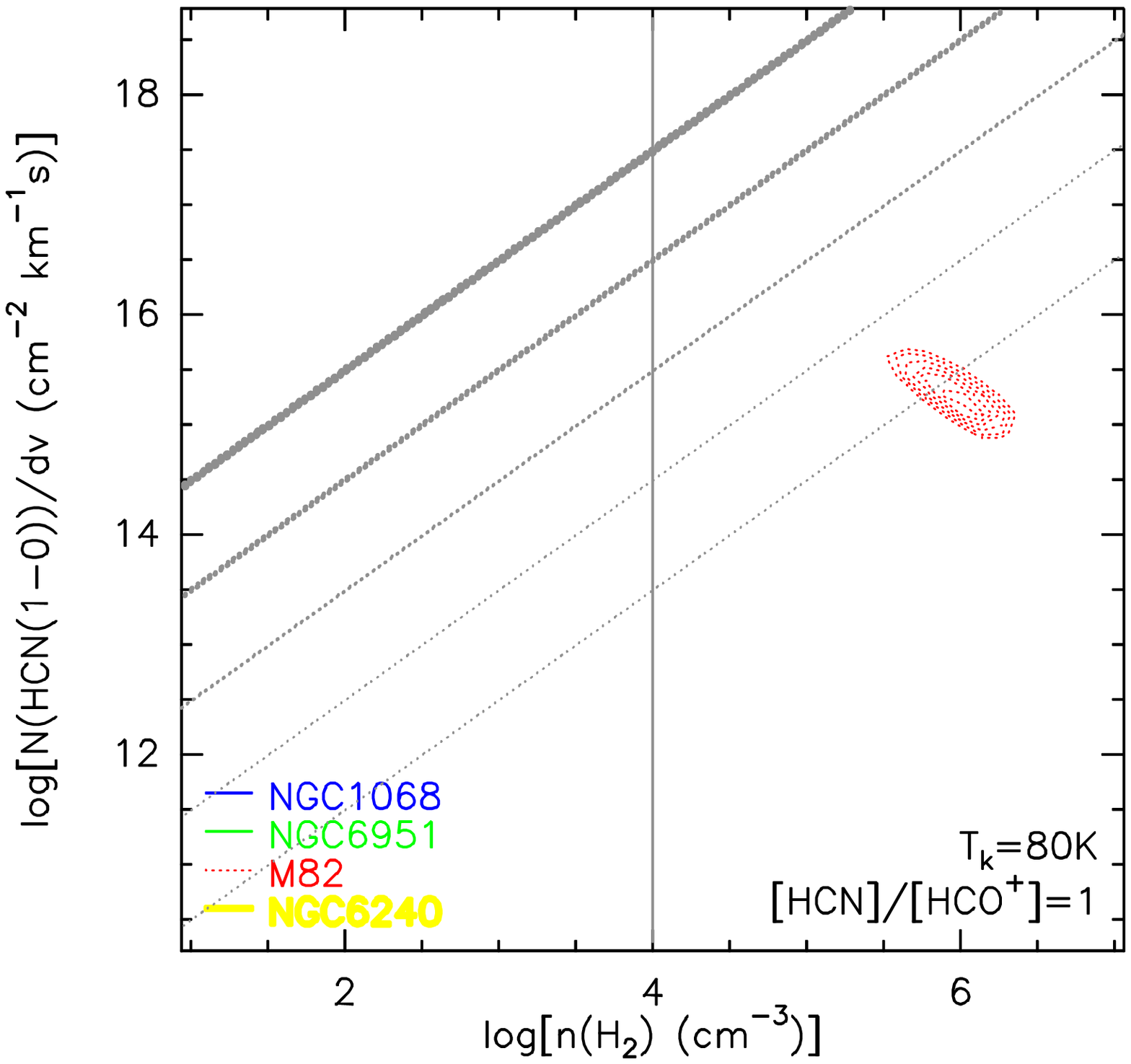}}}
\resizebox{5.9cm}{!}{\rotatebox{-90}{\includegraphics{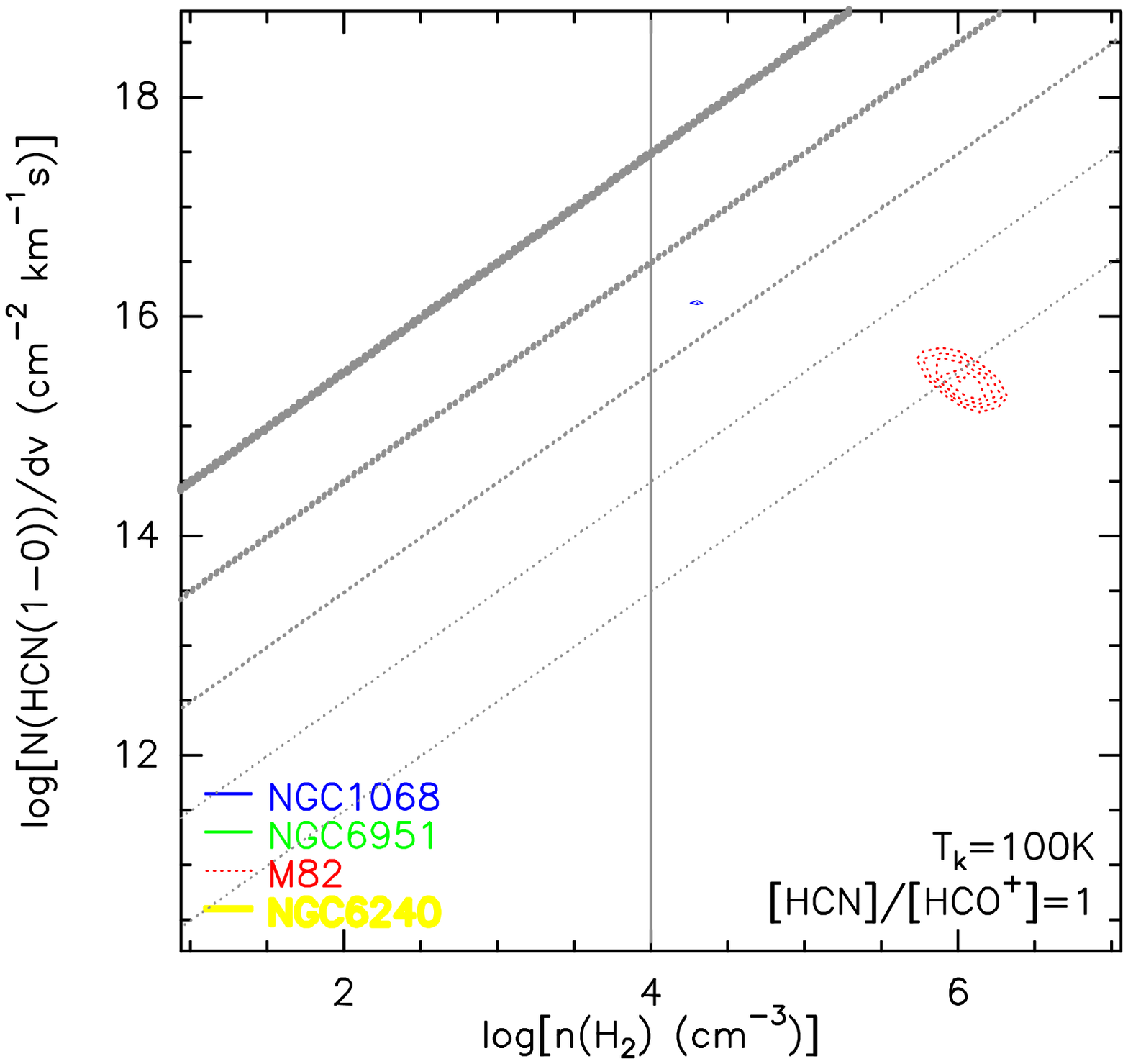}}}
      \caption{$\chi^2$-test results of the LVG analysis (see text for
      explanation) carried out with MIRIAD including the HCN and
      HCO$^+$ data for different kinetic temperatures and abundance
      ratios of [HCN]/[HCO$^+$]=0.01,0.1,1 (see also Fig.~\ref{lvg-r2}
      for the higher abundance ratios of 10 \& 50). The dotted red,
      solid blue, solid green and solid yellow contours correspond to
      M82, NGC~1068, NGC~6951 and NGC~6240 respectively. The area
      enclosed by the contours represent the lowest (reduced)
      $\chi^2$-test with $\chi^2$$\lesssim$1; for these low
      [HCN]/[HCO$^+$] abundance ratios (i.e., $\leq$1) we only find a
      significant result for the SB dominated sources in our
      sample. The grey lines indicate Z(HCN)/(dv/dr) with 10$^{-5}$,
      10$^{-6}$, 10$^{-7}$, 10$^{-8}$ and 10$^{-9}$~km$^{-1}$\ s\ pc
      (from top to bottom, decreasing in line thickness). }
         \label{lvg-r1}
\end{figure*}

\begin{figure*}[!t]
\begin{center}
\resizebox{5.9cm}{!}{\rotatebox{-90}{\includegraphics{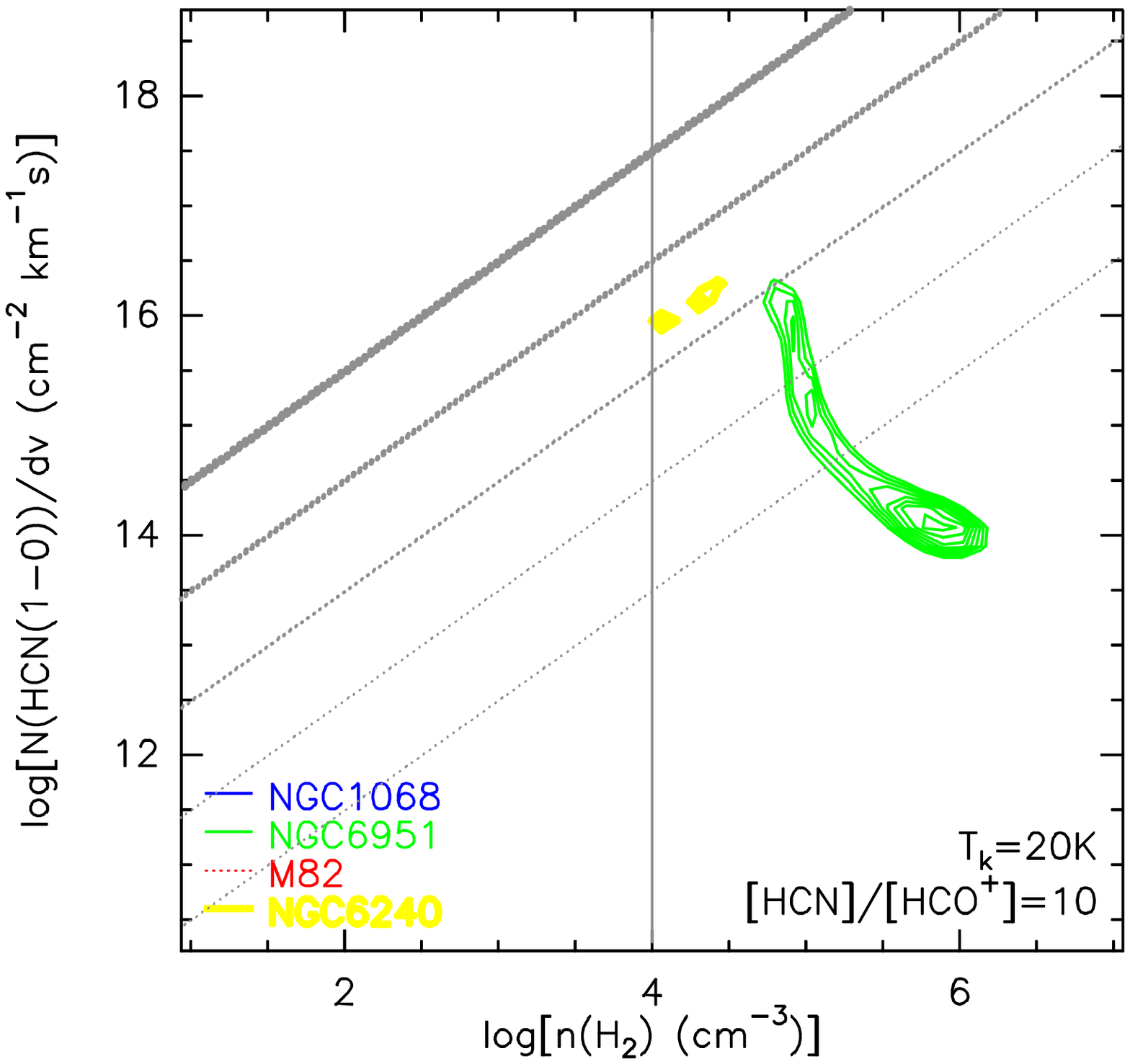}}}
\resizebox{5.9cm}{!}{\rotatebox{-90}{\includegraphics{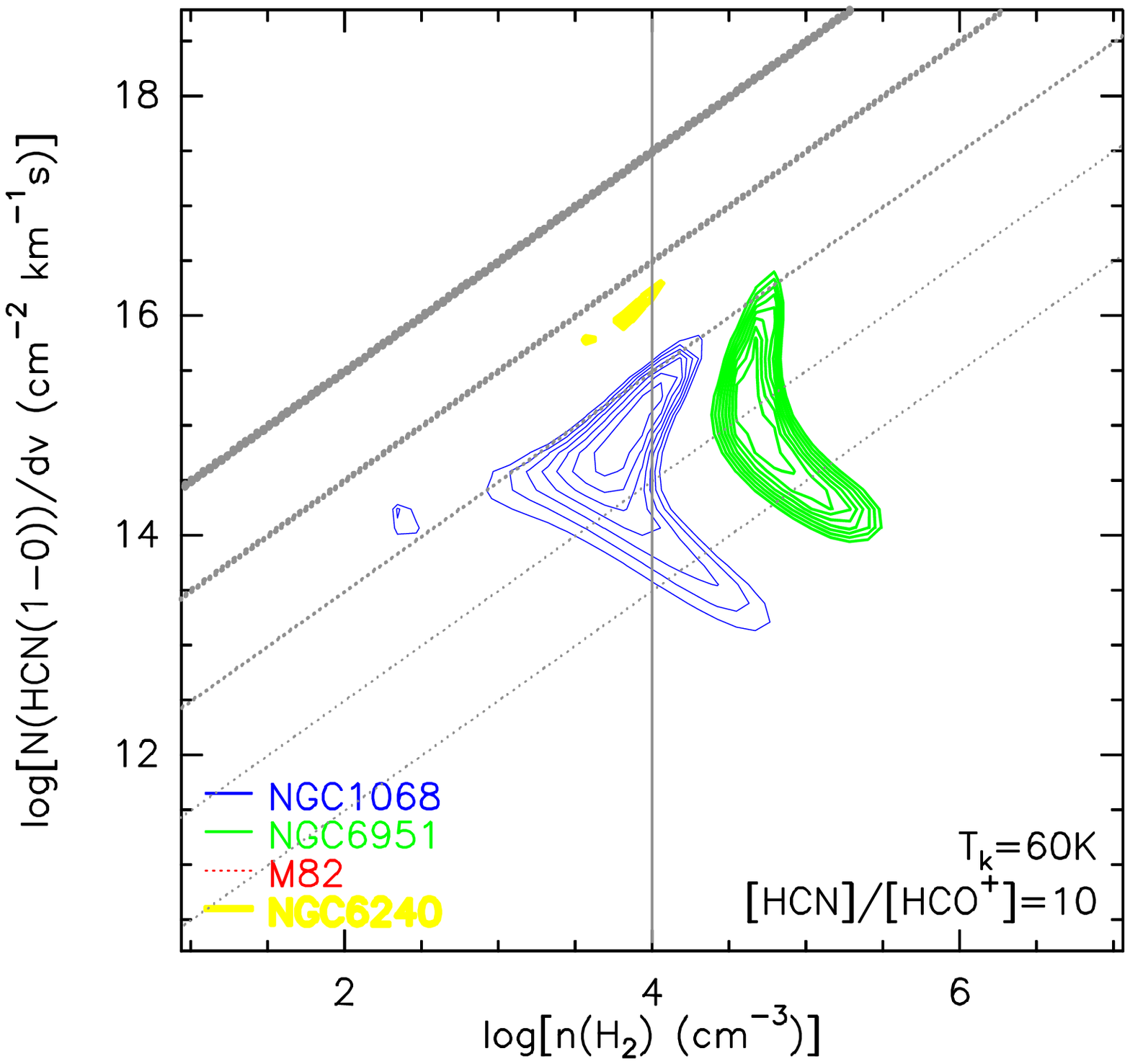}}}
\resizebox{5.9cm}{!}{\rotatebox{-90}{\includegraphics{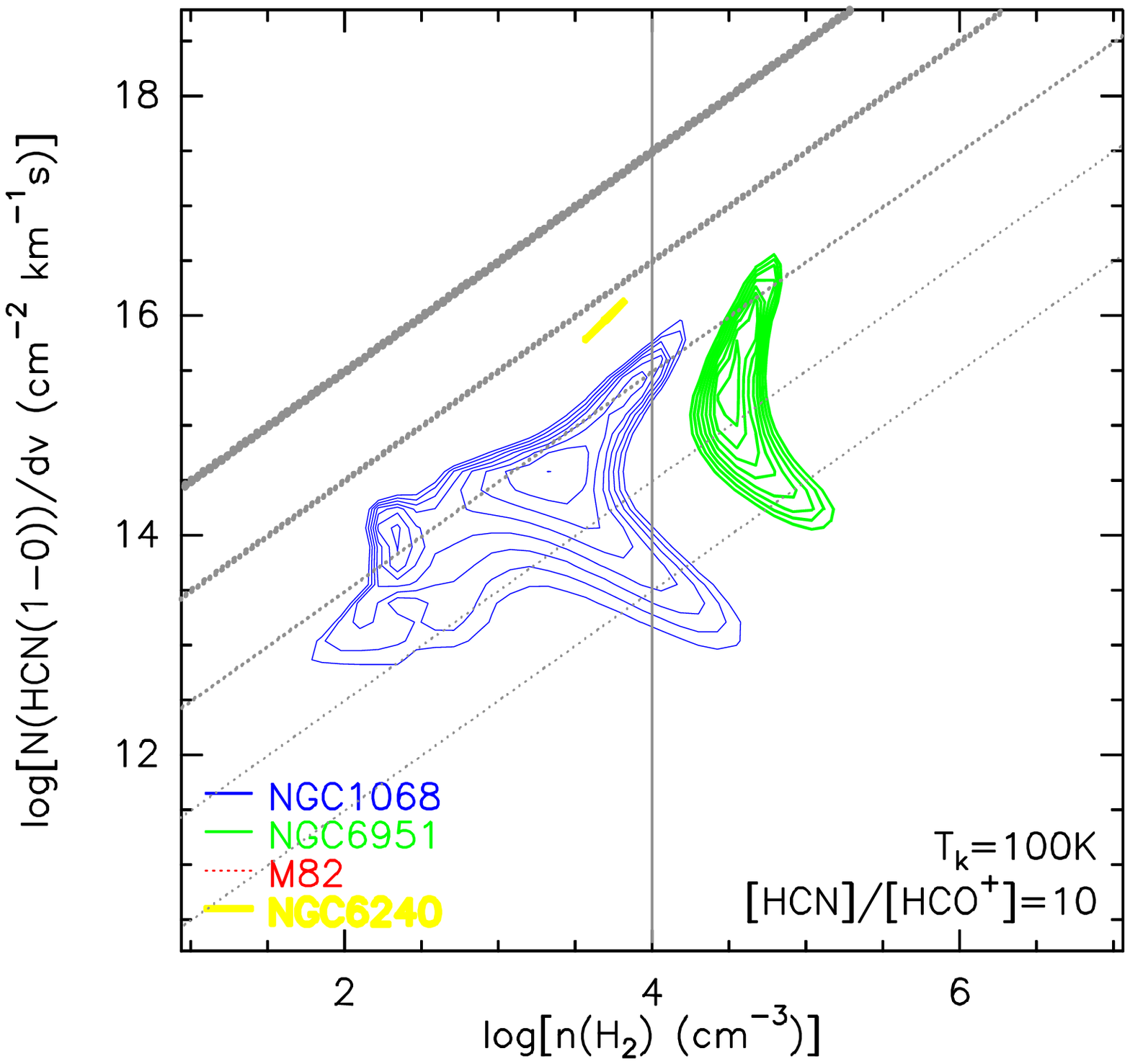}}}
\resizebox{5.9cm}{!}{\rotatebox{-90}{\includegraphics{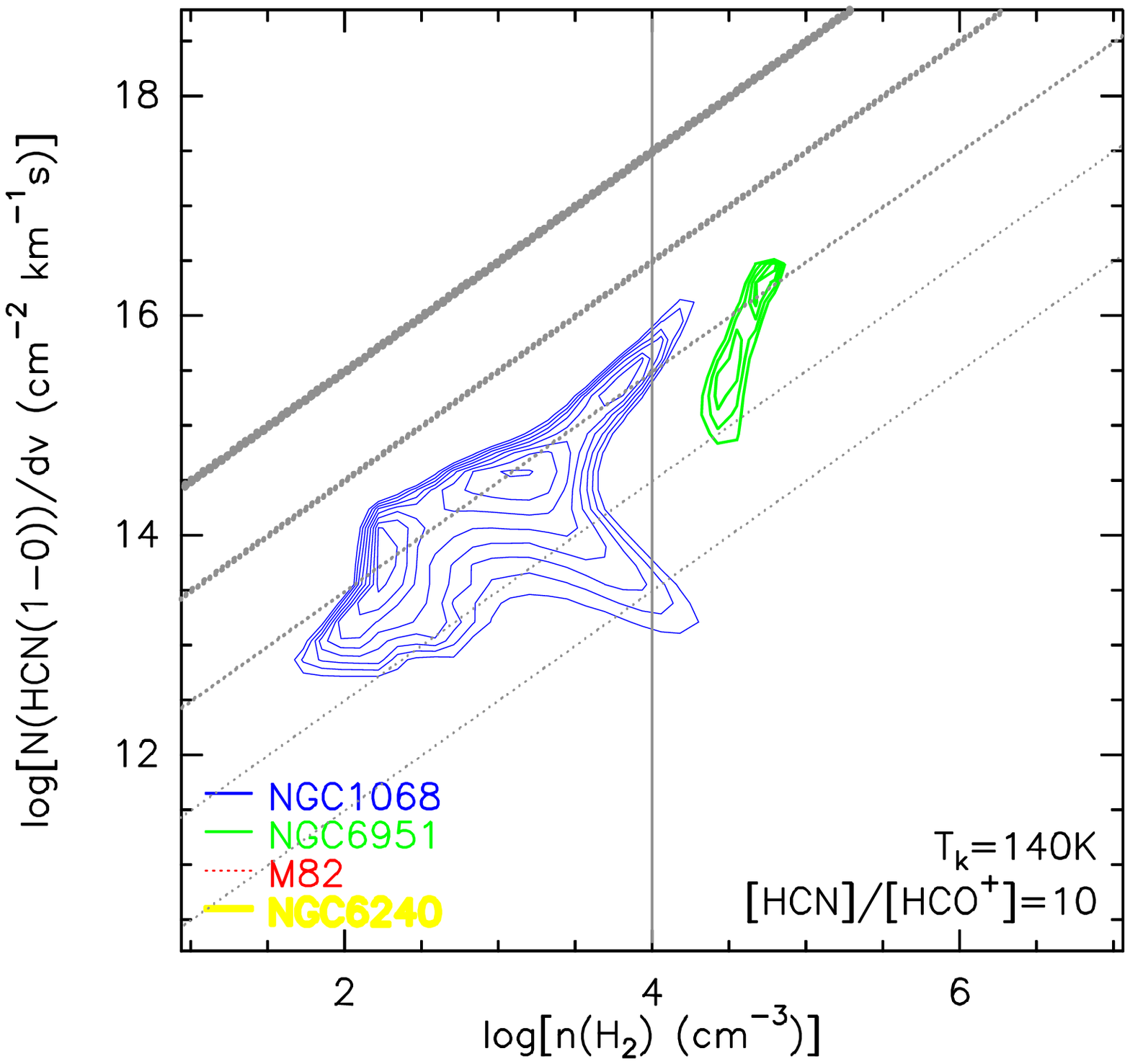}}}
\resizebox{5.9cm}{!}{\rotatebox{-90}{\includegraphics{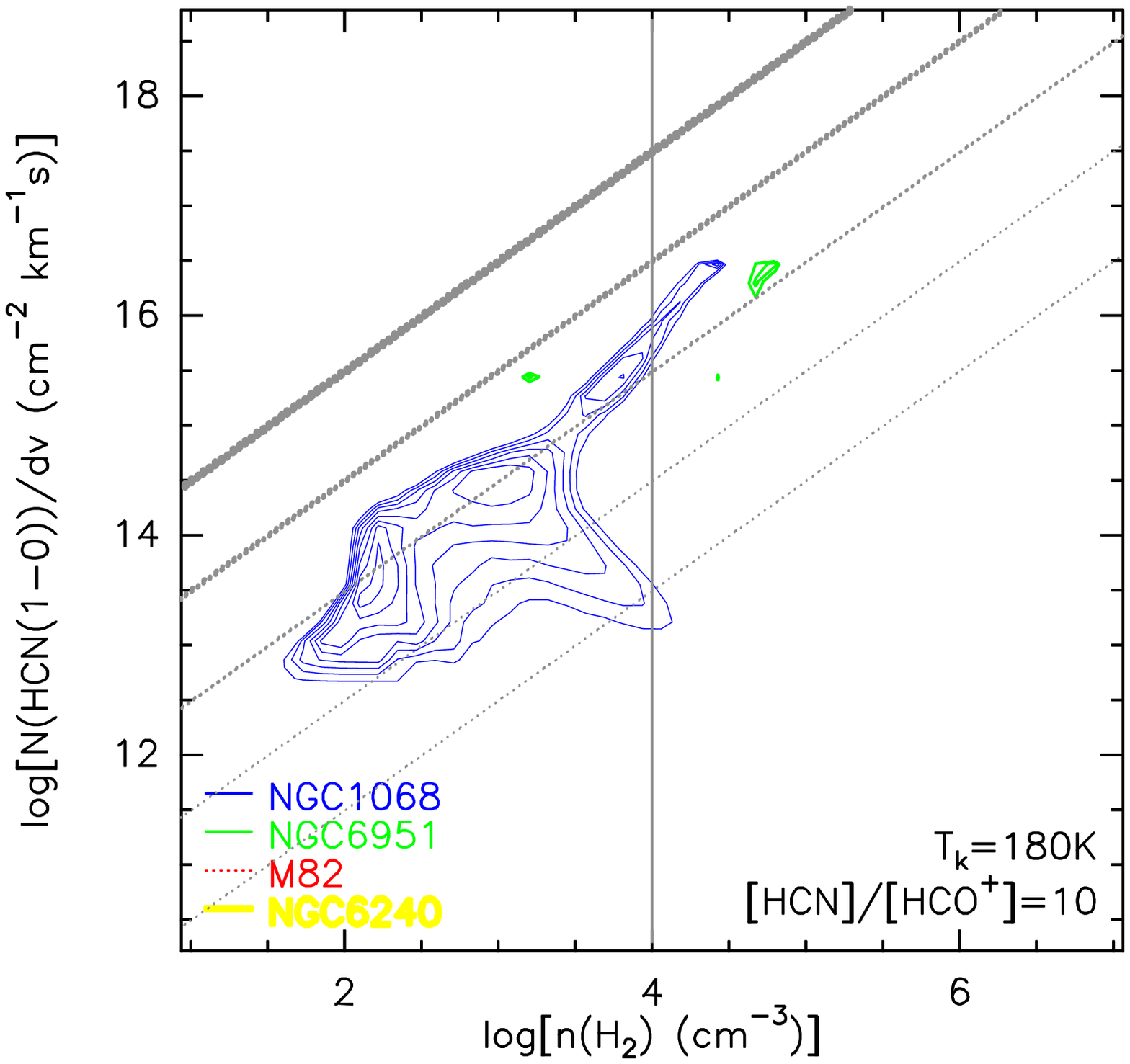}}}
\resizebox{5.9cm}{!}{\rotatebox{-90}{\includegraphics{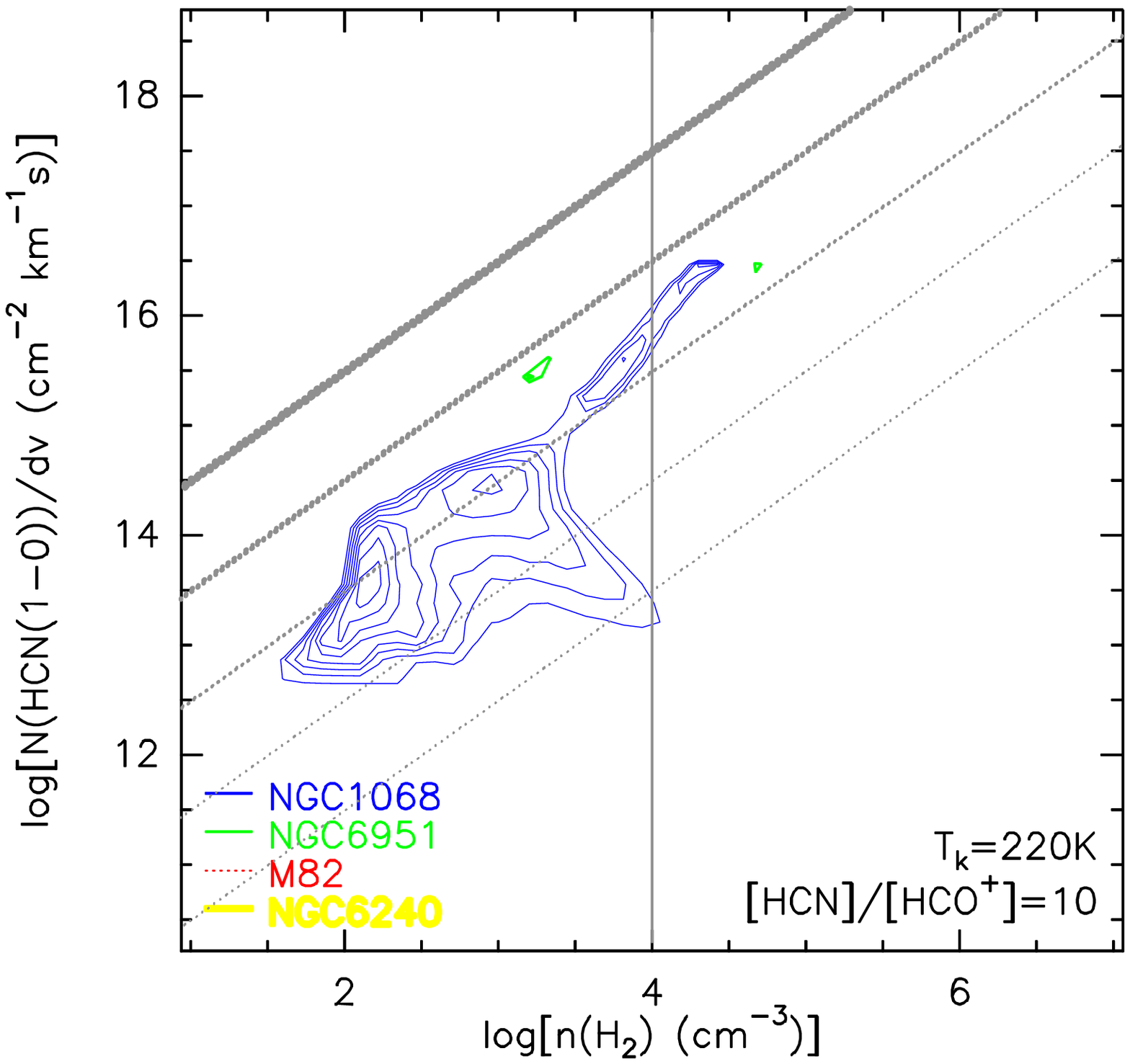}}}
\end{center}
\vskip -0.3cm
\resizebox{5.9cm}{!}{\rotatebox{-90}{\includegraphics{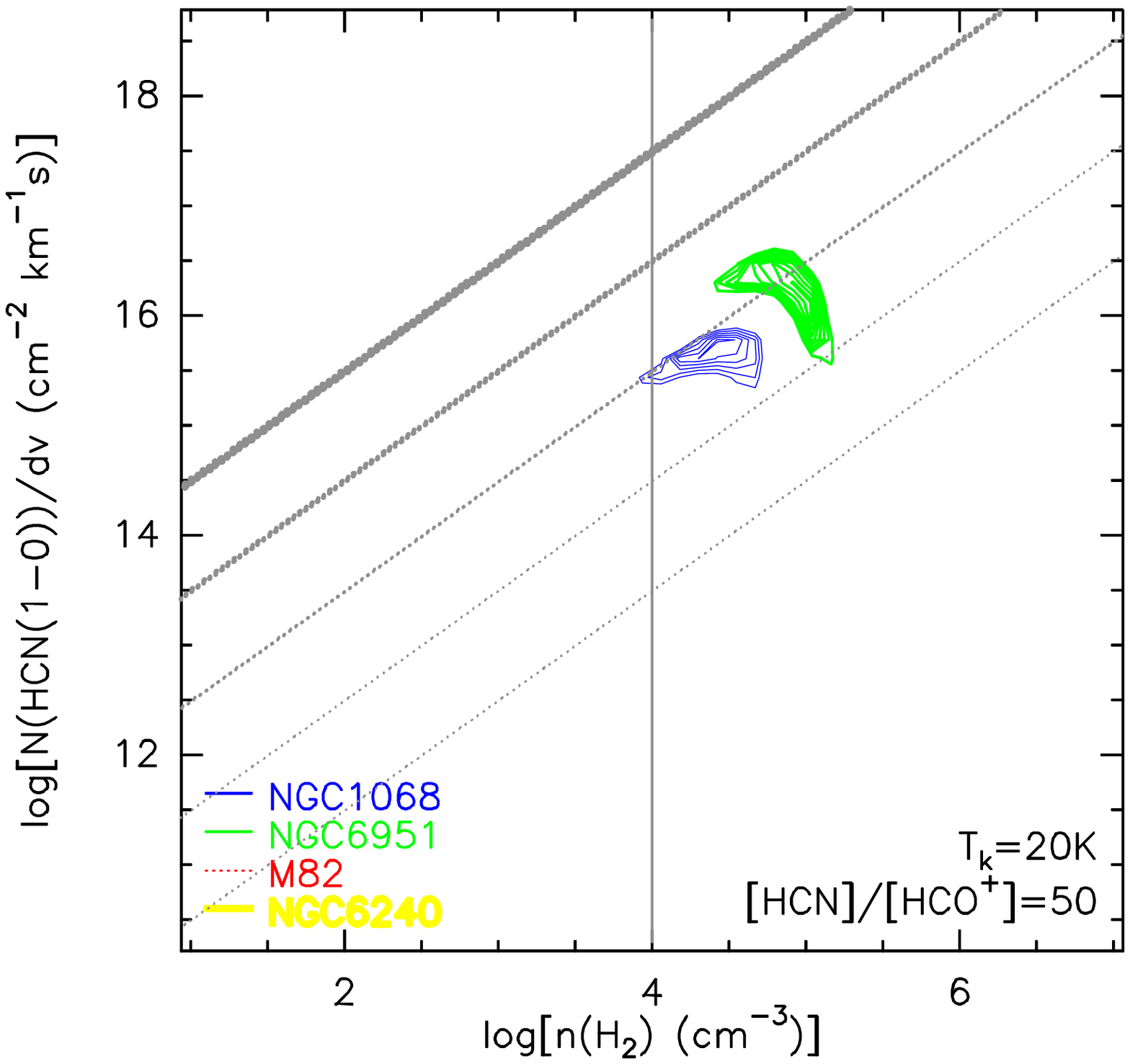}}}
      \caption{Same as Fig.~\ref{lvg-r1} but for abundance ratios of
      [HCN]/[HCO$^+$]=10 \& 50.}
         \label{lvg-r2}
\end{figure*}

Fig.~\ref{lvg-r1} \& \ref{lvg-r2} and Table~\ref{tab5} show the
results of the LVG analysis with the lowest $\chi^2$-values.  Please
note that we find two solutions respectively with similarly low
$\chi^2$-values for NGC~1068 and NGC~6951.  The contours in the plots
encircle regions with $\chi^2$$\leq$1 and are color-coded following
the previous Figures (NGC~1068: blue; NGC~6951: green; M82: red;
NGC~6240: yellow). The HCN abundance is additionally indicated in grey
lines decreasing in thickness from 10$^{-5}$(\kms\,/pc$^2$)$^{-1}$ to
10$^{-9}$(\kms\,/pc$^2$)$^{-1}$ in logarithmic steps of 1 based on
equation~7.  The standard (galactic) value for the HCN abundance found
in galactic clouds is Z$^{\rm s}$(HCN)=2$\times$10$^{-8}$.  Giving a
velocity range of $\sim$100-500~\kms\, (compare Table~\ref{tab2}) for
this sample and assuming typical sizes of $\sim$100-500~pc, we can
constrain dv/dr to be in the range of $\sim$0.2-5\kms/pc. This results
in a range of Z$^{\rm a}$
(HCN)/(dv/dr)$\simeq$10$^{-9}$-10$^{-7}$~(\kms\,/pc$^2$)$^{-1}$ if
assuming standard HCN abundances. However, for the AGN sources in our
sample dv/dr is probably rather $\sim$1-5 (large velocity width
($\sim$300~\kms) and quite compact gas regions ($\leq$100~pc)) as
opposed to $\sim$0.2-1 for the (local) SB(+AGN) dominated sources
(covering a more extended region of a few 100~pc than the AGN; e.g.,
M82 or NGC~6951), i.e.,
\begin{equation}
\rm Z^s(HCN)/(dv/dr)_{\rm AGN}\lesssim2\times10^{-8}~(km~s^{-1}~pc^2)^{-1}
\hspace*{0.4cm} 
\end{equation}
while
\begin{equation}
\rm Z^s(HCN)/(dv/dr)_{\rm SB}\gtrsim 2 \times10^{-8} (km~s^{-1}~pc^2)^{-1}
\hspace*{0.4cm} 
\end{equation}
if taking a standard Z$^{\rm s}$(HCN).

The LVG analysis most strikingly constrains the molecular gas
densities in the AGN and SB(+AGN) sources. While HCN and HCO$^+$
emission in the SB dominated sources, such as M82 and NGC~6951, seem
to emerge from regions with high H$_2$ densities in the range of
n(H$_2$)=(10$^4$-10$^{6.5}$)cm$^{-3}$, the HCN and HCO$^+$ emission in
sources of our sample with a pure AGN seem to be restricted to regions
with gas densities of n(H$_2$)$\leq$10$^{4.5}$cm$^{-3}$. The LVG
analysis also restricts the kinetic gas temperatures to be below
$<$120~K for the SB sources while the pure AGN cases appear to have no
upper limit for T$_{\rm k}$. The possibility of significantly higher
temperatures in AGN has been already discussed for NGC~5194 by
\cite{mat98} to explain the $^{13}$CO emission in this source. A high
kinetic temperature of T$_{\rm k}$$>$70~K has been also reported for
NGC~1068 before by \cite{ste94}. We furthemore find tendentially
larger HCN abundances in the AGN sources than in the SB sources which
may even lie significantly above Z$^{\rm s}$(HCN) by a factor of
$\sim$2-200. M82 thereby seems to denote a different extreme, i.e.,
having an HCN abundance that appears to be lower by at least an order
of magnitude than Z$^{\rm s}$(HCN). We also find an extremely low
[HCN]/[HCO$^+$] abundance ratio of $\leq$1 that implies a very low HCN
abundance and/or an additionally increased HCO$^+$ abundance. The
[HCN]/[HCO$^+$] abundance ratios seem to be around 10 (or more
conservatively: 1$<$[HCN]/[HCO$^+$]$\leq$50) for the rest.

The LVG analysis does not yield a very good solution for the HCO$^+$
emission in NGC~6240, which seems to stand out a little bit with its
intensity ratios, especially in R$^{\rm HCO^+/HCN}_{10/10}$, compared
to the other two ULIRGs. The intensity ratios in Arp~220 and Mrk~231
are more like NGC~6951 in their ratios. This might indicate that a
one-component model may be too simplistic in this case, while it
reproduces well the values of the rest of the sample. \cite{gre06}
present a two-phase LVG model for NGC~6240 (and Arp~220) that seems to
fit nicely their data including also several transition of HCN and
HCO$^+$. However, our results on the gas densities and temperatures
are not inconsistent with their findings, although our n(H$_2$) seems
a little bit lower than theirs; this may be though a consequence of
our one-component model as their two phase model assumes a very dense
plus a moderately dense gas phase which should result in an averaged
and thus lower density in a one-phase model.

The systematically increased HCN abundance in the AGN sources of our
sample implies that a mass determination through HCN in AGN may
significantly overestimate the dense molecular gas mass \citep[compare
  also][]{kra07} when using the same conversion factor as for
SBs/ULIRGs. Based on the LVG analysis and adopting HCN abundances in
the range of
Z(HCN)/(dv/dr)$\approx$(0.1-10)$\times$10$^{-7}$~pc/(\kms) with
T$_{\rm k}$=40-240~K for AGN, we find brightness temperatures (T$_{\rm
  b}$) in the range of $\sim$10-100~K and H$_2$ gas densities of
n(H$_2$)=10$^{2.0}$-10$^{4.5}$~cm$^{-3}$. Taking the molecular mass
(M$_{\rm H_2}$[$M_\odot$]) to HCN luminosity ratio (L$_{\rm
  HCN}$[${\rm (K~km~s^{-1}~pc^2)}$]) of \citep[e.g.,][]{sol92,sol90}:
\begin{equation}
{\rm X(HCN)\equiv M_{\rm H_2}/L_{\rm HCN}=2.1\times n(H_2)^{0.5}/T_{\rm b},} 
\end{equation}
we obtain X(HCN)=10$^{+10}_{-7}$~$M_\odot~({\rm
  K~km~s^{-1}~pc^2})^{-1}$ for the AGN sources. This conversion factor
is $\sim$2 times smaller than that derived for ULIRGs by \cite{sol92}
and \cite{gre07} of X(HCN)=20$^{+30}_{-10}$~$M_\odot~({\rm
  K~km~s^{-1}~pc^2})^{-1}$ but agrees with that favored by
\cite{gaoa04} for their sample of nearby active galaxies and
ULIRGs. An estimate of X(HCN) for the SB dominated sources in our
sample yields a similar value to that of \cite{sol92} and
\cite{gre07}.

By comparing the brightness temperatures with the obtained kinetic
temperatures (for the J=1--0 lines), we can give a very crude estimate
of the optical depth $\tau$, assuming $\tau$=$-$ln(1$-$T$_{\rm
  b}$/T$_{\rm k}$). This yields $\tau$$\approx$0.4-1.0 for the SB
dominated sources, $\tau$$\approx$0.3-1.4 for the AGN sources,
$\tau$$\approx$0.3-1.8 for the composite sources and
$\tau$$\approx$0.7-1.0 for NGC6240, indicating similar opacity
ranges. However, the ratio between T$_{\rm b}$ and T$_{\rm k}$ could
be easily biased if the emission is very clumpy, artificially lowering
the T$_{\rm b}$/T$_{\rm k}$ ratio. As we probably average over many
giant molecular cloud complexes, a clumpy structure may be indeed
suspected in both SB galaxies
\citep[e.g.,][]{zhang01,alonso02,wilson03,galliano05} and also for the
gas/dust emission around AGN.

\section{Discussion}

\subsection{Chemical layout and excitation conditions}

We briefly described various physical processes in the Introduction
that can lead to the observed differences in the line and transition
ratios between AGN and SB dominated galaxies. As first scenario, we
mentioned gas densities and temperature effects (cases (a) and (b) in
the Introduction): higher gas densities and/or temperatures tend to
increase populations of higher-J CO levels and may lead, for a given
column density, to a reduced CO(J=1-0) line intensity (assuming that
the gas densities are not that high to have a similar effect on the
HCN emission). Given the {\it high} HCN-to-CO(J=1--0) intensity ratios
in AGN, this would consequently mean that the gas densities and/or
temperatures in AGN must be higher than around SB activity.  We can
clearly exclude higher (average) gas densities in AGN, as our LVG
analysis yields quite 'low' gas densities of
n(H$_2$)$<$10$^{4.5}$cm$^{-3}$ in the AGN sources. However, within the
30m beam, we potentially average over many giant molecular cloud
complexes (GMCs) which could each actually be denser than their
average, if, for instance, the GMC number density is not very high and
lower than that in the SB sources. Thus, one explanation for the
'lower' average gas densities may be that the molecular gas traced by
the HCN and HCO$^+$ emission does not split up into as many high
density clumps/GMCs in our AGN sources as it seems to be the case in
SB environments
\citep[compare][]{zhang01,alonso02,wilson03,galliano05}.

An increased HCN abundance in AGN compared to SB environments could
equally explain the observed differences in the line ratios.  A
chemical enhancement of HCN can be created in two different ways:
either through far-ultraviolet radiation from O and B stars in young
high-mass star forming regions \citep[e.g.,][]{tie85,bla87,ste95}, or
through strong X-ray (ir)radiation from an AGN
\citep[e.g.,][]{lep96,ma96}. While UV radiation affects primarily the
surfaces of gas clouds in the circumnuclear regions ($\leq$1kpc),
creating Photon Dissociation Regions (PDRs), X-rays penetrate deeply
into the circumnuclear disk (CND), forming huge X-ray Dissociation
Regions (XDRs). As a consequence of this volume versus surface effect,
the X-ray radiation from the AGN might thereby produce higher HCN
abundances relative to CO than the UV radiation of SBs which may
explain the significantly higher HCN-to-CO(J=1--0) luminosity ratios
found in AGN \citep[e.g.,][]{koh01,use04,ima06,gra06}. This scenario
seems to be consistent with the estimated abundance ratios in our
sample. They appear to be higher in the AGN than in the SB dominated
sources in our sample.

Alternatively, the CO abundance might be smaller in AGN than around SB
activity either because of oxygen depletion
\citep[e.g.,][]{ste96,sha96} or the influence of X-rays from the AGN
\citep[e.g.][]{mei05}. This could also result in the apparently higher
HCN-to-CO ratios in AGN. The oxygen depletion has, however, been
already ruled out for the XDR in NGC~1068 because it would produce
different HCO$^+$-to-HCN(J=1--0) ratios than observed (e.g., Usero et
al.\ 2004; Kohno et al.\ 2005). On the other hand, at the presence of
a strong X-ray radiation field, CO dissociation may occur more
frequently predicting similar HCO$^+$-to-HCN intensity ratios to those
observed in NGC~1068 and NGC~5194 and eventually also a decreased CO
abundance \citep[e.g.][]{mei05}. This scenario can neither be
supported nor discarded with our current data and has to be
investigated further.

Supernovae explosions (SNe), or more generally, ionisation effects
from cosmic rays \citep[e.g.,][]{dic80,woo81,ngu92,wil92}, are
suspected to significantly increase the HCO$^+$ abundance while
potentially decreasing the HCN abundance, yielding thus higher
HCO$^+$-to-HCN intensity ratios in evolved SBs (higher frequency of
SNe) than in AGN. The most prominent examples for the role of the
evolutionary stage of a SB are M82 and NGC~253; the starburst in
NGC~253 is supposed to be in an evolutionary stage prior to that one
in M82 and its molecular gas appears to be mainly dominated by shocks
\citep[e.g.,][]{mar06} rather than PDRs/SNe in contrast to M82
\citep[e.g.,][]{gar02}. The HCO$^+$-to-HCN intensity ratio is observed
to be significantly lower in NGC~253 than in the PDR of M82 (see
Fig.~\ref{ratios3}). In this context, the results obtained on the HCN
and HCO$^+$ abundances in the evolved SB, PDR-dominated M82 are in
excellent agreement with what is expected from theoretical
predictions; M82 shows a very low [HCN]/[HCO$^+$] abundance ratio of
$\leq$1. This may also explain the difference to the other SB sources
in our sample who may all be in an earlier SB phase than M82, and
emphasize the dependence of the excitation and chemistry on the
evolutionary stage of the SB, whether reflected in more pronounced
shocks or PDRs/SNe dominance. This is of particular importance as it
sets tight constraints on the comparison to AGN dominated sources.

As a further alternative, dust heating, coupled to gas density and
caused by the UV/X-ray radiation of the AGN, together with infrared
(IR) radiative pumping could also be the origin of the stronger HCN
emission in XDRs although, if true, a tighter correlation of the HCN
luminosity with the IR luminosity than with the FIR luminosity would
be expected as well as a strong correlation between the IR and X-ray
luminosity. Such correlations have not yet been found
\citep[e.g.,][]{lut04,gaoa04,gaob04} but X-ray absorption in compton
thick sources and variability effects might lead to a large scatter in
the data used so far, thus washing out possible correlations. However,
IR pumping should similarly affect the HCN {\it and} HCO$^+$ emission,
which both have vibrational modes at similar wavelengths of
12-14~$\mu$m, as recently discussed by \citep{gue07}. Thus,
HCO$^+$-to-HCN ratios close to unity would be expected, in case IR
pumping is significant, implying that non-collisional excitation of
HCN and HCO$^+$ in sources with low HCO$^+$-to-HCN ratios cannot be a
dominant process. Also, the HCO$^+$-to-HCN intensity ratios in the AGN
sources even decrease with increasing J; this would not be expected if
IR pumping were significant.  However, recent observations of the HNC
emission in some ULIRGs and a high redshift quasar
\citep[e.g.,][]{aal07,gue07} indicate that IR pumping may well be a
significant factor in the excitation of at least HNC in sources with
compact IR nuclei, but HNC (its vibrational state is at 24~$\mu$m)
seems to be more easily excitable through IR pumping than HCN and
HCO$^+$ \citep[e.g.,][]{aal07,gue07}. HNC observations of our sample
will be discussed in a future paper.

We note that, although the statistics in our sample may be still
small, especially for the AGN sources, the trend of different line
ratios between AGN and SB dominated sources, seen in our data, is very
pronounced and also further supported by literature data. Not only
fall sources such as NGC~253 and NGC~4569 in the same line ratio
regions as our SB sources, but the results on the AGN sources are also
encouraged by recent results on the center of NGC~6951 \citep{kra07}
as well as that on NGC~1097 and NGC~5033
\citep[e.g.][]{koh05,koh07}. However, more AGN dominated galaxies in
our sample would certainly be beneficial to substantiate our results.

\subsection{Comparison to theoretical results}

The lower gas densities in our AGN sources may actually resolve a
'conflict' introduced by recent theoretical studies
\citep{mei05,mei06,mei07} on XDR and PDR environments. While these
authors claim that the HCO$^+$-to-HCN intensity ratios are actually
$>$1 in XDRs, we rather find {\it low} HCO$^+$-to-HCN intensity ratios
as opposed to their results. However, as emphasized in Meijerink et
al.\ (2007), high HCO$^+$-to-HCN intensity ratios are expected in high
density regions, which does not seem to be the case for the 100~pc
scale disks/regions in NGC~1068 and NGC~5194. For lower density
regions, Meijerink et al\ (2007) indeed find low HCO$^+$-to-HCN
intensity ratios in agreement with our results. Also, their models
predict higher surface temperatures in XDRs for strong X-ray
radiations fields but low gas densities when compared to PDRs. This
indicates a more efficient gas heating in XDRs than PDRs. The high
kinetic temperatures resulting from our LVG analysis for the AGN
sources fit nicely into this picture and are even more supported by
recent results found in nearby AGN through the CO(J=3--2)
emission. Matsushita et al.\ (2005) detect stronger central emission
through CO(J=3--2) than through CO(J=1--0) in their SMA sample of
several nearby Seyfert galaxies. This may indicate that temperature
effects play a non-negligible role for the high HCN-to-CO(J=1--0)
intensity ratios in AGN. This would also explain the decrease of the
HCN-to-CO intensity ratios with increasing rotational number J in our
sample for the AGN sources. However, we also find significantly
different HCN abundances that may support an HCN enhancement through
the strong X-ray radiation field in AGN in addition to a temperature
effect.

\subsection{Implication for the SFR - dense gas relation in galaxies}
\cite{gaoa04,gaob04} find a strong correlation between the star
formation rate (SFR), being proportional to the infrared luminosity,
and the dense gas mass, traced by the HCN luminosity. However, the
results on the AGN dominated galaxies of our sample may indicate that
this correlation is violated in certain AGN environments
\citep[compare also][]{koh07}. The HCN-to-CO(J=1--0) ratios for
NGC~1068, NGC~5194 and at the center of NGC~6951 \citep{kra07} do not
fall on the correlation of \cite{gaoa04,gaob04}. This may be a
consequence of the overabundance of HCN in the AGN sources of our
sample. As a potential alternative, the HCN-to-CO ratios at higher
transitions may be better suited for AGN environments than the
HCN-to-CO(J=1--0) ratios as they seem to coincide with the SFR-HCN
correlation.

\section{Summary and Conclusions}
\label{con}
We observed the centre of 12 nearby active galaxies in several
transitions of HCN and HCO$^+$ with the IRAM 30m telescope. The
results can be summarised as follows:

\begin{itemize}
\item[1.)] We find that the HCN intensity ratios vary significantly
  with activity type, i.e., depending on which power source dominates
  the central emission. The HCN (R$^{\rm HCN}_{\rm J_u,J_l/10}$),
  HCO$^+$ (R$^{\rm HCO^+}_{\rm J_u,J_l/10}$) and HCO$^+$-to-HCN
  (R$^{\rm HCO^+/HCN}_{\rm J_u,J_l/10}$) intensity ratios seem to
  increase with increasing SB contribution, in agreement with
  predictions from theoretical models. The highest intensity ratios
  are found in the evolved PDR-dominated starburst galaxy M82 while
  the lowest ratios are found in NGC~1068, a pure AGN source with a
  potential central XDR.

\item[2.)] We also find a variation in the intensity ratios among the
  starburst sources of our sample which may be explained by the
  evolutionary phase of the starburst, i.e., a differing dominance of
  shocks (pre-SB), hot cores (young SB), PDRs and SNe/cosmic rays
  (evolved SB).

\item[3.)] An LVG analysis of the HCN and HCO$^+$ data suggests that
  the SB dominated sources in our sample have high molecular gas
  densities around n(H$_2$)$\approx$10$^{4.0}$-10$^{6.5}$~cm$^{-3}$,
  kinetic gas temperatures of T$_{\rm k}$$\approx$20-120~K and HCN
  abundances of Z(HCN)$\approx$(0.001-2)$\times$10$^{-8}$, while the
  AGN dominated regions seem to show
  n(H$_2$)$\leq$10$^{4.5}$~cm$^{-3}$ and temperatures of T$_{\rm
    k}$$>$40~K with HCN abundances of
  Z(HCN)$\approx$(0.1-10)$\times$10$^{-7}$.

\item[4.)] The low HCO$^+$-to-HCN intensity ratios found in the AGN
  sources of our sample seem to make it unlikely that non-collisional
  excitation plays a significant role in AGN for the HCN and HCO$^+$
  emission. This may be further supported by the {\it decreasing}
  HCO$^+$-to-HCN intensity ratios with increasing rotational number J
  in the AGN sources of our sample.

\item[5.)] Assuming thus purely collisional excitation in AGN, we can
  exclude gas density effects as main cause for the higher
  HCN-to-CO(J=1--0) ratios found in AGN favorising hence an increased
  HCN abundance and/or temperature effects in AGN. The latter is
  supported by decreasing HCN-to-CO ratios with increasing rotational
  number J.

\item[6.)] To explain the potential differences in gas densities
  between AGN and SB environments of our sample, we favor a scenario
  in which the molecular gas may be either significantly less clumpy,
  i.e., has a lower 'clump/GMC number density' in AGN than SB
  environments or, alternatively, is more continously smeared over AGN
  environments, i.e., has indeed a lower gas density.

\item[7.)] An estimate of the dense molecuar mass to HCN luminosity (
  M$_{\rm H_2}$-to-L$_{\rm HCN}$) conversion factor X(HCN) results in
  X(HCN)=10$^{+10}_{-7}$~$M_\odot~({\rm K~km~s^{-1}~pc^2})^{-1}$ for
  the AGN sources in our sample. This is a factor of $\sim$2 lower
  than that found by \cite{sol92,gre07} for ULIRGs and starburst
  sources of X(HCN)=20$^{+30}_{-10}$~$M_\odot~({\rm ~
    km~s^{-1}~pc^2})^{-1}$, but consistent with the value favored by
  \cite{gaoa04} for their sample of nearby active galaxies and ULIRGs.

\item[8.)] The overabundance of HCN found in the AGN sources of our
  sample indicates that the correlation between SFR and HCN luminosity
  may be violated in the vicinity of an AGN. Given the decreasing
  nature of the HCN-to-CO ratio with increasing transition, we suggest
  to use the J=3--2 transition of HCN and CO as an alternative.

\end{itemize}

\acknowledgments 

The results in this paper are based on observations carried out with
the IRAM 30m telescope in Spain. IRAM is supported by INSU/CNRS
(France), MPG (Germany) and IGN (Spain). We are grateful to the
dedicated staff at the IRAM 30m telescope who have been very
supportive and helpful during the observations. We thank the anonymous
referee for very helpful suggestions on improving the paper.


\clearpage

\begin{deluxetable}{ccccc} 
\tabletypesize{\small} 
\tablecaption{Basic properties of the sample. } 
\tablewidth{0pt} \tablehead{Source & z & RA & Dec & Type$^a$ } 
\startdata 
NGC1068 & 0.0038 & 02:42:40.7 & -00:00:47.9 & AGN \\ 
NGC5194 & 0.0015 & 13:29:52.7 & +47:11:42.6 & AGN \\ 
\hline
NGC4826 & 0.0014 & 12:56:43.8 & +21:40:58.9 & AGN+SB? \\ 
NGC3627 & 0.0024 & 11:20:15.0 & +12:59:29.5 & AGN+SB \\ 
NGC4569 & 0.0008 & 12:36:49.8 & +13:09:46.3 & AGN+SB \\ 
NGC6951 & 0.0047 & 20:37:14.5 & +66:06:19.7 & AGN+SB \\ 
\hline
NGC6946 & 0.0002 & 20:34:52.3 & +60:09:14.2 & SB \\
NGC2146 & 0.0031 & 06:18:37.7 & +78:21:25.3 & SB \\ 
M82     & 0.0007 & 09:55:52.2 & +69:40:46.9 & SB \\ 
\hline
NGC6240 & 0.0245 & 16:52:58.9 & +02:24:03.4 & ULIRG(AGN+SB) \\ 
Mrk231  & 0.0422 & 12:56:14.2 & +56:52:25.2 & ULIRG(AGN+SB) \\ 
Arp220  & 0.0181 & 15:34:57.1 & +23:30:11.5 & ULIRG(AGN+SB) \\
\enddata 
\tablenotetext{a}{type of activity found in the central
$\pm$15$''$. AGN denotes thereby Seyfert or LINER galaxies. Please
note that NGC~1068 for instance also contains a SB ring but on
much larger scales (i.e., $\sim\pm$20$''$). NGC~6951 also houses a
strong SB which is still within the 30m beam as opposed to
NGC~1068. Most classifications are from the NED; composite nature
(AGN+SB): NGC~3627 \citep{dah94}; NGC~4569 (Gabel et al.\ 2002);
NGC~4826 (Garc\'{\i}a-Burillo et al.\ 2003); NGC~6951 (see text);
NGC~6240 (Tecza et al.\ 2000; Gallimore \& Beswick 2004); Mrk231 (Smith
et al.\ 1999).}
\label{tab1}
\end{deluxetable}

\begin{deluxetable}{cccc} 
\tabletypesize{\small} \tablecaption{Observational parameters as taken
from the IRAM 30m homepage. }
\tablewidth{0pt} \tablehead{Observed & F$_{\rm eff}$ & B$_{\rm eff}$ &
$\theta_{\rm beam}$ \\ 
Frequency & & & ($''$)\\} 
\startdata
89~GHz  & 0.95 & 0.78 & 29.5 \\
177~GHz & 0.93 & 0.65 & 14.0 \\
267~GHz & 0.88 & 0.46 & 9.5\\
\enddata 
\label{tab0}
\end{deluxetable}

\begin{deluxetable}{ccccccccccc}
\tabletypesize{\small}
\centering
\tablecaption{Line parameteres for the HCN and HCO$^+$ data. }
\tablewidth{0pt}
\tablehead{\colhead{Source} 
& \colhead{I$_{10}^{\rm HCN}$\tablenotemark{\,\,\,\,a}} 
& \colhead{I$_{21}^{\rm HCN}$\tablenotemark{\,\,\,\,a}} 
& \colhead{I$_{32}^{\rm HCN}$\tablenotemark{\,\,\,\,a}}  
& \colhead{$\Delta$v\tablenotemark{b}} 
& \colhead{$\theta_s^{10}$$^c$} 
& \colhead{$\theta_s^{21}$$^c$} 
& \colhead{$\theta_s^{32}$$^c$}  
& \colhead{f$_{10}$\tablenotemark{d}} 
& \colhead{f$_{21}$\tablenotemark{d}} 
& \colhead{f$_{32}$\tablenotemark{d}} \\
 & \colhead{K\ \kms} & \colhead{K\ \kms} & \colhead{K~\kms}  & \colhead{\kms} & \colhead{['']} 
& \colhead{['']} &\colhead{['']} & \colhead{} & \colhead{} & \colhead{} }
\startdata
NGC1068 & 24.5$\pm$0.9 & 20.0$\pm$0.4 &  19.0$\pm$0.6 & 220$\pm$10 &  4.5 &  3.0 &  3.0 &  0.09 & 0.16 & 0.29 \\
NGC5194 &  4.7$\pm$0.2 &  2.2$\pm$0.6 &   $<$2.0      & 120$\pm$10 & 15   & 10   & 10   &  0.53 & 0.67 & 0.83 \\
\hline	   
NGC4826 &  6.0$\pm$0.4 &  4.7$\pm$0.4 &   3.4$\pm$0.4 & 300$\pm$20 & 20   & 20   & 20   &  0.67 & 0.91 & 0.91  \\
NGC3627 &  2.7$\pm$0.2 &  $<$2.0      &   $<$3.0      & 290$\pm$30 &  8   &  8   &  8   &  0.23$^f$ & 0.56$^f$ & 0.71$^f$ \\
NGC4569 &  2.8$\pm$0.1 &  2.6$\pm$0.4 &  2.3$\pm$0.5  & 210$\pm$30 & 10   & 10   & 10   &  0.32$^f$ & 0.67$^f$ & 0.83$^f$ \\
NGC6951 &  3.1$\pm$0.1 &  5.0$\pm$0.7 &  3.6$\pm$1.0  & 300$\pm$20 & 17   & 17   & 17   &  0.59 & 0.83 & 0.91 \\
\hline
NGC6946 &  9.9$\pm$0.1 & 11.3$\pm$0.6 &  9.2$\pm$0.6  & 150$\pm$5  & 10   & 10   & 10   &  0.32$^f$ & 0.67$^f$ & 0.83$^f$ \\
NGC2146 &  5.0$\pm$0.1 &  4.4$\pm$0.3 &  4.3$\pm$0.4  & 290$\pm$10 & 20   & 20   & 20   &  0.67$^f$ & 0.91$^f$ & 0.91$^f$ \\
M82\tablenotemark{e} 
        & 29$\pm$0.2   & 26$\pm$0.6   & 27$\pm$1      & 130$\pm$5  & $-$   & $-$   & $-$  &  1 & 1 & 1 \\
\hline
NGC6240 &  3.2$\pm$0.2 &  8.8$\pm$0.3 &  12.3$\pm$0.8 & 410$\pm$20 &  3   &  3   &  3   &  0.07 & 0.24 & 0.42 \\
Mrk231  &  1.9$\pm$0.1 &  4.9$\pm$1.0 &  9.3$\pm$0.6  & 220$\pm$20 &  3   &  3   &  3   &  0.04 & 0.16 & 0.29 \\
Arp220  &  9.7$\pm$0.4 & 28.4$\pm$0.7 & 43.0$\pm$1.0  & 530$\pm$20 &  2   &  2   &  2   &  0.02 & 0.08 & 0.15 \\
\hline
\hline
& \raisebox{-0.2cm}{I$_{10}^{\rm HCO^+}$\tablenotemark{\,\,\,\,a}} 
&  
& \raisebox{-0.2cm}{I$_{32}^{\rm HCO^+}$\tablenotemark{\,\,\,\,a}} 
& \raisebox{-0.2cm}{$\Delta$v\tablenotemark{b}} 
& \multicolumn{6}{c}{\raisebox{-0.2cm}{same as above}} \\
&\raisebox{-0.1cm}{ K\ \kms} & & \raisebox{-0.1cm}{K\ \kms} & \raisebox{-0.1cm}{K~\kms} \\[0.2cm]
\hline

NGC1068 &  14.6$\pm$0.2     & &  7.6$\pm$0.8     & 234$\pm$4  \\
NGC5194 &   2.4$\pm$0.1     & &  $<$1.3          & 134$\pm$7  \\  
\hline	   
NGC4826 &   3.5$\pm$0.1     & &  $<$2.5          & 300$\pm$30 \\
NGC3627 &   2.7$\pm$0.2     & &  $<$2.1          & 230$\pm$15 \\
NGC4569 &   2.3$\pm$0.2     & &  $<$2.1          & 211$\pm$17 \\
NGC6951 &   2.2$\pm$0.1     & &  $<$2.9          & 300$\pm$20 \\
\hline
NGC6946 &   8.5$\pm$0.1     & &  8.6$\pm$0.6     & 150$\pm$3  \\
NGC2146 &   6.3$\pm$0.2     & &  5.2$\pm$0.6     & 300$\pm$20 \\
M82     &  40.2$\pm$0.2     & & 34.0$\pm$2.0     & 130$\pm$10 \\
\hline
NGC6240 &   5.0$\pm$0.5$^g$ & &  8.0$\pm$1.0$^g$ & $-$          \\
Mrk231  &   1.6$\pm$0.2$^g$ & &  3.8$\pm$0.4$^g$ & $-$          \\
Arp220  &   4.6$\pm$0.5$^g$ & &  8.8$\pm$0.9$^g$ & $-$          \\
\enddata

\tablenotetext{a}{Velocity integrated intensities, not yet corrected
  for filling factors; for a definition see equation~2. Error are
  purely statistical and were determined from the gaussian line fits to the
  data.}

\tablenotetext{b}{FWHM of HCN and HCO$^+$ line.}  

\tablenotetext{c}{Most of the source sizes have been determined either
from the HCN maps where accesible (e.g., NGC~6951: Krips et al.\ 2007b)
or from CO maps (NUGA project: Garc\'{\i}a-Burillo et al.\ 2004;
BIMA-SONG: Helfer et al.\ 2003).  Individual galaxies not found in the
BIMA SONG or NUGA survey: NGC~2146 (Greve et al.\ 2006); Mrk231,
Arp220 (Downes \& Solomon 1998); NGC~6946 (Schinnerer et al.\ 2007);
NGC~6240 (Iono et al.\ 2007). }

\tablenotetext{d}{For a definition see Equation~3. While for the
first one the major axis (FWHM) of the emission was chosen, the minor
axis was taken for the elliptical case.}

\tablenotetext{e}{All HCN and HCO$^+$ lines were averaged over the
same area of 30$''$ so that they all should have the same filling
factor assuming that they come from the same emission region. However,
it is not unprobable that the higher-J HCN and HCO$^+$ lines originate
from a more compact region than the \hcn\, and \hcop\, line emission
but this would make the ratios listed in Table~\ref{tab3} for M82 even
higher.}

\tablenotetext{f}{We assumed elliptical source sizes for the filling
  factor here (see definition in Equation~3). The minor axis of the
  emission was taken for the source size.}

\tablenotetext{g}{HCO$^+$ data taken from Graci\'a-Carpio et al.\
2006. We assumed a 10\% error for the integrated intensities here.}

\label{tab2}
\end{deluxetable}

\begin{deluxetable}{lccccc}
\tabletypesize{\small}
\tablewidth{0pt}
\tablecaption{Intensity and luminosity ratios.}
\tablehead{\colhead{Source} 
& \colhead{$R_{\rm 21/10}^{\rm HCN}$\tablenotemark{a}} 
& \colhead{$R_{\rm 32/10}^{\rm HCN}$\tablenotemark{a} } 
& \colhead{$R_{\rm 32/10}^{\rm HCO^+}$\tablenotemark{a}} 
& \colhead{$R_{\rm 10/10}^{\rm HCO^+/HCN}$\tablenotemark{a}}  
& \colhead{$R_{\rm 32/32}^{\rm HCO^+/HCN}$\tablenotemark{a}} }
\startdata
NGC1068 & 0.44$\pm$0.02 & 0.21$\pm$0.01 & 0.14$\pm$0.02     & 0.60$\pm$0.01     & 0.38$\pm$0.07 \\
NGC5194 & 0.3$\pm$0.1   & $<$0.23       & $<$0.2            & 0.72$\pm$0.06     & $<$0.7        \\
\hline
NGC4826 & 0.38$\pm$0.04 & 0.23$\pm$0.01 & $<$0.3            & 0.59$\pm$0.04     & $<$0.7        \\
NGC3627 & $<$0.4        & $<$0.4        & $<$0.3            & 1.0$\pm$0.1       & $<$0.7        \\
NGC4569 & 0.5$\pm$0.1   & 0.4$\pm$0.2   & $<$0.4            & 0.85$\pm$0.07     & $<$0.9        \\
NGC6951 & 0.7$\pm$0.1   & 0.4$\pm$0.1   & 0.44$\pm$0.02     & 0.70$\pm$0.05     & $<$0.8        \\
\hline
NGC6946 & 0.64$\pm$0.05 & 0.42$\pm$0.03 & 0.45$\pm$0.03     & 0.86$\pm$0.03     & 0.94$\pm$0.08 \\
NGC2146 & 0.67$\pm$0.04 & 0.57$\pm$0.05 & 0.51$\pm$0.06     & 1.30$\pm$0.08     & 1.2$\pm$0.2   \\
M82 (aver)$^b$        
        & 0.88$\pm$0.04 & 0.92$\pm$0.07 & 0.96$\pm$0.05     & 1.40$\pm$0.10     & 1.30$\pm$0.10 \\
M82 (ePDR)$^b$ 
        & 1.00$\pm$0.06 & 1.00$\pm$0.10 & 1.10$\pm$0.10     & 1.40$\pm$0.10     & 1.20$\pm$0.30 \\
M82 (centre)$^b$    
        & 0.70$\pm$0.1  & 0.50$\pm$0.05 & 0.40$\pm$0.10     & 1.50$\pm$0.04     & 1.30$\pm$0.10  \\
\hline
NGC6240 & 0.67$\pm$0.05 & 0.45$\pm$0.04 & 0.19$\pm$0.07$^c$ & 1.60$\pm$0.20$^c$ & 0.70$\pm$0.20$^c$ \\
Mrk231  & 0.62$\pm$0.07 & 0.58$\pm$0.05 & 0.30$\pm$0.10$^c$ & 0.87$\pm$0.09$^c$ & 0.40$\pm$0.10$^c$ \\
Arp220  & 0.69$\pm$0.03 & 0.50$\pm$0.02 & 0.21$\pm$0.06$^c$ & 0.48$\pm$0.05$^c$ & 0.20$\pm$0.05$^c$ \\
\hline
\multicolumn{6}{c}{\raisebox{-0.15cm}{LITERATURE DATA }}\\[0.2cm]
\hline
NGC253$^d$  & $-$          & $<$0.7       & 0.9$\pm$0.2 & 0.8$\pm$0.1 & $>$1.0 \\
IC342$^e$   & 0.8$\pm$0.05 & 0.4$\pm$0.04 & $-$         & 0.9$\pm$0.1 & $-$    \\
NGC4945$^f$ & $-$          & 0.7$\pm$0.1  & 0.4$\pm$0.1 & 1.1$\pm$0.1 & 0.7$\pm$0.1 \\
\enddata
\tablenotetext{a}{corrected for beam filling effects using
Equation~1 (see also Table~\ref{tab2}).}
\tablenotetext{b}{Intensity ratios of M82 averaged over the entire
disk ($\equiv$aver), at the position ($\equiv$($+$12$''$,$+$8$''$)) of
the eastern PDR ($\equiv$ePDR) and at the center
($\equiv$(0$''$,0$''$)). The ratios for the eastern PDR are given
because of consistency reason.}

\tablenotetext{c}{HCO$^+$ values taken from Graci\'a-Carpio et al.\
2006.}
\tablenotetext{d}{SB galaxy. Data taken from Martin et al.\ in (prep.?).}
\tablenotetext{e}{SB galaxy. Data taken from \cite{nguyen94} and \cite{schulz01}.}
\tablenotetext{f}{SB galaxy. taken from \cite{wang04}}
\label{tab3}
\end{deluxetable}

\begin{deluxetable}{ccccc}
\tabletypesize{\small}
\tablewidth{0pt}
\tablecaption{Intensity and luminosity ratios.}
\tablehead{\colhead{Source} 
& \colhead{$R_{\rm 21/10}^{\rm CO}$\tablenotemark{a}} 
& \colhead{$\frac{L_{\rm HCN10}}{L_{\rm CO10}}$\tablenotemark{b} }
& \colhead{$\frac{L_{\rm HCN21}}{L_{\rm CO21}}$\tablenotemark{b}}  
& \colhead{$\frac{L_{\rm HCN32}}{L_{\rm CO32}}$\tablenotemark{b}} }
\startdata
NGC1068 & 0.4 & 1.9  & 1.1    & 0.5    \\
NGC5194 & 0.6 & 0.5  & 0.2    & $<$0.3 \\
\hline
NGC4826 & 0.5 & 0.4  & 0.3    & -      \\
NGC3627 & 0.6 & 0.1  & $<$0.1 & $<$0.1 \\
NGC4569 & 0.6 & 0.2  & 0.1    & -      \\
NGC6951 & 0.7 & 0.2  & 0.2    & -      \\
\hline
NGC6946 & 0.4 & 0.2  & 0.3    & 0.3    \\
NGC2146 & 0.3 & 0.01 & 0.02   & 0.03   \\
M82     & 0.9 & 0.2  & 0.2    & 0.1    \\
\hline
NGC6240 & 0.3 & 0.1  & 0.1    & -      \\
Mrk231  & 0.1 & 0.1  & 0.3    & -      \\
Arp220  & 0.2 & 0.1  & 0.3    & 0.3    \\
\enddata
\tablenotetext{a}{corrected for beam filling effects using
Equation~1 (see also Table~\ref{tab2}).}

\tablenotetext{b}{CO luminosities were taken from IRAM 30m
observations of NGC~3627, NGC~4569, NGC~4826 and NGC~6951 in context
of the NUGA project. The CO(1--0) and CO(2--1) fluxes for NGC~1068
were derived from interferometric maps (taken from Schinnerer et al.\
2000 and Krips et al., 2007a) while the CO(3--2) emission has been
taken with APEX during its commissioning which is consistent with
published values. The rest has been taken from the literature: Arp220
and Mrk231 from Radford et al.\ 1991 (CO(1--0) and CO(2--1)) and
Narayanan et al.\ 2005 (CO(3--2)); NGC~6240 from Solomon et al.\ 1997
(CO(1--0)) and Tacconi et al.\ 1999 (CO(2--1)); NGC~2146 and NGC~6946
from Gao \& Solomon 2004a (CO(1--0)), Braine et al.\ 1993 (CO(2--1)),
Casoli et al.\ 1991 (CO(2--1)) and Dumke et al.\ 2001 (CO(3--2));
NGC~5149 from Nakai et al.\ 1994 (CO(1--0) and CO(2--1)) and Dumke et
al.\ 2001 (CO(3--2)); M82 from Walter et al.\ 2002 (CO(1--0)), Thuma
et al.\ 2000, (CO(2--1)) and Dumke et al.\ 1994 (CO(3--2)). CO(3--2)
data on NGC~3627 are from Petitpas et al.\ (2005).}
\label{tab4}
\end{deluxetable}

\begin{deluxetable}{lccccc}
\tabletypesize{\small}
\tablewidth{0pt}
\tablecaption{Results of the LVG analysis in dependency of the 
dominant activity type.}
\tablehead{\colhead{Example}
& \colhead{T$_{\rm k}$\tablenotemark{a}}
& \colhead{n(H$_2$)\tablenotemark{b}}
& \colhead{N(HCN)/(dv)\tablenotemark{c}}
& \colhead{Z(HCN)/(dv/dr)\tablenotemark{d}}
& \colhead{[HCN]/[HCO$^+$]\tablenotemark{e}}\\
  \colhead{source}
& \colhead{K} 
& \colhead{cm$^{-3}$}
& \colhead{~cm$^{-2}$\, km$^{-1}$\ s} 
& \colhead{~km$^{-1}$\ s\ pc} 
& \colhead{}} 
\startdata
NGC1068 &  20     & 10$^{4.0}$$-$10$^{4.5}$ & 10$^{15}$$-$10$^{16}$ & 10$^{-8 }$$-$10$^{-7}$ & 50 \\
        &  60-240 & 10$^{2.0}$$-$10$^{4.5}$ & 10$^{13}$$-$10$^{16}$ & 10$^{-8 }$$-$10$^{-6}$ & 10 \\
\hline			     		      	      	 		      	      
NGC6951 &  20     & 10$^{4.5}$$-$10$^{5.0}$ & 10$^{15}$$-$10$^{17}$ & 10$^{-8 }$$-$10$^{-7}$ & 50 \\
        &  20-120 & 10$^{4.0}$$-$10$^{6.0}$ & 10$^{13}$$-$10$^{16}$ & 10$^{-9 }$$-$10$^{-7}$ & 10 \\
\hline			     		      	      	 		      	      
M82(aver) &  60-100 & 10$^{5.0}$$-$10$^{6.5}$ & 10$^{15}$$-$10$^{16}$ & 10$^{-10}$$-$10$^{-8}$ & 0.01-1 \\
\hline			     		      	      	 		      	      
NGC6240  & 20-120 & 10$^{3.5}$$-$10$^{4.5}$ & 10$^{15}$$-$10$^{16}$ &  10$^{-7 }$$-$10$^{-6}$ & 10 \\
\enddata
\tablenotetext{a}{Kinetic gas temperature.}
\tablenotetext{b}{H$_2$ density.}
\tablenotetext{c}{HCN column density.}
\tablenotetext{d}{HCN abundance per velocity gradient.}
\tablenotetext{e}{Abundance ratio between HCN and HCO$^+$.}
\label{tab5}
\end{deluxetable}

\clearpage

\end{document}